\documentclass[a4paper,11pt]{article}

\usepackage[utf8]{inputenc} 
\usepackage[T1]{fontenc}  
\usepackage{ae,aecompl,aeguill}   

\usepackage{verbatim} 

\usepackage{graphicx}
\usepackage{dcolumn}
\usepackage{rotating}
\usepackage{epstopdf}
\usepackage{array}
\newcolumntype{H}{>{\setbox0=\hbox\bgroup}c<{\egroup}@{}}
\usepackage{mdframed}

\usepackage{amstext,amsmath,amssymb,amsfonts}
\usepackage{bm}
\usepackage{empheq}

\usepackage{fancyhdr}

\newcommand{\captionfonts}{\small}
\makeatletter  
\long\def\@makecaption#1#2{%
  \vskip\abovecaptionskip
  \sbox\@tempboxa{{\captionfonts #1: #2}}%
  \ifdim \wd\@tempboxa >\hsize
    {\captionfonts #1: #2\par}
  \else
    \hbox to\hsize{\hfil\box\@tempboxa\hfil}%
  \fi
  \vskip\belowcaptionskip}
\makeatother   

\usepackage{cite} 
\usepackage{hyperref}

\newlength{\bibitemsep}
\newlength{\bibparskip}
\let\oldthebibliography\thebibliography
\renewcommand\thebibliography[1]{%
  \oldthebibliography{#1}%
  \setlength{\parskip}{\bibitemsep}%
  \setlength{\itemsep}{\bibparskip}%
}

\usepackage{color}
\definecolor{lightblue}{rgb}{0.2,0.2,0.7}
\definecolor{darkblue}{rgb}{0,0.25,0.5}
\definecolor{redbrown}{rgb}{0.875,0.25,0.125}
\definecolor{darkgreen}{rgb}{0,0.5,0}

\newcommand{\bra}[1]{\ensuremath{\langle #1 \vert}}
\newcommand{\ket}[1]{\ensuremath{\vert #1  \rangle}}
\newcommand{\braket}[2]{\ensuremath{\langle  #1 \vert #2  \rangle}}
\renewcommand{\b}[1]{\ensuremath{\mathbf{#1}}}
\renewcommand{\H}{\ensuremath{\text{H}}}
\newcommand{\s}{\ensuremath{\text{s}}}

\newcommand{\Hx}{\ensuremath{\text{Hx}}}
\newcommand{\Hxc}{\ensuremath{\text{Hxc}}}
\newcommand{\xc}{\ensuremath{\text{xc}}}
\newcommand{\x}{\ensuremath{\text{x}}}
\renewcommand{\c}{\ensuremath{\text{c}}}

\renewcommand{\l}{\ensuremath{\lambda}}
\newcommand{\w}{\ensuremath{\omega}}
\newcommand{\tne}{\ensuremath{\text{ne}}}
\newcommand{\ee}{\ensuremath{\text{ee}}}

\newcommand{\lr}{\ensuremath{\text{lr}}}
\newcommand{\sr}{\ensuremath{\text{sr}}}
\newcommand{\KS}{\ensuremath{\text{KS}}}

\newcommand{\UEG}{\ensuremath{\text{UEG}}}
\newcommand{\LDA}{\ensuremath{\text{LDA}}}
\newcommand{\LSDA}{\ensuremath{\text{LSDA}}}

\newcommand{\GGA}{\ensuremath{\text{GGA}}}
\newcommand{\mGGA}{\ensuremath{\text{mGGA}}}
\newcommand{\mKS}{\ensuremath{\text{mKS}}}
\newcommand{\HF}{\ensuremath{\text{HF}}}
\newcommand{\MP}{\ensuremath{\text{MP2}}}
\newcommand{\GL}{\ensuremath{\text{GL2}}}

\newcommand{\dRPA}{\ensuremath{\text{dRPA}}}

\renewcommand{\d}{\ensuremath{\text{d}}}

\renewcommand{\d}{\ensuremath{\text{d}}}
\DeclareMathOperator{\erf}{erf}

\newcommand{\mut}{\ensuremath{\tilde{\mu}}}

\newcommand{\Psil}{\ensuremath{\Psi^{\l}}}

\newcommand{\g}{\ensuremath{\gamma}}

\newcommand{\Tr}{\ensuremath{\text{Tr}}}
\newcommand{\isEquivTo}[1]{\underset{#1}{\sim}}
\newcommand{\isPropTo}[1]{\underset{#1}{\propto}}

\usepackage{fancyhdr}

\usepackage{indentfirst}
\numberwithin{equation}{section}
\newcounter{myequation}
\makeatletter
\@addtoreset{equation}{myequation}
\makeatother

\begin{document}

\title{Review of approximations for the exchange-correlation energy in density-functional theory}

\author{Julien Toulouse\\
Laboratoire de Chimie Th\'eorique (LCT), Sorbonne Universit\'e and CNRS, F-75005 Paris, France\\
Institut Universitaire de France, F-75005 Paris, France}

\date{31 August, 2022}

\maketitle
\thispagestyle{fancy}

In this chapter, we provide a review of ground-state Kohn--Sham density-functional theory of electronic systems and some of its extensions, we present exact expressions and constraints for the exchange and correlation density functionals, and we discuss the main families of approximations for the exchange-correlation energy: semilocal approximations, single-determinant hybrid approximations, multideterminant hybrid approximations, dispersion-corrected approximations, as well as orbital-dependent exchange-correlation density functionals. The chapter aims at providing both a consistent bird's-eye view of the field and a detailed description of some of the most used approximations. It is intended to be readable by chemists/physicists and applied mathematicians. For more coverage of the subject, the reader may consult for example the books of Refs.~\cite{ParYan-BOOK-89,DreGro-BOOK-90,PerEngDreGroGodNogCasMar-BOOK-03,EngDre-BOOK-11,Tsu-BOOK-14} and the review articles of Refs.~\cite{PerSch-AIPCP-01,ScuSta-INC-05,Cap-BJP-06,CohMorYan-CR-12,Bur-JCP-12,Sta-INC-13,Bec-JCP-14,Jon-RMP-15,SuLiTru-JCP-16,MarHea-MP-17}.

\tableofcontents

\section{Basics of density-functional theory}
\stepcounter{myequation}

\subsection{The many-body problem}

We consider an $N$-electron system (atom or molecule) in the Born--Oppenheimer and non-relativistic approximations. The electronic Hamiltonian in the position representation is, in atomic units,
\begin{equation}
H = -\frac{1}{2} \sum_{i=1}^{N} \nabla^2_{\b{r}_i}  + \frac{1}{2} \sum_{i=1}^N \sum_{\substack{j=1\\i\ne j}}^{N} \frac{1}{|\b{r}_i -\b{r}_j|}+ \sum_{i=1}^{N} v_{\tne}(\b{r}_i),
\label{Hr}
\end{equation}
where $\nabla^2_{\b{r}_i} = \Delta_{\b{r}_i}$ is the Laplacian with respect to the electron coordinate $\b{r}_i$ and $v_{\tne}(\b{r}_i)=- \sum_{\alpha=1}^{N_\text{n}} Z_{\alpha}/|\b{r}_i - \b{R}_{\alpha}|$ is the nuclei-electron interaction depending on the positions $\{\b{R}_{\alpha}\}$ and charges $\{Z_{\alpha}\}$ of the $N_\text{n}$ nuclei. The stationary electronic states are determined by the time-independent Schrödinger equation,
\begin{equation}
H \Psi(\b{x}_1,\b{x}_2,...,\b{x}_N) = E \Psi(\b{x}_1,\b{x}_2,...,\b{x}_N),
\label{HrPsir}
\end{equation}
where $\Psi(\b{x}_1,\b{x}_2,...,\b{x}_N)$ is a wave function written with space-spin coordinates $\b{x}_i = (\b{r}_i,\sigma_i) \in \mathbb{R}^3 \times \{ \uparrow,\downarrow\}$ (with $\{ \uparrow,\downarrow\} \cong \mathbb{Z}_2$ being the set of spin coordinates) which is antisymmetric with respect to the exchange of two coordinates, and $E$ is the associated energy.

Using Dirac notation, the Schrödinger equation~(\ref{HrPsir}) can be rewritten in a representation-independent formalism,
\begin{equation}
\hat{H} \ket{\Psi} = E \ket{\Psi},
\label{HPsi}
\end{equation}
where the Hamiltonian is formally written as
\begin{equation}
\hat{H} = \hat{T} + \hat{W}_{\ee} + \hat{V}_{\tne},
\end{equation}
with the kinetic-energy operator $\hat{T}$, the electron-electron interaction operator $\hat{W}_{\ee}$, and the nuclei-electron interaction operator $\hat{V}_{\tne}$. 

The quantity of primary interest is the ground-state energy $E_0$. The variational theorem establishes that $E_0$ can be expressed as an infimum,
\begin{equation}
E_{0} = \inf_{\Psi \in {\cal W}^N} \bra{\Psi} \hat{H} \ket{\Psi},
\label{theovar}
\end{equation}
where the search is over the set of $N$-electron antisymmetric normalized wave functions $\Psi$ having a finite kinetic energy,
\begin{equation}
{\cal W}^N = \Big\{ \Psi \in \bigwedge^N L^2 (\mathbb{R}^3 \times \{ \uparrow,\downarrow\}; \mathbb{C}), \Psi \in H^1((\mathbb{R}^3 \times \{ \uparrow,\downarrow\})^N; \mathbb{C}), \braket{\Psi}{\Psi}=1 \Big\},
\label{W}
\end{equation}
where $\bigwedge^N$ is the $N$-fold antisymmetrized tensor product, $L^2$ and $H^1$ are the standard Lebesgue and Sobolev spaces (i.e., respectively, the space of functions that are square integrable and the space of functions that are square integrable together with their first-order derivatives), and $\braket{\cdot}{\cdot}$ designates the $L^2$ inner product. Density-functional theory (DFT) is based on a reformulation of the variational theorem in terms of the one-electron density defined as\footnote{In Eq.~(\ref{nfromPsi}), an integration over a spin coordinate $\sigma$ just means a sum over the two values $\sigma\in\{\uparrow,\downarrow\}$.}
\begin{equation}
\rho_\Psi(\b{r}) = N \int_{\{ \uparrow,\downarrow\} \times (\mathbb{R}^3 \times \{ \uparrow,\downarrow\})^{N-1}} \left| \Psi(\b{x},\b{x}_2,...,\b{x}_N) \right|^2 \d \sigma \d \b{x}_2 ... \d \b{x}_N,
\label{nfromPsi}
\end{equation}
which is normalized to the electron number, $\int_{\mathbb{R}^3} \rho_\Psi(\b{r}) \d\b{r} = N$.

\subsection{The universal density functional}

Building on the work of Hohenberg and Kohn~\cite{HohKoh-PR-64}, Levy~\cite{Lev-PNAS-79} and Lieb~\cite{Lie-IJQC-83} proposed to define the following universal density functional $F[\rho]$ using a {\it constrained-search} approach,
\begin{equation}
F[\rho] = \min_{\Psi \in {\cal W}^{N}_{\rho}}\bra{\Psi} \hat{T} + \hat{W}_\ee \ket{\Psi} = \bra{\Psi[\rho]} \hat{T} + \hat{W}_\ee \ket{\Psi[\rho]},
\label{FnLevy}
\end{equation}
where the minimization is done over the set of $N$-electron wave functions $\Psi$ yielding the fixed density $\rho$ [via Eq.~(\ref{nfromPsi})],
\begin{equation}
{\cal W}^N_\rho = \{ \Psi \in {\cal W}^N, \rho_\Psi = \rho\}.
\end{equation}
In Eq.~(\ref{FnLevy}), for a given density $\rho$, $\Psi[\rho]$ denotes a minimizing wave function, which is known to exist~\cite{Lie-IJQC-83} but is possibly not unique. This so-called Levy--Lieb functional $F[\rho]$ is defined on the set of $N$-representable densities~\cite{Lie-IJQC-83}:
\begin{eqnarray}
{\cal D}^N &=& \{ \rho \; | \; \exists \Psi \in {\cal W}^N \; \text{s.t.} \;  \rho_\Psi = \rho\}
\nonumber\\
&=& \{ \rho \in L^1(\mathbb{R}^3) \; | \; \rho \geq 0,  \int_{\mathbb{R}^3} \rho(\b{r}) \d \b{r} = N, \sqrt{\rho} \in H^1(\mathbb{R}^3) \}.
\end{eqnarray}

We note that an alternative universal density functional can be defined by a Legendre--Fenchel transformation, or equivalently by a constrained-search over $N$-electron ensemble density matrices~\cite{Lie-IJQC-83}. This so-called Lieb functional has the advantage of being convex but in this chapter we will simply use the Levy--Lieb functional of Eq.~(\ref{FnLevy}).

The exact ground-state energy can then be expressed as
\begin{eqnarray}
E_0 &=&  \inf_{\rho \in {\cal D}^N} \left\{ F[\rho] + \int_{\mathbb{R}^3} v_\tne(\b{r}) \rho(\b{r}) \d \b{r} \right\},
\label{E0minnLevy}
\end{eqnarray}
and if a minimizer exists then it is a ground-state density $\rho_0(\b{r})$ for the potential $v_\tne(\b{r})$. Hence, the ground-state energy can be in principle obtained by minimizing over the density $\rho$, i.e. a simple function of 3 real variables, which is a tremendous simplification compared to the minimization over a complicated many-body wave function $\Psi$. However, the explicit expression of $F[\rho]$ in terms of the density is not known, and the direct approximations for $F[\rho]$ that have been tried so far turn out not to be accurate enough.

If there is a unique wave function $\Psi[\rho]$ (up to a phase factor) in Eq.~(\ref{FnLevy}), we can define kinetic and potential contributions to $F[\rho]$,
\begin{equation}
F[\rho] = T[\rho] + W_\ee[\rho],
\label{}
\end{equation}
where $T[\rho] = \bra{\Psi[\rho]} \hat{T} \ket{\Psi[\rho]}$ and $W_\ee[\rho]=\bra{\Psi[\rho]} \hat{W}_\ee \ket{\Psi[\rho]}$. The kinetic-energy functional $T[\rho]$ is the contribution which is particularly difficult to approximate as an explicit functional of the density.

\subsection{The Kohn--Sham scheme}

\subsubsection{Decomposition of the universal functional}
\label{sec:decompF}

Following the idea of Kohn and Sham (KS)~\cite{KohSha-PR-65}, the difficulty of approximating $F[\rho]$ directly can be circumvented by decomposing $F[\rho]$ as
\begin{equation}
F[\rho] = T_\s[\rho] + E_\Hxc[\rho],
\label{FKS}
\end{equation}
where $T_\s[\rho]$ is the non-interacting kinetic-energy functional which can be defined with a constrained search\footnote{It is also possible to define the non-interacting kinetic-energy functional analogously to the Levy--Lieb functional in Eq.~(\ref{FnLevy}) by minimizing over wave functions $\Psi \in {\cal W}^N_\rho$, i.e. $T_{\s,\text{LL}}[\rho] = \min_{\Psi \in {\cal W}^N_\rho}\bra{\Psi} \hat{T} \ket{\Psi}$~\cite{Lie-IJQC-83}. In this case, the corresponding minimizing KS wave function can generally be a linear combination of Slater determinants. However, we often have $T_{\s,\text{LL}}[\rho]=T_{\s}[\rho]$, in particular for densities $\rho$ that come from a non-interacting ground-state wave function which is not degenerate. In this chapter, we will usually assume this nondegeneracy condition.
},
\begin{equation}
T_\s[\rho] = \min_{\Phi \in {\cal S}^N_\rho}\bra{\Phi} \hat{T} \ket{\Phi} = \bra{\Phi[\rho]} \hat{T} \ket{\Phi[\rho]},
\label{Ts}
\end{equation}
where the minimization is over the set of $N$-electron single-determinant wave functions $\Phi$ yielding the fixed density $\rho$:
\begin{equation}
{\cal S}^N_\rho = \{ \Phi \in {\cal S}^N, \rho_\Phi = \rho\}.
\end{equation}
Here, ${\cal S}^N$ is the set of $N$-electron single-determinant wave functions built from orthonormal spin orbitals
\begin{equation}
{\cal S}^N = \{ \Phi= \phi_{1} \wedge \phi_{2} \wedge ... \wedge \phi_{N} \; | \; \forall i \; \phi_{i} \in H^1(\mathbb{R}^3\times \{ \uparrow,\downarrow\}; \mathbb{C}), \; \forall i,j \;  \braket{\phi_{i}}{\phi_{j}}=\delta_{i,j} \},
\label{}
\end{equation}
where $\phi_{1} \wedge \phi_{2} \wedge ... \wedge \phi_{N}$ designates the normalized $N$-fold antisymmetrized tensor product of $N$ spin orbitals.
The functional $T_\s[\rho]$ is defined over the entire set of $N$-representable densities ${\cal D}^N$ since any $N$-representable density can be obtained from a single-determinant wave function~\cite{Gil-PRB-75,Har-PRA-81,Lie-IJQC-83}. In Eq.~(\ref{Ts}), for a given density $\rho$, $\Phi[\rho]$ denotes a minimizing single-determinant wave function (again known to exist~\cite{Lie-IJQC-83} but possibly not unique), also referred to as {\it KS wave function}. The remaining functional $E_\Hxc[\rho]$ that Eq.~(\ref{FKS}) defines is called the Hartree-exchange-correlation functional. The idea of the KS scheme is then to use the exact expression of $T_\s[\rho]$ by reformulating the minimization over densities in Eq.~(\ref{E0minnLevy}) as a minimization over single-determinant wave functions $\Phi$,
\begin{equation}
E_0 = \inf_{\Phi \in {\cal S}^N} \left\{ \bra{\Phi} \hat{T} + \hat{V}_\tne \ket{\Phi} + E_\Hxc[\rho_\Phi] \right\},
\label{E0minKS}
\end{equation}
and if a minimum exists then any minimizing single-determinant wave function in Eq.~(\ref{E0minKS}) gives a ground-state density $\rho_0(\b{r})$. Thus, the exact ground-state energy can in principle be obtained by minimizing over single-determinant wave functions only. Even though a wave function has been reintroduced compared to Eq.~(\ref{E0minnLevy}), it is only a single-determinant wave function $\Phi$ and therefore it still represents a tremendous simplification over the usual variational theorem involving a correlated (multideterminant) wave function $\Psi$. The advantage of Eq.~(\ref{E0minKS}) over Eq.~(\ref{E0minnLevy}) is that a major part of the kinetic energy can be treated exactly with the single-determinant wave function $\Phi$, and only $E_\Hxc[\rho]$ needs to be approximated as an explicit functional of the density.

In practice, $E_\Hxc[\rho]$ is decomposed as
\begin{equation}
E_\Hxc[\rho] = E_\H[\rho] + E_\xc[\rho],
\label{EHxc}
\end{equation}
where $E_\H[\rho]$ is the Hartree energy functional,
\begin{equation}
E_\H[\rho] = \frac{1}{2} \int_{\mathbb{R}^3\times\mathbb{R}^3} \frac{\rho(\b{r}_1) \rho(\b{r}_2)}{|\b{r}_1 -\b{r}_2|} \d \b{r}_1 \d \b{r}_2,
\label{EHn}
\end{equation}
representing the classical electrostatic repulsion energy for the charge distribution $\rho(\b{r})$ and which is calculated exactly, and $E_\xc[\rho]$ is the exchange-correlation energy functional that remains to be approximated. If there is a unique KS wave function $\Phi[\rho]$ (up to a phase factor), we can further decompose $E_\xc[\rho]$ as
\begin{equation}
E_\xc[\rho] = E_\x[\rho] + E_\c[\rho],
\label{Exc}
\end{equation}
where $E_\x[\rho]$ is the exchange energy functional,
\begin{eqnarray}
E_\x[\rho] &=& \bra{\Phi[\rho]} \hat{W}_\ee \ket{\Phi[\rho]} - E_\H[\rho],
\label{Exn}
\end{eqnarray}
and $E_\c[\rho]$ is the correlation energy functional,
\begin{equation}
E_\c[\rho] = \bra{\Psi[\rho]} \hat{T} + \hat{W}_\ee \ket{\Psi[\rho]}  -  \bra{\Phi[\rho]} \hat{T} + \hat{W}_\ee \ket{\Phi[\rho]}  = T_\c[\rho] + U_\c[\rho],
\label{}
\end{equation}
which contains a kinetic contribution $T_\c[\rho] =  \bra{\Psi[\rho]} \hat{T} \ket{\Psi[\rho]}  -  \bra{\Phi[\rho]} \hat{T} \ket{\Phi[\rho]}$ and a potential contribution $U_\c[\rho] = \bra{\Psi[\rho]} \hat{W}_\ee \ket{\Psi[\rho]}  -  \bra{\Phi[\rho]} \hat{W}_\ee \ket{\Phi[\rho]}$. Using the fact that $\Phi[\rho]$ is a single-determinant wave function, it can be shown that the exchange functional can be expressed as
\begin{equation}
E_\x[\rho]  = -\frac{1}{2} \sum_{\sigma \in \{\uparrow,\downarrow\}} \int_{\mathbb{R}^3\times\mathbb{R}^3} \frac{|\gamma_{\sigma}(\b{r}_1,\b{r}_2)|^2}{|\b{r}_1-\b{r}_2|} \d\b{r}_1 \d\b{r}_2,
\label{Exgamma}
\end{equation}
where $\gamma_{\sigma}$, for $\sigma\in\{\uparrow,\downarrow\}$, is the spin-dependent one-particle KS density matrix,
\begin{equation}
\gamma_{\sigma}(\b{r},\b{r}') = N \int_{(\mathbb{R}^3 \times \{ \uparrow,\downarrow\})^{N-1}} \Phi[\rho](\b{r}',\sigma,\b{x}_2,...,\b{x}_N)^* \; \Phi[\rho](\b{r},\sigma,\b{x}_2,...,\b{x}_N)  \d \b{x}_2 ... \d \b{x}_N,
\label{gammafromPhi}
\end{equation}
which shows that $E_\x[\rho] \leq 0$. Moreover, from the variational definition of $F[\rho]$, we see that $E_\c[\rho] \leq 0$.

\subsubsection{The Kohn--Sham equations}

The single-determinant wave function $\Phi$ in Eq.~(\ref{E0minKS}) is constructed from a set of $N$ orthonormal occupied spin-orbitals $\{\phi_i\}_{i=1,...,N}$. To enforce ${S}_z$ spin symmetry, each spin-orbital is factorized as $\phi_i(\b{x}) = \varphi_i(\b{r}) \chi_{\sigma_i}(\sigma)$ where $\varphi_i \in H^1(\mathbb{R}^3,\mathbb{C})$ is a spatial orbital and $\chi_{\sigma_i}$ is a spin function from $\{ \uparrow,\downarrow\}$ to $\{ 0,1\}$ such that $\forall \sigma_i,\sigma \in \{ \uparrow,\downarrow\}, \, \chi_{\sigma_i}(\sigma)=\delta_{\sigma_i,\sigma}$ ($\sigma_i$ is the spin of the spin-orbital $i$). Alternatively, when this is convenient, we will sometimes reindex the spatial orbitals, $\{\varphi_{i}\} \longrightarrow \{\varphi_{i \sigma}\}$, including explicitly the spin $\sigma$ in the index. Writing the total electronic energy in Eq.~(\ref{E0minKS}) in terms of spin-orbitals and integrating over the spin variables, we obtain:
\begin{equation}
E[\{\varphi_i\}] = \frac{1}{2} \sum_{i=1}^N \int_{\mathbb{R}^3} |\nabla \varphi_i(\b{r})|^2 \d\b{r} + \int_{\mathbb{R}^3} v_{\tne}(\b{r}) \rho(\b{r}) \d\b{r} + E_\Hxc[\rho],
\label{EKS}
\end{equation}
where the density is expressed in terms of the orbitals as
\begin{equation}
\rho(\b{r}) = \sum_{i=1}^N \left| \varphi_i(\b{r}) \right|^2.
\label{nphi}
\end{equation}
The minimization over $\Phi$ can then be recast into a minimization of $E[\{\varphi_i\}]$ with respect to the spatial orbitals $\{\varphi_i\}$ with the constraint of keeping the orbitals orthonormalized. The stationary condition with respect to variations of $\varphi_i(\b{r})$ leads to the \textit{KS equations}~\cite{KohSha-PR-65},
\begin{equation}
\left( -\frac{1}{2}\nabla^2 +v_{\tne}(\b{r}) + v_\Hxc(\b{r}) \right) \varphi_i(\b{r}) = \varepsilon_i \varphi_i(\b{r}),
\label{KSeqs}
\end{equation}
where $\varepsilon_i$ is the Lagrange multiplier associated to the normalization condition of $\varphi_i$ and $v_\Hxc(\b{r})$ is the Hartree-exchange-correlation potential defined as the functional derivative of $E_\Hxc[\rho]$ with respect to $\rho(\b{r})$,
\begin{equation}
v_\Hxc(\b{r}) = \frac{\delta E_\Hxc[\rho]}{\delta \rho(\b{r})},
\label{VHxc}
\end{equation}
which is itself a functional of the density. The orbitals satisfying Eq.~(\ref{KSeqs}) are called the KS orbitals. They are the eigenfunctions of the KS one-electron Hamiltonian,
\begin{eqnarray}
h_\s(\b{r}) = -\frac{1}{2} \nabla^2 + v_\s(\b{r}),
\label{hs}
\end{eqnarray}
where 
\begin{eqnarray}
v_\s(\b{r})=v_{\tne}(\b{r}) + v_\Hxc(\b{r})
\label{vKS}
\end{eqnarray}
is the KS potential, and $\varepsilon_i$ are then the KS orbital energies. Note that Eq.~(\ref{KSeqs}) constitutes a set of coupled self-consistent equations since the potential depends on all the occupied orbitals $\{\varphi_i\}_{i=1,...,N}$ through the density [Eq.~(\ref{nphi})]. The operator $h_\s(\b{r})$ defines the KS system which is a system of $N$ non-interacting electrons in an effective external potential $v_\s(\b{r})$ ensuring that the density $\rho(\b{r})$ in Eq.~(\ref{nphi}) is the same as the exact ground-state density $\rho_0(\b{r})$ of the physical system of $N$ interacting electrons. The exact ground-state energy $E_0$ is then obtained by injecting the KS orbitals in Eq.~(\ref{EKS}). The other (unoccupied) eigenfunctions in Eq.~(\ref{KSeqs}) define virtual KS orbitals $\{\varphi_a\}_{a \geq N+1}$.

Note that to define the potential $v_\Hxc(\b{r})$ in Eq.~(\ref{VHxc}) a form of differentiability of the functional $E_\Hxc[\rho]$, also referred to as $v$-representability of the density, has been assumed. Justifying this is in fact subtle and has been debated~\cite{EngEng-PSS-84a,EngEng-PSS-84b,LinSal-AQC-03,Lam-IJQC-07,KvaEksTeaHel-JCP-14,Hel-TALK-17,LaePenTelRugKvaHel-JCP-18} (see also the chapter by Kvaal in the volume). Here, we will simply assume that a form of differentiability of $E_\Hxc[\rho]$ holds on at least a restricted set of densities that allows one to define the potential $v_\Hxc(\b{r})$ up to an additive constant. For a further restricted set of densities that should include ground-state densities of electronic Hamiltonians of molecular systems [Eq.~(\ref{Hr})], it is expected that the KS potential $v_\s(\b{r})$ tends to a constant as $|\b{r}| \to \infty$ and we choose this constant to be zero. Note also that the assumption of the existence of the KS potential $v_\s(\b{r})$ in Eq.~(\ref{hs}), which does not depend on spin coordinates, implies that each spin-orbital must indeed have a definite ${S}_z$ spin value.

Following the decomposition of $E_\Hxc[\rho]$ in Eq.~(\ref{EHxc}), the potential $v_\Hxc(\b{r})$ is written as
\begin{eqnarray}
v_\Hxc(\b{r}) = v_\H(\b{r}) + v_\xc(\b{r}),
\end{eqnarray}
where $v_\H(\b{r}) = \delta E_\H[\rho]/\delta \rho(\b{r}) = \int_{\mathbb{R}^3} \rho(\b{r}')/|\b{r} -\b{r}'| \d \b{r}'$ is the Hartree potential and $v_\xc(\b{r}) = \delta E_\xc[\rho]/\delta \rho(\b{r})$ is the exchange-correlation potential. Likewise, following the decomposition of $E_\xc[\rho]$ in Eq.~(\ref{Exc}), and assuming that both $E_\x[\rho]$ and $E_\c[\rho]$ are differentiable with respect to $\rho$, the potential $v_\xc(\b{r})$ can be further decomposed as
\begin{eqnarray}
v_\xc(\b{r}) = v_\x(\b{r}) + v_\c(\b{r}),
\end{eqnarray}
where $v_\x(\b{r}) = \delta E_\x[\rho]/\delta \rho(\b{r})$ is the exchange potential and $v_\c(\b{r}) = \delta E_\c[\rho]/\delta \rho(\b{r})$ is the correlation potential. Thus, the KS equations are similar to the Hartree-Fock (HF) equations, with the difference that they involve a local exchange potential $v_\x(\b{r})$ instead of the nonlocal HF exchange potential, and an additional correlation potential. 
At least for ground-state densities of finite molecular systems, the exchange potential has the long-range asymptotic behavior (see, e.g., Ref.~\cite{GraKreKurGro-INC-00}),
\begin{eqnarray}
v_\x(\b{r}) \isEquivTo{|\b{r}| \to \infty} -\frac{1}{|\b{r}|},
\label{vxrinf}
\end{eqnarray}
whereas the correlation potential decays faster~\cite{AlmBar-PRB-85}.

\subsubsection{Extension to spin density-functional theory}
\label{extensionspin}

To deal with an external magnetic field, DFT has been extended from the total density to spin-resolved densities~\cite{BarHed-JPC-72,RajCal-PRB-73}. Without external magnetic fields, this spin density-functional theory is in principle not necessary, even for open-shell systems. However, the dependence on the spin densities allows one to construct approximate exchange-correlation functionals that are more accurate, and is therefore almost always used in practice for open-shell systems.

The spin density $\rho_{\sigma,\Psi}$ with $\sigma\in \{ \uparrow,\downarrow\}$ associated to a wave function $\Psi$ is defined as
\begin{equation}
\rho_{\sigma,\Psi}(\b{r}) = N \int_{(\mathbb{R}^3\times \{ \uparrow,\downarrow\})^{N-1}} \left| \Psi(\b{x},\b{x}_2,...,\b{x}_N) \right|^2 \d \b{x}_2 ... \d \b{x}_N,
\end{equation}
and integrates  to the number of $\sigma$-spin electrons $N_{\sigma}$, i.e. $\int_{\mathbb{R}^3} \rho_{\sigma,\Psi}(\b{r}) \d \b{r} = N_{\sigma}$. For $\rho_\uparrow \in {\cal D}^{N_\uparrow}$ and $\rho_\downarrow \in {\cal D}^{N_\downarrow}$, the universal density functional is now defined as~\cite{PerZun-PRB-81},
\begin{equation}
F[\rho_\uparrow,\rho_\downarrow] = \min_{\Psi \in {\cal W}^{N}_{\rho_\uparrow,\rho_\downarrow}}\bra{\Psi} \hat{T} + \hat{W}_\ee \ket{\Psi} = \bra{\Psi[\rho_\uparrow,\rho_\downarrow]} \hat{T} + \hat{W}_\ee \ket{\Psi[\rho_\uparrow,\rho_\downarrow]},
\label{FnLevyspin}
\end{equation}
where the search is over the set of normalized antisymmetric wave functions $\Psi$ with $N= N_\uparrow + N_\downarrow$ electrons and yielding the fixed spin densities $\rho_\uparrow$ and $\rho_\downarrow$
\begin{equation}
{\cal W}^{N}_{\rho_\uparrow,\rho_\downarrow}= \{ \Psi \in {\cal W}^{N}, \; \rho_{\uparrow,\Psi} = \rho_\uparrow, \; \rho_{\downarrow,\Psi} = \rho_\downarrow\}.
\end{equation}
In Eq.~(\ref{FnLevyspin}), $\Psi[\rho_\uparrow,\rho_\downarrow]$ designates a minimizing wave function.

A spin-dependent KS scheme is obtained by decomposing $F[\rho_\uparrow,\rho_\downarrow]$ as
\begin{equation}
F[\rho_\uparrow,\rho_\downarrow] = T_\s[\rho_\uparrow,\rho_\downarrow] + E_\H[\rho] + E_\xc[\rho_\uparrow,\rho_\downarrow],
\label{FKSspin}
\end{equation}
where $T_\s[\rho_\uparrow,\rho_\downarrow]$ is defined as 
\begin{equation}
T_\s[\rho_\uparrow,\rho_\downarrow]  = \min_{\Phi \in {\cal S}^{N}_{\rho_\uparrow,\rho_\downarrow}} \bra{\Phi} \hat{T} \ket{\Phi} = \bra{\Phi[\rho_\uparrow,\rho_\downarrow]} \hat{T} \ket{\Phi[\rho_\uparrow,\rho_\downarrow]},
\label{Tsspin}
\end{equation}
with a constrained search over the set of single-determinant wave functions $\Phi$ yielding the fixed spin densities $\rho_\uparrow$ and $\rho_\downarrow$
\begin{equation}
{\cal S}^{N}_{\rho_\uparrow,\rho_\downarrow} = \{ \Phi \in {\cal S}^{N}, \; \rho_{\uparrow,\Phi} = \rho_\uparrow, \; \rho_{\downarrow,\Phi} = \rho_\downarrow\}.
\end{equation}
Here, $\Phi[\rho_\uparrow,\rho_\downarrow]$ denotes a minimizing KS single-determinant wave function,
$E_\H[\rho]$ is the Hartree energy which is a functional of the total density $\rho=\rho_\uparrow+\rho_\downarrow$ only [Eq.~(\ref{EHn})], and $E_\xc[\rho_\uparrow,\rho_\downarrow]$ is the spin-dependent exchange-correlation energy functional. The ground-state energy is then obtained as
\begin{equation}
E_0 = \inf_{\Phi \in {\cal S}^N} \left\{ \bra{\Phi} \hat{T} + \hat{V}_\tne \ket{\Phi} + E_\H[\rho_\Phi] + E_\xc[\rho_{\uparrow,\Phi},\rho_{\downarrow,\Phi}] \right\}.
\label{E0minKSspin}
\end{equation}

Writing the spatial orbitals of the spin-unrestricted determinant as $\{\varphi_{i \sigma}\}_{i=1,...,N}$ (with the index explicitly including the spin $\sigma$ now for clarity), we arrive at the spin-dependent KS equations,
\begin{equation}
\left( -\frac{1}{2}\nabla^2 +v_{\tne}(\b{r}) + v_\H(\b{r}) + v_{\xc,\sigma}(\b{r}) \right) \varphi_{i\sigma}(\b{r}) = \varepsilon_{i\sigma} \varphi_{i\sigma}(\b{r}),
\label{KSeqsspin}
\end{equation}
with the spin-dependent exchange-correlation potential,
\begin{eqnarray}
v_{\xc,\sigma}(\b{r}) =  \frac{\delta E_\xc[\rho_\uparrow,\rho_\downarrow]}{\delta \rho_\sigma(\b{r})},
\end{eqnarray}
and the spin density,
\begin{eqnarray}
\rho_\sigma(\b{r}) = \sum_{i=1}^{N_\sigma} \left| \varphi_{i\sigma}(\b{r}) \right|^2.
\end{eqnarray}

As before, if there is a unique KS wave function $\Phi[\rho_\uparrow,\rho_\downarrow]$ (up to a phase factor), we can decompose $E_\xc[\rho_\uparrow,\rho_\downarrow]$ into exchange and correlation contributions,
\begin{equation}
E_\xc[\rho_\uparrow,\rho_\downarrow] = E_\x[\rho_\uparrow,\rho_\downarrow] + E_\c[\rho_\uparrow,\rho_\downarrow],
\label{Excspin}
\end{equation}
with $E_\x[\rho_\uparrow,\rho_\downarrow] =\bra{\Phi[\rho_\uparrow,\rho_\downarrow]} \hat{W}_\ee \ket{\Phi[\rho_\uparrow,\rho_\downarrow]} - E_\H[\rho]$. It turns out that the spin-dependent exchange functional $E_\x[\rho_\uparrow,\rho_\downarrow]$ can be exactly expressed in terms of the spin-independent exchange functional $E_\x[\rho]$~\cite{OliPer-PRA-79},
\begin{equation}
E_\x[\rho_\uparrow,\rho_\downarrow]  = \frac{1}{2} \left( E_\x[2\rho_\uparrow] + E_\x[2\rho_\downarrow] \right),
\label{Exspinscaling}
\end{equation}
which is known as the \textit{spin-scaling relation} and stems directly from the fact the $\uparrow$- and $\downarrow$-spin electrons are uncoupled in the exchange energy [see Eq.~(\ref{Exgamma})]. Therefore, any approximation for $E_\x[\rho]$ can be easily extended to an approximation for $E_\x[\rho_\uparrow,\rho_\downarrow]$. Unfortunately, there is no such relation for the spin-dependent correlation functional $E_\c[\rho_\uparrow,\rho_\downarrow]$.

Obviously, in the spin-unpolarized case, i.e. $\rho_\uparrow = \rho_\downarrow = \rho/2$, this spin-dependent formalism reduces to the spin-independent one.

\subsection{The generalized Kohn--Sham scheme}
\label{GKS}

An important extension of the KS scheme is the so-called {\it generalized Kohn--Sham} (GKS) scheme~\cite{SeiGorVogMajLev-PRB-96}, which recognizes that the universal density functional $F[\rho]$ of Eq.~(\ref{FnLevy}) can be decomposed in other ways than the KS decomposition of Eq.~(\ref{FKS}). In particular, we can decompose $F[\rho]$ as
\begin{eqnarray}
F[\rho] = \min_{\Phi \in {\cal S}^{N}_{\rho}} \left\{ \bra{\Phi} \hat{T} \ket{\Phi} + E_\H[\rho_\Phi] + S[\Phi] \right\} + \bar{S}[\rho],
\label{FGKS}
\end{eqnarray}
where $S[\Phi]$ is any functional of a single-determinant wave function $\Phi\in {\cal S}^{N}$ leading to a minimum in Eq.~(\ref{FGKS}), and $\bar{S}[\rho]$ is the corresponding complementary density functional that makes Eq.~(\ref{FGKS}) exact. Defining the $S$-dependent GKS exchange-correlation functional as
\begin{eqnarray}
E_{\xc}^{S}[\Phi] = S[\Phi] + \bar{S}[\rho_\Phi],
\label{ExcGKS}
\end{eqnarray}
we can express the exact ground-state energy as
\begin{eqnarray}
E_0 = \inf_{\Phi \in {\cal S}^N} \left\{ \bra{\Phi} \hat{T} + \hat{V}_\tne \ket{\Phi} + E_\H[\rho_\Phi] + E_\xc^S[\Phi] \right\},
\label{E0GKS}
\end{eqnarray}
and if a minimum exists then any minimizing single-determinant wave function in Eq.~(\ref{E0GKS}) gives a ground-state density $\rho_0(\b{r})$. Similarly to the KS equations [Eq.~(\ref{KSeqs})], Eq.~(\ref{E0GKS}) leads to the one-electron GKS equations,
\begin{equation}
\left( -\frac{1}{2}\nabla^2 +v_{\tne}(\b{r}) + v_\H(\b{r}) + v_{\bar{S}}(\b{r}) \right) \varphi_{i\sigma}(\b{r}) + \frac{\delta S[\Phi]}{\delta \varphi_{i\sigma}^*(\b{r})} = \varepsilon_{i\sigma} \varphi_{i\sigma}(\b{r}),
\label{GKSeqs}
\end{equation}
where $v_{\bar{S}}(\b{r}) = \delta \bar{S}[\rho]/\delta \rho(\b{r})$ is a local potential and $\delta S[\Phi]/\delta \varphi_{i\sigma}^*(\b{r})$ generates a one-electron (possibly nonlocal) operator.

In the special case $S[\Phi]=0$, we recover the KS exchange-correlation density functional:
\begin{eqnarray}
E_{\xc}^{S=0}[\Phi] =  E_{\xc}[\rho_\Phi].
\label{ExcS=0}
\end{eqnarray}
Due to the freedom in the choice of $S[\Phi]$, there is an infinity of GKS exchange-correlation functionals $E_{\xc}^{S}[\Phi]$ giving the exact ground-state energy via Eq.~(\ref{E0GKS}). This freedom and the fact that $\Phi$ carries more information than $\rho_\Phi$ gives the possibility to design more accurate approximations for the exchange-correlation energy.

Of course, by starting from the density functional $F[\rho_\uparrow,\rho_\downarrow]$ in Eq.~(\ref{FnLevyspin}), this GKS scheme can be extended to the spin-dependent case, leading to GKS exchange-correlation functionals of the form $E_{\xc}^{S}[\Phi] = S[\Phi] + \bar{S}[\rho_{\uparrow,\Phi},\rho_{\downarrow,\Phi}]$.

\section{Exact expressions and constraints for the Kohn--Sham exchange and correlation functionals}
\stepcounter{myequation}

\subsection{The exchange and correlation holes}
\label{exchangecorrelationholes}

Let us consider the pair density associated with the wave function $\Psi[\rho]$ defined in Eq.~(\ref{FnLevy}),
\begin{equation}
\rho_2(\b{r}_1,\b{r}_2) = N(N-1) \int_{\{ \uparrow,\downarrow\}^2 \times (\mathbb{R}^3\times\{ \uparrow,\downarrow\})^{N-2}} \left| \Psi[\rho](\b{x}_1,\b{x}_2,...,\b{x}_N) \right|^2 \d \sigma_1 \d \sigma_2 \d \b{x}_3 ... \d \b{x}_N,
\label{n2def}
\end{equation}
which is a functional of the density, and normalized to the number of electron pairs,\\ $\int_{\mathbb{R}^3\times\mathbb{R}^3} \rho_2(\b{r}_1,\b{r}_2) \d \b{r}_1 \d \b{r}_2 = N(N-1)$. The pair density is proportional to the probability density of finding two electrons at positions $(\b{r}_1,\b{r}_2)$ with all the other electrons being anywhere. The pair density is useful to express the expectation value of the electron-electron interaction operator,
\begin{equation}
\bra{\Psi[\rho]} \hat{W}_{\ee} \ket{\Psi[\rho]}  = \frac{1}{2} \int_{\mathbb{R}^3\times\mathbb{R}^3} \frac{\rho_2(\b{r}_1,\b{r}_2)}{|\b{r}_1-\b{r}_2|} \d\b{r}_1 \d\b{r}_2.
\label{PsiWeen2}
\end{equation}
Mirroring the decomposition of the Hartree-exchange-correlation energy performed in the KS scheme [Eq.~(\ref{EHxc})], the pair density can be decomposed as
\begin{equation}
\rho_2(\b{r}_1,\b{r}_2) = \rho(\b{r}_1) \rho(\b{r}_2) + \rho_{2,\xc}(\b{r}_1,\b{r}_2).
\label{n2decomp}
\end{equation}
The product of the densities $\rho(\b{r}_1) \rho(\b{r}_2)$ corresponds to the case of independent electrons [up to a change of normalization, i.e. $\int_{\mathbb{R}^3\times\mathbb{R}^3} \rho(\b{r}_1) \rho(\b{r}_2) \d \b{r}_1 \d \b{r}_2 = N^2$ instead of $N(N-1)$] and the exchange-correlation pair density $\rho_{2,\xc}(\b{r}_1,\b{r}_2)$ represents the modification of the pair density due to exchange and correlation effects between the electrons. It can be further written as
\begin{equation}
\rho_{2,\xc}(\b{r}_1,\b{r}_2) = \rho(\b{r}_1) h_{\xc}(\b{r}_1,\b{r}_2),
\label{n2xceqnnxc}
\end{equation}
where $h_{\xc}(\b{r}_1,\b{r}_2)$ is the {\it exchange-correlation hole}. Introducing the conditional density $\rho_{}^\text{cond}(\b{r}_1,\b{r}_2) =\rho_2(\b{r}_1,\b{r}_2)/\rho(\b{r}_1)$ of the remaining $N-1$ electrons at $\b{r}_2$ given that one electron has been found at $\b{r}_1$, the exchange-correlation hole can be interpreted as the modification of $\rho_{}^\text{cond}(\b{r}_1,\b{r}_2)$ due to exchange and correlation effects:
\begin{equation}
\rho_{}^\text{cond}(\b{r}_1,\b{r}_2) = \rho(\b{r}_2) + h_{\xc}(\b{r}_1,\b{r}_2).
\label{rhocond}
\end{equation}
The positivity of $\rho_2(\b{r}_1,\b{r}_2)$ implies that 
\begin{equation}
h_{\xc}(\b{r}_1,\b{r}_2) \ge -\rho(\b{r}_2).
\end{equation}
Moreover, from Eq.~(\ref{rhocond}), we have the following sum rule:
\begin{equation}
\forall \b{r}_1 \in \mathbb{R}^3, \; \int_{\mathbb{R}^3} h_{\xc}(\b{r}_1,\b{r}_2) \d\b{r}_2 = -1.
\label{intnxc}
\end{equation}

We can separate the exchange and correlation contributions in the exchange-correlation hole. For this, consider the pair density $\rho_{2,\KS}(\b{r}_1,\b{r}_2)$ associated with the KS single-determinant wave function $\Phi[\rho]$ defined in Eq.~(\ref{Ts}). It can be decomposed as
\begin{equation}
\rho_{2,\KS}(\b{r}_1,\b{r}_2) = \rho(\b{r}_1) \rho(\b{r}_2) + \rho_{2,\x}(\b{r}_1,\b{r}_2),
\label{n2KS}
\end{equation}
where $\rho_{2,\x}(\b{r}_1,\b{r}_2)$ is the exchange pair density, which is further written as
\begin{equation}
\rho_{2,\x}(\b{r}_1,\b{r}_2) = \rho(\b{r}_1) h_{\x}(\b{r}_1,\b{r}_2),
\label{n2x}
\end{equation}
where $h_{\x}(\b{r}_1,\b{r}_2)$ is the {\it exchange hole}. Just like the exchange-correlation hole, the exchange hole satisfies the conditions
\begin{equation}
h_{\x}(\b{r}_1,\b{r}_2) \ge -\rho(\b{r}_2),
\end{equation}
and
\begin{equation}
\forall \b{r}_1 \in \mathbb{R}^3, \; \int_{\mathbb{R}^3} h_{\x}(\b{r}_1,\b{r}_2) \d\b{r}_2 =  -1.
\label{intnx}
\end{equation}
Moreover, since the exchange hole can be written as [compare with Eq.~(\ref{Exgamma})]
\begin{equation}
h_{\x}(\b{r}_1,\b{r}_2)=-\frac{1}{\rho(\b{r}_1)}\sum_{\sigma \in \{\uparrow,\downarrow\}}|\gamma_{\sigma}(\b{r}_1,\b{r}_2)|^2,
\label{hxKSorb}
\end{equation}
where $\gamma_{\sigma}(\b{r}_1,\b{r}_2) = \sum_{i=1}^{N_\sigma} \varphi_{i \sigma}^*(\b{r}_2) \varphi_{i \sigma}(\b{r}_1)$ is the spin-dependent one-particle KS density matrix, it thus appears that the exchange hole is always non-positive,
\begin{equation}
h_{\x}(\b{r}_1,\b{r}_2) \le 0.
\label{nxneg}
\end{equation}
From Eqs.~(\ref{Exn}),~(\ref{PsiWeen2}),~(\ref{n2KS}), and~(\ref{n2x}), it can be seen that the exchange energy functional can be written in terms of the exchange hole,
\begin{equation}
E_\x[\rho]  = \frac{1}{2} \int_{\mathbb{R}^3\times\mathbb{R}^3} \frac{\rho(\b{r}_1) h_{\x}(\b{r}_1,\b{r}_2)}{|\b{r}_1-\b{r}_2|} \d\b{r}_1 \d\b{r}_2,
\label{Exint}
\end{equation}
leading to the interpretation of $E_{\x}$ as the electrostatic interaction energy of an electron and its exchange hole. It is useful to write the exchange energy functional as
\begin{eqnarray}
E_\x[\rho]  = \int_{\mathbb{R}^3} \rho(\b{r}_1) \varepsilon_\x [\rho](\b{r}_1) \d\b{r}_1,
\label{Exinteps}
\end{eqnarray}
where $\varepsilon_\x [\rho](\b{r}_1)$ is the exchange energy density per particle
\begin{eqnarray}
\varepsilon_\x [\rho](\b{r}_1) = \frac{1}{2} \int_{\mathbb{R}^3} \frac{h_{\x}(\b{r}_1,\b{r}_2)}{|\b{r}_1-\b{r}_2|} \d\b{r}_2,
\label{epsx}
\end{eqnarray}
which is itself a functional of the density. It is also convenient to define the exchange energy density $e_\x[\rho](\b{r}) = \rho(\b{r}) \varepsilon_\x [\rho](\b{r})$. 
For finite systems, we have the exact asymptotic behavior~\cite{Mar-PRA-87,Bec-PRA-88}
\begin{eqnarray}
\varepsilon_\x [\rho](\b{r}) \isEquivTo{|\b{r}| \to + \infty} -\frac{1}{2|\b{r}|}.
\label{epsxinf}
\end{eqnarray}

The {\it correlation hole} is defined as the difference
\begin{equation}
h_{\c}(\b{r}_1,\b{r}_2) = h_{\xc}(\b{r}_1,\b{r}_2) - h_{\x}(\b{r}_1,\b{r}_2),
\end{equation}
and, from Eqs.~(\ref{intnxc}) and~(\ref{intnx}), satisfies the sum rule
\begin{equation}
\forall \b{r}_1 \in \mathbb{R}^3, \; \int_{\mathbb{R}^3} h_{\c}(\b{r}_1,\b{r}_2) \d\b{r}_2 = 0,
\label{intnc}
\end{equation}
which implies that the correlation hole has negative and positive contributions\footnote{Therefore, the correlation hole is really a ``hole'' only in some region of space, and a ``bump'' in other regions.}. In contrast with the exchange hole which is a smooth function of the interelectronic coordinate $\b{r}_{12} =\b{r}_2 - \b{r}_1$, the correlation hole satisfies the electron-electron cusp condition (i.e., it has a derivative discontinuity in $\b{r}_{12}$)~\cite{Kim-PRA-73,Tew-JCP-08},
\begin{equation}
\forall \b{r}_1 \in \mathbb{R}^3, \; h_\c'(\b{r}_1,\b{r}_1) = h_\c(\b{r}_1,\b{r}_1),
\label{hccusp}
\end{equation}
where $h_\c'(\b{r}_1,\b{r}_1) = (\partial \tilde{h}_\c(\b{r}_1,r_{12})/ \partial r_{12})_{r_{12}=0}$ is the first-order derivative of the spherically averaged correlation hole $\tilde{h}_\c(\b{r}_1,r_{12}) = (1/(4\pi r_{12}^2)) \int_{S(\b{0},r_{12})} h_\c(\b{r}_1,\b{r}_1 + \b{r}_{12}) \d \b{r}_{12}$ and
$S(\b{0},r_{12})$ designates the sphere centered at $\b{0}$ and of radius $r_{12}=|\b{r}_{12}|$. The potential contribution to the correlation energy can be written in terms of the correlation hole:
\begin{equation}
U_\c[\rho]  = \frac{1}{2} \int_{\mathbb{R}^3\times\mathbb{R}^3} \frac{\rho(\b{r}_1) h_{\c}(\b{r}_1,\b{r}_2)}{|\b{r}_1-\b{r}_2|} \d\b{r}_1 \d\b{r}_2.
\label{Ucint}
\end{equation}
In order to express the total correlation energy $E_\c[\rho]=T_\c[\rho]+U_\c[\rho]$ in a form similar to Eq.~(\ref{Ucint}), we need to introduce the adiabatic-connection formalism.

\subsection{The adiabatic connection}
\label{sec:adiabatic}

The idea of the {\it adiabatic connection}~\cite{LanPer-SSC-75,GunLun-PRB-76,LanPer-PRB-77} (see, also, Ref.~\cite{HarJon-JPF-74}) is to have a continuous path between the non-interacting KS system and the physical system while keeping the ground-state density constant. This allows one to obtain a convenient expression for the correlation functional $E_\c[\rho]$ as an integral over this path. An infinity of such paths are possible, but the one most often considered consists in switching on the electron-electron interaction linearly with a coupling constant $\l$. The Hamiltonian along this adiabatic connection is
\begin{equation}
\hat{H}^{\l} = \hat{T} + \l \hat{W}_{\ee} + \hat{V}^{\l},
\label{Hl}
\end{equation}
where $\hat{V}^{\l}$ is the external local potential operator imposing that the ground-state density is the same as the ground-state density of the physical system for all $\l \in \mathbb{R}$. Of course, Eq.~(\ref{Hl}) relies on a $v$-representability assumption, i.e. the external potential is assumed to exist for all $\l$. The Hamiltonian~(\ref{Hl}) reduces to the KS non-interacting Hamiltonian for $\l=0$ and to the physical Hamiltonian for $\l=1$.

Just as for the physical system, it is possible to define a universal functional associated with the system of Eq.~(\ref{Hl}) for each value of the parameter $\l$,
\begin{eqnarray}
F^{\l}[\rho] &=& \min_{\Psi \in {\cal W}^N_\rho} \bra{\Psi} \hat{T} + \l \hat{W}_{\ee} \ket{\Psi} = \bra{\Psi^\l[\rho]} \hat{T} + \l \hat{W}_{\ee} \ket{\Psi^\l[\rho]},
\label{Fln}
\end{eqnarray}
where $\Psi^{\l}[\rho]$ denotes a minimizing wave function. This functional can be decomposed as
\begin{eqnarray}
F^{\l}[\rho] = T_\s[\rho] + E_{\Hxc}^{\l}[\rho],
\label{Flndecomp}
\end{eqnarray}
where $E_{\Hxc}^{\l}[\rho]$ is the Hartree-exchange-correlation functional associated with the interaction $\l \hat{W}_{\ee}$. One can write this functional as  $E_{\Hxc}^{\l}[\rho]= E_{\H}^{\l}[\rho] + E_{\x}^{\l}[\rho] + E_{\c}^{\l}[\rho]$, where the Hartree and exchange contributions are simply linear in $\l$,
\begin{equation}
E_\H^\l[\rho] = \frac{1}{2} \int_{\mathbb{R}^3\times\mathbb{R}^3} \rho(\b{r}_1) \rho(\b{r}_2)\frac{\l}{|\b{r}_1 -\b{r}_2|} \d \b{r}_1 \d \b{r}_2 = \l E_{\H}[\rho],
\end{equation}
and
\begin{equation}
E_{\x}^{\l}[\rho] = \bra{\Phi[\rho]} \l \hat{W}_{\ee} \ket{\Phi[\rho]} - E_{\H}^{\l}[\rho] = \l E_{\x}[\rho].
\end{equation}
The correlation contribution is nonlinear in $\l$,
\begin{equation}
E_{\c}^{\l}[\rho] = \bra{\Psil[\rho]} \hat{T} + \l \hat{W}_{\ee} \ket{\Psil[\rho]} - \bra{\Phi[\rho]} \hat{T} + \l \hat{W}_{\ee} \ket{\Phi[\rho]}.
\label{Ecl}
\end{equation}

We will assume that $F^{\l}[\rho]$ is of class $C^1$ as a function of $\l$ for $\l \in [0,1]$ and that $F^{\l=0}[\rho]=T_\s[\rho]$, the latter condition being guaranteed for nondegenerate KS systems [see footnote on the definition of $T_\s[\rho]$ just before Eq.~(\ref{Ts})]. Taking the derivative of Eq.~(\ref{Ecl}) with respect to $\l$ and using the Hellmann-Feynman theorem for the wave function $\Psil[\rho]$\footnote{In this context, the Hellmann-Feynman theorem states that in the derivative $\frac{\partial F^\l[\rho]}{\partial \l} = \bra{\frac{\partial\Psil[\rho]}{\partial \l}} \hat{T} + \l \hat{W}_{\ee} \ket{\Psil[\rho]} + \bra{\Psil[\rho]} \hat{W}_{\ee} \ket{\Psil[\rho]} + \bra{\Psil[\rho]} \hat{T} + \l \hat{W}_{\ee} \ket{\frac{\partial\Psil[\rho]}{\partial \l}}$ the first and third terms involving the derivative of $\Psil[\rho]$ vanish. This is due to the fact that $\Psil[\rho]$ is obtained via the minimization of Eq.~(\ref{Fln}) and thus any variation of $\Psil[\rho]$ which keeps the density constant (which is the case for a variation with respect to $\l$) gives a vanishing variation of $F^\l[\rho]$.}, we obtain
\begin{eqnarray}
\frac{\partial E_{\c}^{\l}[\rho]}{\partial \l} &=& \bra{\Psil[\rho]} \hat{W}_{\ee} \ket{\Psil[\rho]} - \bra{\Phi[\rho]} \hat{W}_{\ee} \ket{\Phi[\rho]}.
\label{dEcldl}
\end{eqnarray}
Integrating over $\l$ from $0$ to $1$, and using $E_{\c}^{\l=1}[\rho]=E_{\c}[\rho]$ and $E_{\c}^{\l=0}[\rho]=0$, we arrive at the {\it adiabatic-connection formula} for the correlation energy functional of the physical system
\begin{equation}
E_{\c}[\rho] = \int_{0}^{1} \d\l \; \bra{\Psil[\rho]} \hat{W}_{\ee} \ket{\Psil[\rho]} - \bra{\Phi[\rho]} \hat{W}_{\ee} \ket{\Phi[\rho]}.
\label{EcintPsi}
\end{equation}
By introducing the correlation hole $h_{\c}^{\l}(\b{r}_1,\b{r}_2)$ associated to the wave function $\Psil[\rho]$, the adiabatic-connection formula for the correlation energy can also be written as
\begin{equation}
E_{\c}[\rho] = \frac{1}{2} \int_{0}^{1} \d\l \int_{\mathbb{R}^3\times\mathbb{R}^3} \frac{\rho(\b{r}_1) h_{\c}^{\l}(\b{r}_1,\b{r}_2)}{|\b{r}_1 -\b{r}_2|} \d\b{r}_1  \d\b{r}_2,
\label{Ecint}
\end{equation}
or, noting that $h_{\c}^{\l}(\b{r}_1,\b{r}_2)$ is the only quantity that depends on $\l$ in Eq.~(\ref{Ecint}), in a more compact way,
\begin{equation}
E_{\c}[\rho] = \frac{1}{2} \int_{\mathbb{R}^3\times\mathbb{R}^3} \frac{\rho(\b{r}_1) \bar{h}_{\c}(\b{r}_1,\b{r}_2)}{|\b{r}_1 -\b{r}_2|} \d\b{r}_1  \d\b{r}_2,
\label{Ecint2}
\end{equation}
where $\bar{h}_{\c}(\b{r}_1,\b{r}_2) = \int_{0}^{1} \d\l \; h_{\c}^{\l}(\b{r}_1,\b{r}_2)$ is the coupling-constant-integrated correlation hole. This leads to the interpretation of $E_\c$ as the electrostatic interaction energy of an electron with its coupling-constant-integrated correlation hole. As for the exchange energy, the correlation energy functional can be written as
\begin{equation}
E_{\c}[\rho] = \int_{\mathbb{R}^3} \rho(\b{r}_1) \varepsilon_\c[\rho](\b{r}_1) \d\b{r}_1,
\label{Ecinteps}
\end{equation}
where $\varepsilon_\c[\rho](\b{r}_1)$ is the correlation energy density per particle
\begin{equation}
\varepsilon_\c[\rho](\b{r}_1) = \frac{1}{2} \int_{\mathbb{R}^3} \frac{\bar{h}_{\c}(\b{r}_1,\b{r}_2)}{|\b{r}_1 -\b{r}_2|} \d\b{r}_2,
\label{epsc}
\end{equation}
which is a functional of the density. We can also define the correlation energy density $e_\c[\rho](\b{r}) = \rho(\b{r}) \varepsilon_\c [\rho](\b{r})$. 

Finally, note that the sum-rule and cusp conditions of Eqs.~(\ref{intnc}) and~(\ref{hccusp}) apply to the $\lambda$-dependent correlation hole in the form 
\begin{equation}
\forall \b{r}_1 \in \mathbb{R}^3, \; \int_{\mathbb{R}^3} h_{\c}^\l(\b{r}_1,\b{r}_2) \d\b{r}_2 = 0,
\label{inthcl}
\end{equation}
and
\begin{equation}
\forall \b{r}_1 \in \mathbb{R}^3, \; h_\c^{\l\;\prime}(\b{r}_1,\b{r}_1) = \l \; h_\c^\l(\b{r}_1,\b{r}_1).
\label{hclcusp}
\end{equation}

\subsection{One-orbital and one-electron spatial regions}
\label{sec:onetwoelectron}

For systems composed of only one spin-$\uparrow$ (or, symmetrically, one spin-$\downarrow$) electron (e.g., the hydrogen atom) with ground-state density $\rho_\text{1e}(\b{r}) = |\varphi_{1\uparrow}(\b{r})|^2$ where $\varphi_{1\uparrow}(\b{r})$ is the unique occupied KS orbital, the exchange hole in Eq.~(\ref{hxKSorb}) simplifies to $h_\x(\b{r}_1,\b{r}_2) = -\rho(\b{r}_2)$, and consequently the exchange energy cancels out the Hartree energy:
\begin{equation}
E_\x[\rho_\text{1e}] = - E_\H[\rho_\text{1e}].
\label{Ex1e}
\end{equation}
Furthermore, the correlation energy vanishes:
\begin{equation}
E_\c[\rho_\text{1e}] = 0.
\label{Ec1e}
\end{equation}
This must of course also be true for the spin-dependent version of the functionals introduced in Section~\ref{extensionspin}, i.e. 
\begin{equation}
E_\x[\rho_\text{1e},0] = - E_\H[\rho_\text{1e}]
\label{Ex1espin}
\end{equation}
and
\begin{equation}
E_\c[\rho_\text{1e},0] = 0.
\label{Ec1espin}
\end{equation}
For systems composed of two opposite-spin electrons (e.g., the helium atom or the dihydrogen molecule) in a unique doubly occupied KS orbital $\varphi_1(\b{r})=\varphi_{1\uparrow}(\b{r}) = \varphi_{1\downarrow}(\b{r})$ with ground-state density $\rho_\text{2e}^{\uparrow\downarrow}(\b{r}) = 2 |\varphi_1(\b{r})|^2$, the exchange hole simplifies to $h_\x(\b{r}_1,\b{r}_2) = -\rho(\b{r}_2)/2$, and consequently the exchange energy is equal to half the opposite of the Hartree energy:
\begin{equation}
E_\x[\rho_\text{2e}^{\uparrow\downarrow}] = -\frac{1}{2} E_\H[\rho_\text{2e}^{\uparrow\downarrow}].
\label{Ex2e}
\end{equation}
These are constraints for the exchange and correlation density functionals in the special cases $N=1$ and $N=2$.

These special cases can be extended to more general systems. For systems with $N\geq 1$ electrons containing a spatial region $\Omega_\text{1o}^\uparrow$ in which, among the occupied KS orbitals, only one spin-$\uparrow$ (or, symmetrically, one spin-$\downarrow$) orbital is not zero (or, more generally, takes non-negligible values), we have again in this region
\begin{equation}
\forall \b{r}_1,\b{r}_2 \in \Omega_\text{1o}^\uparrow, \; h_{\x}(\b{r}_1,\b{r}_2) = - \rho(\b{r}_2),
\end{equation}
and therefore the contribution to the exchange energy density per particle coming from this region must locally cancel out the contribution to the Hartree energy density per particle coming from the same region,
\begin{eqnarray}
\forall \b{r}_1 \in \Omega_\text{1o}^\uparrow, \; \varepsilon_\x^{\Omega_\text{1o}^\uparrow} (\b{r}_1) = - \varepsilon_\H^{\Omega_\text{1o}^\uparrow} (\b{r}_1),
\label{epsxOmega1o}
\end{eqnarray}
where $\varepsilon_\H^{\Omega} (\b{r}_1) = (1/2) \int_{\Omega} \rho(\b{r}_2)/|\b{r}_1-\b{r}_2| \d\b{r}_2$ and $\varepsilon_\x^{\Omega} (\b{r}_1) = (1/2) \int_{\Omega} h_{\x}(\b{r}_1,\b{r}_2)/|\b{r}_1-\b{r}_2| \d\b{r}_2$. Similarly, for systems with $N \geq 2$ electrons containing a spatial region $\Omega_\text{1o}^{\uparrow\downarrow}$ in which, among the occupied KS orbitals, only one doubly occupied orbital is not zero, we have in this region
\begin{equation}
\forall \b{r}_1,\b{r}_2 \in \Omega_\text{1o}^{\uparrow\downarrow}, \; h_{\x}(\b{r}_1,\b{r}_2) = - \frac{1}{2} \rho(\b{r}_2),
\end{equation}
and therefore the contribution to the exchange energy density per particle coming from this region must locally be equal to half the opposite of the contribution to the Hartree energy density per particle coming from the same region,
\begin{eqnarray}
\forall \b{r}_1 \in \Omega_\text{1o}^{\uparrow\downarrow}, \; \varepsilon_\x^{\Omega_\text{1o}^{\uparrow\downarrow}} (\b{r}_1) = - \frac{1}{2} \varepsilon_\H^{\Omega_\text{1o}^{\uparrow\downarrow}} (\b{r}_1).
\label{}
\label{epsxOmega2o}
\end{eqnarray}
Thus, we see, particularly clearly for these $\Omega_\text{1o}^\uparrow$ or $\Omega_\text{1o}^{\uparrow\downarrow}$ regions, that the Hartree functional introduces a spurious \textit{self-interaction} contribution which must be eliminated by the exchange functional. Even though the concepts of $\Omega_\text{1o}^\uparrow$ and $\Omega_\text{1o}^{\uparrow\downarrow}$ regions are formal, in practice they can be approximately realized in chemical systems. For example, the unpaired electron in a radical approximately corresponds to a $\Omega_\text{1o}^\uparrow$, and an electron pair in a single covalent bond, in a lone pair, or in a core orbital approximately corresponds to a $\Omega_\text{1o}^{\uparrow\downarrow}$ region.

We can also consider one-electron regions $\Omega_\text{1e}$ that we define as\footnote{In the definition of Eq.~(\ref{Omega1e}) we exclude the point $\l=0$ in order to allow for the possibility of a discontinuity in $\l$ there due to a degeneracy.}
\begin{eqnarray}
\forall \b{r}_1,\b{r}_2 \in \Omega_\text{1e}, \; \forall \l \in (0,1], \; \rho_2^\l(\b{r}_1,\b{r}_2) = 0,
\label{}
\label{Omega1e}
\end{eqnarray}
where $\rho_2^\l(\b{r}_1,\b{r}_2)$ is the pair density associated to the wave function $\Psil[\rho]$ along the adiabatic connection. This implies 
\begin{equation}
\forall \b{r}_1,\b{r}_2 \in \Omega_\text{1e}, \; \bar{h}_{\xc}(\b{r}_1,\b{r}_2) = - \rho(\b{r}_2),
\end{equation}
where $\bar{h}_{\xc}(\b{r}_1,\b{r}_2) = h_{\x}(\b{r}_1,\b{r}_2) + \bar{h}_{\c}(\b{r}_1,\b{r}_2)$ and, consequently, the contribution to the exchange-correlation energy density per particle coming from this region must locally cancel out the contribution to the Hartree energy density per particle coming from the same region,
\begin{eqnarray}
\forall \b{r}_1 \in \Omega_\text{1e}, \; \varepsilon_\xc^{\Omega_\text{1e}} (\b{r}_1) = - \varepsilon_\H^{\Omega_\text{1e}} (\b{r}_1),
\label{epsxcOmega1e}
\end{eqnarray}
where $\varepsilon_\xc^{\Omega} (\b{r}_1) = (1/2) \int_{\Omega} \bar{h}_{\xc}(\b{r}_1,\b{r}_2)/|\b{r}_1 -\b{r}_2| \d\b{r}_2$. For regions that are simultaneously one-electron and one-orbital regions, this simply implies that the contribution to the correlation energy must vanish,
\begin{eqnarray}
\forall \b{r}_1 \in \Omega_\text{1e} \cap \Omega_\text{1o}^\uparrow, \; \varepsilon_\c^{\Omega_\text{1e}\cap \Omega_\text{1o}^\uparrow} (\b{r}_1) =0,
\label{epscOmega1oOmega1o}
\end{eqnarray}
where $\varepsilon_\c^{\Omega} (\b{r}_1) = (1/2) \int_{\Omega} \bar{h}_{\c}(\b{r}_1,\b{r}_2)/|\b{r}_1 -\b{r}_2| \d\b{r}_2$, and we say that the correlation functional must not introduce a self-interaction error. However, the definition of $\Omega_\text{1e}$ regions also includes the case of an electron entangled in several orbitals, such as the region around one hydrogen atom in the dissociated dihydrogen molecule. In this latter case, the Hartree functional introduces an additional spurious contribution (beyond the spurious self-interaction) which must be compensated by a \textit{static correlation} (or \textit{strong correlation}) contribution in the exchange-correlation functional.

\subsection{Coordinate scaling}
\label{sec:scaling}

\subsubsection{Uniform coordinate scaling}

We consider a norm-preserving uniform scaling of the spatial coordinates in the $N$-electron wave function along the adiabatic connection $\Psi^\l[\rho]$ [introduced in Eq.~(\ref{Fln})] while leaving untouched the spin coordinates~\cite{LevPer-PRA-85,Lev-PRA-91,Lev-INC-95},
\begin{equation}
\Psi_\gamma^{\l}[\rho](\b{r}_1, \sigma_1,...,\b{r}_N, \sigma_N) = \g^{3N/2} \Psi^\l[\rho](\g \b{r}_1, \sigma_1,,...,\g\b{r}_N, \sigma_N),
\end{equation}
where $\g \in (0,+\infty)$ is a scaling factor. The scaled wave function $\Psi_{\g}^{\l}[\rho]$ yields the scaled density
\begin{equation}
\rho_\g(\b{r}) = \g^3 \rho(\g \b{r}),
\end{equation}
with $\int_{\mathbb{R}^3} \! \rho_\g(\b{r}) \d\b{r} = \int_{\mathbb{R}^3} \! \rho(\b{r}) \d\b{r} = N$, 
and minimizes $\bra{\Psi} \hat{T} + \l\g \hat{W}_\ee \ket{\Psi}$ since
\begin{equation}
\bra{\Psi^{\l}_{\g}[\rho]} \hat{T} + \l\g\hat{W}_\ee \ket{\Psi^{\l}_{\g}[\rho]} = \g^2 \bra{\Psi^\l[\rho]} \hat{T} + \l\hat{W}_\ee \ket{\Psi^\l[\rho]}.
\end{equation}
We thus conclude that the scaled wave function at the density $\rho$ and coupling constant $\l$ corresponds to the wave function at the scaled density $\rho_\g$ and coupling constant $\l\g$,
\begin{equation}
\Psi_{\g}^{\l}[\rho] = \Psi^{\l\g}[\rho_\g],
\end{equation}
or, equivalently,
\begin{equation}
\Psi_{\g}^{\l/\g}[\rho] = \Psi^{\l}[\rho_\g],
\end{equation}
and that the universal density functional satisfies the scaling relation
\begin{equation}
F^{\l\g}[\rho_\g] = \g^2 F^{\l}[\rho],
\end{equation}
or, equivalently,
\begin{equation}
F^{\l}[\rho_\g] = \g^2 F^{\l/\g}[\rho].
\label{Frhog}
\end{equation}

At $\l=0$, we find the scaling relation of the KS wave function $\Phi[\rho]$ introduced in Section~\ref{sec:decompF}:
\begin{equation}
\Phi[\rho_\g] = \Phi_\g[\rho].
\end{equation}
This directly leads to the scaling relation for the non-interacting kinetic density functional [see Eq.~(\ref{Ts})],
\begin{equation}
T_\s[\rho_\g] = \g^2 T_\s[\rho],
\end{equation}
for the Hartree density functional [see Eq.~(\ref{EHn})],
\begin{equation}
E_\H[\rho_\g] = \g E_\H[\rho],
\end{equation}
and for the exchange density functional [see Eq.~(\ref{Exn})],
\begin{equation}
E_\x[\rho_\g] = \g E_\x[\rho].
\label{Exrhog}
\end{equation}

However, the correlation density functional $E_\c[\rho]$ has the more complicated scaling (as $F[\rho]$),
\begin{equation}
E_\c^{\l}[\rho_\g] = \g^2 E_\c^{\l/\g}[\rho],
\end{equation}
and, in particular for $\l=1$,
\begin{equation}
E_\c[\rho_\g] = \g^2 E_\c^{1/\g}[\rho].
\label{Ecrhog}
\end{equation}

These scaling relations allow one to find the behavior of the density functionals in the high- and low-density limits. In the \textit{high-density limit} ($\g \to \infty$), it can be shown from Eq.~(\ref{Ecrhog}) that, for nondegenerate KS systems, the correlation functional $E_\c[\rho]$ goes to a constant,
\begin{equation}
\lim_{\g\to \infty} E_\c[\rho_\g] = E_\c^{\GL}[\rho],
\label{Ecrhoginf}
\end{equation}
where $E_\c^{\GL}[\rho]$ is the second-order G\"orling--Levy (GL2) correlation energy~\cite{GorLev-PRB-93,GorLev-PRA-94} (see Section~\ref{sec:GLPT2}). This is also called the \textit{weak-correlation limit} since in this limit the correlation energy is negligible with respect to the exchange energy which is itself negligible with respect to the non-interacting kinetic energy: $|E_\c[\rho_\g] |= O(\g^0) \ll |E_\x[\rho_\g]| = O(\g) \ll T_\s[\rho_\g] = O(\g^2)$. Equation~(\ref{Ecrhoginf}) is an important constraint since atomic and molecular correlation energies are often close to the high-density limit. For example, for the ground-state density of the helium atom, we have $E_\c[\rho]=-0.0421$ hartree and $\lim_{\g\to \infty} E_\c[\rho_\g]=-0.0467$ hartree~\cite{HuaUmr-PRA-97}.

In the \textit{low-density limit} ($\g\to 0$), it can be shown from Eq.~(\ref{Frhog}) that the Hartree-exchange-correlation energy $E_\Hxc[\rho]$ goes to zero linearly in $\g$,
\begin{equation}
E_\Hxc[\rho_\g] \isEquivTo{\g \to 0} \g \; W_\ee^\text{SCE}[\rho],
\label{Ecrhog0}
\end{equation}
where $\displaystyle W_\ee^\text{SCE}[\rho]=\inf_{\Psi\in{\cal W}^N_\rho} \bra{\Psi} \hat{W}_\ee \ket{\Psi}$ is the strictly-correlated-electron (SCE) functional~\cite{SeiPerLev-PRA-99,Sei-PRA-99,SeiGorSav-PRA-07,GorSei-PCCP-10}. This is also called the \textit{strong-interaction limit} since in this limit the Hartree-exchange-correlation energy dominates over the non-interacting kinetic energy: $E_\Hxc[\rho_\g] = O(\g) \gg T_\s[\rho_\g] = O(\g^2)$. In this limit, the electrons strictly localize relatively to each other. In particular, for the uniform-electron gas, this corresponds to the Wigner crystallization. Thus, in this limit, each electron is within a one-electron region $\Omega_\text{1e}$ [as defined in Eq.~(\ref{Omega1e})].
For more information on the SCE functional, see the chapter by Friesecke, Gerolin, and Gori-Giorgi in this volume.

\subsubsection{Non-uniform coordinate scaling}

We can also consider non-uniform one-dimensional or two-dimensional coordinate scalings of the density~\cite{OuyLev-PRA-90,LevOuy-PRA-90},
\begin{eqnarray}
\rho_\gamma^{(1)}(x,y,z) = \gamma \rho(\gamma x,y,z)
\end{eqnarray}
and
\begin{eqnarray}
\rho_\gamma^{(2)}(x,y,z) = \gamma^2 \rho(\gamma x,\gamma y,z),
\end{eqnarray}
which also preserve the number of the electrons. These non-uniform density scalings provide constraints for the exchange and correlation functionals. In particular, in the non-uniform one-dimensional high-density limit, the exchange functional remains finite and the correlation functional vanishes~\cite{Lev-PRA-91,GorLev-PRA-92}
\begin{eqnarray}
\lim_{\gamma \to \infty} E_\x[\rho_\gamma^{(1)}] > - \infty
\label{Exgamma1inf}
\end{eqnarray}
and
\begin{eqnarray}
\lim_{\gamma \to \infty} E_\c[\rho_\gamma^{(1)}] =0.
\label{Ecgamma1inf}
\end{eqnarray}
Also, in the non-uniform two-dimensional low-density limit, we have~\cite{Lev-PRA-91,GorLev-PRA-92}:
\begin{eqnarray}
\lim_{\gamma \to 0} \frac{1}{\gamma} E_\x[\rho_\gamma^{(2)}] > - \infty
\label{Exgamma20}
\end{eqnarray}
and
\begin{eqnarray}
\lim_{\gamma \to 0} \frac{1}{\gamma} E_\c[\rho_\gamma^{(2)}] =0.
\label{Ecgamma20}
\end{eqnarray}
The conditions of Eqs.~(\ref{Exgamma1inf})-(\ref{Ecgamma20}) are particularly useful because they also correspond to the limit of rapidly varying densities~\cite{LevPer-PRB-93}.

\subsection{Atoms in the limit of large nuclear charge}
\label{sec:largeZ}

A practical realization of the uniform high-density limit is provided by atomic ions in the limit of large nuclear charge, $Z\to \infty$, at fixed electron number $N$ (see Refs.~\cite{IvaLev-JPCA-98,StaScuPerTaoDav-PRA-04,FriGod-SIAM-09,FriGod-PRA-10}). In this limit, the exact ground-state atomic density $\rho_{N,Z}(\b{r})$ becomes the density of the isoelectronic hydrogenic (i.e., without electron-electron interaction) atom $\rho_{N,Z}^\text{H}(\b{r})$ which obeys a simple scaling with $Z$:
\begin{equation}
\rho_{N,Z}(\b{r}) \isEquivTo{Z\to\infty} \rho_{N,Z}^\text{H}(\b{r}) = Z^3 \rho_{N,Z=1}^\text{H}(Z \b{r}).
\end{equation}
One can thus apply Eqs.~(\ref{Exrhog}) and~(\ref{Ecrhoginf}) with $\gamma=Z$, which reveals that in an isoelectronic series the exchange functional scales linearly with $Z$,
\begin{equation}
E_\x[\rho_{N,Z}] \isEquivTo{Z\to\infty} E_\x[\rho_{N,Z=1}^\text{H}] Z,
\label{ExZinf}
\end{equation}
and, for nondegenerate KS systems, the correlation functional saturates to a constant,
\begin{equation}
\lim_{Z\to\infty} E_\c[\rho_{N,Z}] = E_\c^{\GL}[\rho_{N,Z=1}^\text{H}].
\label{EcZinf}
\end{equation}
Equations~(\ref{ExZinf}) and~(\ref{EcZinf}) are constraints for the exchange and correlation functionals, particularly relevant for highly ionized atoms but also for the core-electron regions of heavy atoms in neutral systems.

Another very interesting limit is the one of large nuclear charge of neutral atoms, $N=Z\to \infty$ (see, e.g., Ref.~\cite{KapSanBhaWagShaChoBheYuTanBurLevPer-JCP-20}). In this semiclassical limit, the exact ground-state atomic density $\rho_{N,Z}(\b{r})$ tends to the Thomas--Fermi (TF) density of a neutral atom $\rho_{Z}^\text{TF}(\b{r})$ which has a known scaling with $Z$~\cite{LieSim-PRL-73,LieSim-AM-77}:
\begin{equation}
\rho_{Z,Z}(\b{r}) \isEquivTo{Z\to\infty} \rho_{Z}^\text{TF}(\b{r}) = Z^2 \rho_{Z=1}^\text{TF}(Z^{1/3}\b{r}).
\end{equation}
In this limit, it was suggested that the exact exchange and correlation energies have the approximate large-$Z$ asymptotic expansions~\cite{EllBur-CJC-09,BurCanGouPit-JCP-16,CanCheKruBur-JCP-18}
\begin{equation}
E_\x[\rho_{Z,Z}] \isEquivTo{Z\to\infty} -A_\x Z^{5/3} + B_\x Z +\cdots
\label{ExrhoZZinf}
\end{equation}
and
\begin{equation}
E_\c[\rho_{Z,Z}] \isEquivTo{Z\to\infty}  -A_\c Z \ln Z + B_\c Z + \cdots,
\label{EcrhoZZinf}
\end{equation}
with the coefficients $A_\x = 0.220827$, $A_\c=0.020727$, $B_\x \approx 0.224$, $B_\c \approx 0.0372$. 
Recently, it was argued that there is in fact a missing term in $Z \ln Z$ in the expansion of the exchange energy in Eq.~(\ref{ExrhoZZinf})~\cite{DaaKooGroSeiGor-JCTC-22,ArgRedCanBur-ARX-22}.

\subsection{Lieb--Oxford lower bound}
\label{sec:LO}

Lieb and Oxford derived a lower bound for the indirect Coulomb energy (i.e., the two-particle Coulomb potential energy beyond the Hartree energy)~\cite{LieOxf-IJQC-81}, which, when expressed in terms of the exchange or exchange-correlation functional, takes the form~\cite{Per-INC-91}
\begin{equation}
E_\x[\rho] \geq E_\xc[\rho] \geq -C_\text{LO} \int_{\mathbb{R}^3} \rho(\b{r})^{4/3} \d\b{r},
\label{LObound}
\end{equation}
where the optimal (i.e., smallest) constant $C_\text{LO}$ (independent of the electron number $N$) was originally shown to be in the range $1.23 \leq C_\text{LO} \leq 1.68$~\cite{LieOxf-IJQC-81}. The range was later successively narrowed to $1.4442 \leq C_\text{LO} \leq 1.5765$~\cite{Per-INC-91,ChaHan-PRA-99,CotPet-ARX-19,LewLieSei-ARX-22}.  This bound is approached only in the low-density limit where the correlation energy becomes comparable to the exchange energy. 
Numerical results suggest that for densities of most physical systems the Lieb--Oxford lower bound on the exchange-correlation energy is far from being reached~\cite{OdaCap-JCP-07}.

For two-electron densities, there is a specific tighter bound,
\begin{equation}
E_\x[\rho_\text{2e}] \geq E_\xc[\rho_\text{2e}] \geq -C_2 \int_{\mathbb{R}^3} \rho_\text{2e}(\b{r})^{4/3} \d\b{r},
\label{LObound2eC2}
\end{equation}
with the best known constant $C_2=1.234$~\cite{LieOxf-IJQC-81}. For one-electron densities, an even tighter bound is known for the exchange functional~\cite{GadBarHan-JCP-80,LieOxf-IJQC-81},
\begin{equation}
E_\x[\rho_\text{1e}] \geq -C_1 \int_{\mathbb{R}^3} \rho_\text{1e}(\b{r})^{4/3} \d\b{r},
\label{LObound1e}
\end{equation}
with the optimal constant $C_1=1.092$. For two-electron spin-unpolarized densities, we have $E_\x[\rho_\text{2e}^{\uparrow\downarrow}] =  2 E_\x[\rho_\text{1e}]$ with $\rho_\text{1e} = \rho_\text{2e}^{\uparrow\downarrow}/2$, and Eq.~(\ref{LObound1e}) implies~\cite{PerRuzSunBur-JCP-14}
\begin{equation}
E_\x[\rho_\text{2e}^{\uparrow\downarrow}] \geq -\frac{C_1}{2^{1/3}} \int_{\mathbb{R}^3} \rho_\text{2e}^{\uparrow\downarrow}(\b{r})^{4/3} \d\b{r},
\label{LObound2eC1}
\end{equation}
which is a much tighter bound than the bounds of Eqs.~(\ref{LObound}) and~(\ref{LObound2eC2}).

\section{Semilocal approximations for the exchange-correlation energy}
\stepcounter{myequation}
\label{sec:approx}

We review here the different classes of \textit{semilocal approximations} for the exchange-correlation energy. 

\subsection{The local-density approximation}
\label{sec:lda}

In the {\it local-density approximation} (LDA), introduced by Kohn and Sham~\cite{KohSha-PR-65}, the exchange-correlation functional is approximated as
\begin{equation}
E_\xc^\LDA[\rho] = \int_{\mathbb{R}^3} e_\xc^{\text{UEG}}(\rho(\b{r})) \d \b{r},
\end{equation}
where $e_\xc^{\text{UEG}}(\rho)$ is the exchange-correlation energy density of the infinite {\it uniform electron gas} (UEG) with the density $\rho$. The UEG represents a family of systems of interacting electrons with an arbitrary spatially constant density $\rho \in [0,+\infty)$ that acts as a parameter. Thus, in the LDA, the exchange-correlation energy density of an inhomogeneous system at a spatial point of density $\rho(\b{r})$ is approximated as the exchange-correlation energy density of the UEG of the same density. 

In the spin-dependent version of LDA, sometimes specifically referred to as the local-spin-density approximation (LSDA), the exchange-correlation functional is approximated as~\cite{BarHed-JPC-72}
\begin{equation}
E_\xc^\LSDA[\rho_\uparrow,\rho_\downarrow] = \int_{\mathbb{R}^3} e_\xc^{\text{UEG}}(\rho_\uparrow(\b{r}),\rho_\downarrow(\b{r})) \d \b{r},
\end{equation}
where $e_\xc^{\text{UEG}}(\rho_\uparrow,\rho_\downarrow)$ is the exchange-correlation energy density of the UEG with spin densities $\rho_\uparrow$ and $\rho_\downarrow$. For spin-unpolarized systems, we recover the spin-independent LDA as $E_\xc^\LDA[\rho] = E_\xc^\LSDA[\rho/2,\rho/2]$.

The function $e_\xc^{\text{UEG}}$ is a sum of exchange and correlation contributions, $e_\xc^{\text{UEG}} = e_\x^{\text{UEG}} + e_\c^{\text{UEG}}$, and it is convenient to introduce exchange and correlation energies per particle, $\varepsilon_\x^{\text{UEG}}$ and $\varepsilon_\c^{\text{UEG}}$, such that $e_\x^{\text{UEG}} = \rho \; \varepsilon_\x^{\text{UEG}}$ and $e_\c^{\text{UEG}} = \rho \; \varepsilon_\c^{\text{UEG}}$. The expression of the exchange energy per particle of the spin-unpolarized UEG is
\begin{equation}
\varepsilon_\x^{\text{UEG}}(\rho) = C_\x \; \rho^{1/3},
\label{epsxUEG}
\end{equation}
where $C_\x=-(3/4)(3/\pi)^{1/3}$, and the spin-polarized version is simply obtained from the spin-scaling relation [Eq.~(\ref{Exspinscaling})], leading to
\begin{equation}
\varepsilon_\x^{\text{UEG}}(\rho_\uparrow,\rho_\downarrow) = \varepsilon_\x^{\text{UEG}}(\rho) \phi_4(\zeta),
\end{equation}
where $\zeta=(\rho_\uparrow-\rho_\downarrow)/\rho$ is the spin polarization and $\phi_4(\zeta)$ is defined by the general spin-scaling function
\begin{equation}
\phi_n(\zeta)=\frac{(1+\zeta)^{n/3}+(1-\zeta)^{n/3}}{2}.
\label{phinzeta}
\end{equation}
The LDA exchange functional is associated with the names of Dirac~\cite{Dir-PCPRS-30} and Slater~\cite{Sla-PR-51}. For a rigorous mathematical derivation of Eq.~(\ref{epsxUEG}), see Ref.~\cite{Fri-CMP-97}.

The correlation energy per particle $\varepsilon_\c^{\text{UEG}}(\rho_\uparrow,\rho_\downarrow)$ of the UEG cannot be calculated analytically. This quantity has been obtained numerically for a sample of densities and fitted to a parametrized function satisfying the known high- and low-density expansions. Expressed in terms of the Wigner--Seitz radius $r_\s=(3/(4\pi \rho))^{1/3}$, the first terms of the high-density expansion ($r_\s \to 0$) have the form
\begin{equation}
\varepsilon_\c^{\text{UEG}}(\rho_\uparrow,\rho_\downarrow) = A(\zeta) \ln r_\s + B(\zeta) + C(\zeta) r_\s \ln r_\s + O(r_\s),
\label{epscunifsmallrs}
\end{equation}
with spin-unpolarized coefficients $A(0)=(1-\ln 2)/\pi^2$, $B(0)=-0.046921$, $C(0)=0.009229$, and fully spin-polarized coefficients $A(1)=A(0)/2$, $B(1)=-0.025738$, $C(1)=0.004792$. The first terms of the low-density expansion ($r_\s\to +\infty$) have the form
\begin{equation}
\varepsilon_\c^{\text{UEG}}(\rho_\uparrow,\rho_\downarrow) = \frac{a}{r_\s} + \frac{b}{r_\s^{3/2}} + \frac{c}{r_\s^2} + O\left(\frac{1}{r_\s^{5/2}}\right),
\label{epscuniflargers}
\end{equation}
where the coefficients $a=-0.895930$, $b=1.325$, and $c=-0.365$ are assumed to be independent of $\zeta$. The low-density limit of the UEG corresponds to the Wigner crystallization. For a recent review of results on the UEG, see Ref.~\cite{LooGil-WIRES-16}. 

The two most used parametrizations are the one of Vosko, Wilk, and Nusair (VWN)~\cite{VosWilNus-CJP-80} and the more recent one of Perdew and Wang (PW92)~\cite{PerWan-PRB-92} which we give here. In this parametrization, the UEG correlation energy per particle is estimated using the approximate spin-interpolation formula
\begin{eqnarray}
\varepsilon_\c^\text{PW92}(\rho_\uparrow,\rho_\downarrow) = \varepsilon_\c(r_\s,0) + \alpha_\c(r_\s) \frac{f(\zeta)}{f''(0)} (1-\zeta^4) + [\varepsilon_\c(r_\s,1) - \varepsilon_\c(r_\s,0)] f(\zeta) \zeta^4,
\label{PW92}
\end{eqnarray}
where $\varepsilon_\c(r_\s,\zeta)$ is the UEG correlation energy per particle as a function of $r_\s$ and $\zeta$, $f(\zeta) = [(1+\zeta)^{4/3}+(1-\zeta)^{4/3}-2]/(2^{4/3}-2)$ is a spin-scaling function borrowed from the exchange energy, and $\alpha_\c(r_\s) = (\partial^2 \varepsilon_\c(r_\s,\zeta)/\partial \zeta^2)_{\zeta=0}$ is the spin stiffness. This spin-interpolation formula was first proposed in the VWN parametrization based on a study of the $\zeta$ dependence of the UEG correlation energy per particle at the random-phase approximation (RPA) level. A unique parametrization function
\begin{eqnarray}
G(r_\s,A,\alpha_1,\beta_1,\beta_2,\beta_3,\beta_4) = -2 (1 + \alpha_1 r_\s) A \ln \left[ 1+ \frac{1}{2 A \left( \beta_1 r_\s^{1/2} + \beta_2 r_\s + \beta_3 r_s^{3/2} + \beta_4 r_\s^2 \right)}\right],
\end{eqnarray}
is then used for approximating $\varepsilon_\c(r_\s,0)$, $\varepsilon_\c(r_\s,1)$, and $-\alpha_\c(r_\s)$, where
\begin{subequations}
\begin{eqnarray}
\varepsilon_\c(r_\s,0) = G(r_\s,A_0,\alpha_{1,0},\beta_{1,0},\beta_{2,0},\beta_{3,0},\beta_{4,0}),
\end{eqnarray}
\begin{eqnarray}
\varepsilon_\c(r_\s,1) = G(r_\s,A_1,\alpha_{1,1},\beta_{1,1},\beta_{2,1},\beta_{3,1},\beta_{4,1}),
\end{eqnarray}
\begin{eqnarray}
-\alpha_\c(r_\s) = G(r_\s,A_2,\alpha_{1,2},\beta_{1,2},\beta_{2,2},\beta_{3,2},\beta_{4,2}).
\end{eqnarray}
\end{subequations}
The form of $G$ was chosen to reproduce the form of the high- and low-density expansions. The parameters $A_i$, $\beta_{1,i}$, and $\beta_{2,i}$ (with $i\in\{0,1,2\}$) are fixed by the first two terms of the high-density expansion, while the parameters $\alpha_{1,i}$, $\beta_{3,i}$, and $\beta_{4,i}$ are fitted to quantum Monte Carlo (QMC) data~\cite{CepAld-PRL-80} for $\varepsilon_\c(r_\s,0)$ and $\varepsilon_\c(r_\s,1)$, and to an estimation of $-\alpha_\c(r_\s)$ extrapolated from RPA data. The parameters are given in Table I of Ref.~\cite{PerWan-PRB-92}. 

We now discuss the merits and deficiencies of the LDA. By construction, the LDA is of course exact in the limit of uniform densities. More relevant to atomic and molecular systems is that the LDA exchange and correlation energies are asymptotically exact in the limit of large nuclear charge of neutral atoms $N=Z\to \infty$. Indeed, in this semiclassical Thomas--Fermi limit, the LDA gives the exact coefficients $A_\x$ and $A_\c$ of the leading terms in the asymptotic expansions of Eqs.~(\ref{ExrhoZZinf}) and~(\ref{ExrhoZZinf})~\cite{PerConSagBur-PRL-06}. However, the coefficients of the next terms are very different: $B_\x^\text{LDA}\approx 0$ instead of $B_\x \approx 0.224$ and $B_\c^\text{LDA}\approx-0.00451$ instead of $B_\c \approx 0.0372$~\cite{BurCanGouPit-JCP-16}.

Due to the scaling of the UEG exchange energy per particle,
\begin{equation}
\varepsilon_\x^\text{UEG}(\gamma^3 \rho_\uparrow, \gamma^3 \rho_\downarrow) = \gamma \varepsilon_\x^\text{UEG}(\rho_\uparrow,\rho_\downarrow),
\end{equation}
the LDA exchange functional correctly scales linearly under uniform coordinate scaling of the density [Eq.~(\ref{Exrhog})]. Similarly, due the scaling of the UEG correlation energy per particle in the low-density limit [Eq.~(\ref{epscuniflargers})],
\begin{equation}
\varepsilon_\c^\text{UEG}(\gamma^3 \rho_\uparrow, \gamma^3 \rho_\downarrow) \isEquivTo{\g \to 0} \g \; \frac{a}{r_\s},
\label{epscUEGgamma0}
\end{equation}
the LDA correlation functional correctly scales linearly under uniform coordinate scaling to the low-density limit [Eq.~(\ref{Ecrhog0})]. However, from the behavior of $\varepsilon_\c^\text{UEG}$ in the high-density limit [Eq.~(\ref{epscunifsmallrs})],
\begin{equation}
\varepsilon_\c^\text{UEG}(\gamma^3 \rho_\uparrow, \gamma^3 \rho_\downarrow) \isEquivTo{\g \to \infty} - A(\zeta) \; \ln \g,
\label{epscUEGgammainf}
\end{equation}
we see that the LDA correlation functional diverges logarithmically under uniform coordinate scaling to the high-density limit whereas the exact correlation functional goes to a constant for nondegenerate KS systems [Eq.~(\ref{Ecrhoginf})]. Consequently, in the limit of large nuclear charge, $Z\to \infty$, at fixed electron number $N$, the LDA exchange energy correctly scales linearly with $Z$ [Eq.~(\ref{ExZinf})], albeit with an incorrect coefficient, and the LDA correlation energy does not reproduce the exact saturation behavior [Eq.~(\ref{EcZinf})] for a nondegenerate isoelectronic series but incorrectly diverges~\cite{PerMcmZun-PRA-81}. Also, the LDA exchange and correlation functionals do not satisfy the non-uniform scaling conditions of Eqs.~(\ref{Exgamma1inf})-(\ref{Ecgamma20}), but instead both diverge in these limits.

The LDA can also be thought of as approximating the exchange and the (coupling-constant-integrated) correlation holes of an inhomogeneous system in Eqs.~(\ref{epsx}) and~(\ref{epsc}) by the corresponding exchange and correlation holes of the UEG. Namely, considering the spin-independent version for simplicity, the LDA exchange hole is
\begin{equation}
h_\x^\LDA(\b{r}_1,\b{r}_2) = h_\x^\text{UEG}(\rho(\b{r}_1),r_{12}),
\label{hxLDA}
\end{equation}
with
\begin{equation}
h_\x^\text{UEG}(\rho,r_{12})=- \rho \; \frac{9}{2} \left( \frac{j_1(k_\text{F} r_{12})}{k_\text{F} r_{12}} \right)^2,
\label{hxUEG}
\end{equation}
where $r_{12} =|\b{r}_2 - \b{r}_1|$ is the interelectronic distance, $k_\text{F} = (3 \pi^2 \rho)^{1/3}$ is the Fermi wave vector, and $j_1$ is the spherical Bessel function of the first kind. Similarly, the LDA correlation hole is
\begin{equation}
\bar{h}_\c^\LDA(\b{r}_1,\b{r}_2) = \bar{h}_\c^\text{UEG}(\rho(\b{r}_1),r_{12}) = \int_0^1 \d\l \; h_\c^{\l,\text{UEG}}(\rho(\b{r}_1),r_{12}).
\end{equation}
Since the UEG is a physical system, the LDA exchange hole correctly fulfills the negativity and sum-rule condition [Eqs.~(\ref{intnx}) and~(\ref{nxneg})]
and the LDA correlation hole correctly fulfills the sum-rule and electron-electron cusp condition [Eqs.~(\ref{inthcl}) and~(\ref{hclcusp})].
This constitutes a significant merit of the LDA. However, because the LDA exchange hole $h_\x^\LDA(\b{r}_1,\b{r}_2)$ only depends on $\rho(\b{r}_1)$ and not on $\rho(\b{r}_2)$, the LDA exchange functional does not entirely eliminate the self-interaction contribution of the Hartree functional, in particular in one and two-electron systems [Eqs.~(\ref{Ex1e}) or (\ref{Ex1espin}), and~(\ref{Ex2e})], or in one-orbital spatial regions of many-electron systems [Eqs.~(\ref{epsxOmega1o}) and~(\ref{epsxOmega2o})]. Similarly, the LDA correlation functional does not vanish in one-electron systems [Eq.~(\ref{Ec1e}) or (\ref{Ec1espin})], or more generally in one-orbital one-electron regions [Eq.~(\ref{epscOmega1oOmega1o})]. Thus, the LDA introduces a self-interaction error. Moreover, the LDA exchange-correlation functional does not entirely cancel out the Hartree energy in entangled one-electron spatial regions [Eq.~(\ref{epsxcOmega1e}], i.e. it introduces a static-correlation error.

Another deficiency of the LDA is that the (spin-independent) LDA exchange potential
\begin{equation}
v_\x^\LDA(\b{r}) = \frac{\delta E_\x^\LDA[\rho]}{\delta \rho(\b{r})} = \frac{4}{3} C_\x \; \rho(\b{r})^{1/3},
\end{equation}
decays exponentially at infinity for finite molecular systems (since the density $\rho(\b{r})$ decays exponentially), i.e. much too fast in comparison to the $-1/|\b{r}|$ asymptotic behavior of the exact exchange potential [Eq.~(\ref{vxrinf})]. Since asymptotic spatial regions are dominated by the highest occupied molecular orbital (HOMO) and are thus one-orbital regions (assuming the HOMO is not degenerate), this is another signature of the incorrectness of the LDA exchange functional in these one-orbital regions.

For a review of mathematical results on the LDA, see the chapter by Lewin, Lieb, and Seiringer in this volume.

\subsection{The gradient-expansion approximation}

The next logical step beyond the LDA is the {\it gradient-expansion approximation} (GEA)~\cite{KohSha-PR-65}, in which the exchange-correlation functional is systematically expanded in the gradient and higher-order derivatives of the density. One way of deriving the GEA is to start from the UEG, introduce a weak and slowly-varying external potential $\delta v(\b{r})$, and expand the exchange-correlation energy in terms of the gradients of the density (see, e.g., Refs.~\cite{MaBru-PR-68,KleLee-PRB-88,EngVos-PRB-90,SveBar-PRB-96}). Alternatively, one can perform a semiclassical expansion (i.e., an expansion in powers of the reduced Planck constant $\hbar$) of the exact $E_\xc[\rho]$ in terms of the gradients of the external potential and use the mapping between the potential and the density to express it in terms of the gradients of the density (see, e.g., Ref.~\cite{DreGro-BOOK-90}).

The spin-independent gradient expansion of the exchange functional is known up to fourth order (GEA4)~\cite{SveBar-PRB-96},
\begin{eqnarray}
E_\x^\text{GEA4}[\rho] &=& E_\x^\LDA[\rho] + C_\x^{(2)}\int_{\mathbb{R}^3} \frac{|\nabla \rho(\b{r})|^2}{\rho(\b{r})^{4/3}} \d\b{r} 
\nonumber\\
&& + C_{\x,1}^{(4)}\int_{\mathbb{R}^3} \frac{|\nabla^2 \rho(\b{r})|^2}{\rho(\b{r})^{2}} \d\b{r} 
+ C_{\x,2}^{(4)}\int_{\mathbb{R}^3} \frac{|\nabla \rho(\b{r})|^2 \, \nabla^2 \rho(\b{r})}{\rho(\b{r})^{3}} \d\b{r},
\label{ExGEA}
\end{eqnarray}
involving the density gradient $\nabla \rho(\b{r})$ and Laplacian $\nabla^2 \rho(\b{r})$. Sham~\cite{Sha-INC-71} obtained the second-order coefficient $C_{\x,\text{S}}^{(2)} = -7/(432\pi(3\pi^2)^{1/3})\approx -0.001667$. The calculation was done by starting with the screened Yukawa interaction $e^{-\kappa r_{12}}/r_{12}$ and taking the limit $\kappa \to 0$ at the end of the calculation. It was later shown that this calculation contains an order-of-limit problem and that the correct Coulombic second-order coefficient is $C_\x^{(2)}=-5/(216\pi(3\pi^2)^{1/3})\approx-0.002382$~\cite{KleLee-PRB-88,EngVos-PRB-90}. The fourth-order coefficients are $C_{\x,1}^{(4)}=-73/(64800\pi^3)\approx-0.000036$, and $C_{\x,2}^{(4)}\approx 0.00009$, where the last one has been numerically estimated~\cite{SveBar-PRB-96}. Note that each term in Eq.~(\ref{ExGEA}) correctly fulfills the scaling relation of Eq.~(\ref{Exrhog}). The spin-dependent gradient exchange expansion is simply obtained from the spin-scaling relation [Eq.~(\ref{Exspinscaling})]. 

Similarly, Ma and Brueckner~\cite{MaBru-PR-68} obtained the spin-independent second-order gradient expansion (GEA2) of the correlation functional,
\begin{equation}
E_\c^\text{GEA2}[\rho] = E_\c^\LDA[\rho] + \int_{\mathbb{R}^3} C_\c^{(2)}(r_\s(\b{r})) \frac{|\nabla \rho(\b{r})|^2}{\rho(\b{r})^{4/3}} \d\b{r},
\label{EcGEA}
\end{equation}
with a second-order coefficient in the high-density limit $C_{\c,\text{MB}}^{(2)}(r_\s\to 0) = 0.004235$. It is believed~\cite{LanVos-PRL-87} that this calculation contains a similar order-of-limit problem as in Sham's coefficient $C_{\x,\text{S}}^{(2)}$, in such a way that these two coefficients must be combined to obtain the correct second-order exchange-correlation coefficient in the high-density limit $C_\xc^{(2)}(r_\s\to 0) = C_{\x,\text{S}}^{(2)} + C_{\c,\text{MB}}^{(2)}(r_\s\to 0)$. The correct second-order correlation coefficient in the high-density limit is then $C_\c^{(2)}(r_\s\to 0) = C_\xc^{(2)}(r_\s\to 0) - C_\x^{(2)} = 0.004950$. Similarly, the second-order correlation coefficient as a function of $r_\s$ can be obtained by $C_\c^{(2)}(r_\s) = C_\xc^{(2)}(r_\s) - C_\x^{(2)}$ where $C_\xc^{(2)}(r_\s)$ has been parametrized in Ref.~\cite{RasGel-PRB-86}. The spin-dependent generalization has the form~\cite{Ras-PRB-77,WanPer-PRB-91}
\begin{equation}
E_\c^\text{GEA2}[\rho_\uparrow,\rho_\downarrow] = E_\c^\LSDA[\rho_\uparrow,\rho_\downarrow] + \sum_{\sigma,\sigma'\in \{\uparrow,\downarrow\}} \int_{\mathbb{R}^3} C_\c^{\sigma,\sigma',(2)}(r_\s(\b{r}),\zeta(\b{r})) \frac{\nabla \rho_\sigma(\b{r})}{\rho_\sigma(\b{r})^{2/3}} \cdot \frac{\nabla \rho_{\sigma'}(\b{r})}{\rho_{\sigma'}(\b{r})^{2/3}} \d\b{r},
\label{EcGEAspin}
\end{equation}
where the functions $C_{\c}^{\sigma,\sigma',(2)}(r_\s,\zeta)$ have been numerically calculated in the high-density limit~\cite{Ras-PRB-77,RasDav-PL-81}.

The GEA should improve over the LDA for sufficiently slowly varying densities. Since the spin-independent GEA2 exchange energy per particle has the form 
\begin{equation}
\varepsilon_\x^\text{GEA2}(\rho,\nabla \rho) = \rho^{1/3} (C_\x + C_\x^{(2)} x^2),
\end{equation}
where $x=|\nabla \rho|/\rho^{4/3}$ is a dimensionless reduced density gradient, the precise condition for exchange is $x \ll 1$. Unfortunately, for real systems like atoms and molecules, the reduced density gradient $x$ can be large in some regions of space. In particular, in the exponential density tail, $\rho(\b{r}) \isPropTo{|\b{r}|\to\infty}  e^{-\alpha |\b{r}|}$, the reduced density gradient diverges $x(\b{r}) \xrightarrow[|\b{r}|\to\infty]{} \infty$. But this is not as bad as it seems since $\rho \; \varepsilon_\x^\text{GEA2}$ goes to zero anyway in this limit. The situation is more catastrophic for correlation. Indeed, in the high-density limit, the spin-independent GEA2 correlation energy per particle behaves as
\begin{equation}
\varepsilon_\c^\text{GEA2}(\gamma^3 \rho, \gamma^4 \nabla \rho) \isEquivTo{\g \to \infty} - A(0) \ln \gamma + \gamma^{1/2} C_\c^{(2)}(r_\s\to 0) y^2,
\end{equation}
where $y=|\nabla \rho|/\rho^{7/6}$ is another reduced density gradient adapted to correlation. Therefore, in this limit, the GEA2 correlation correction diverges to $+\infty$ even faster than the LDA diverges to $-\infty$.

Another aspect of the deficiency of the GEA is that the corresponding GEA exchange and correlation holes have unphysical long-range parts which break the negativity [Eq.~(\ref{nxneg})] and sum-rule conditions [Eqs.~(\ref{intnx}) and~(\ref{intnc})]. 

In practice, the GEA tends to deteriorate the results obtained at the LDA level. Truncated gradient expansions should not be directly used but need to be resummed.

\subsection{Generalized-gradient approximations}

The failures of the GEA led to the development, which really started in the 1980s, of {\it generalized-gradient approximations} (GGAs) with the generic form
\begin{equation}
E_\xc^\GGA[\rho_\uparrow,\rho_\downarrow] = \int_{\mathbb{R}^3} e_\text{xc}^\text{GGA}(\rho_\uparrow(\b{r}),\rho_\downarrow(\b{r}),\nabla \rho_\uparrow(\b{r}),\nabla \rho_\downarrow(\b{r})) \d\b{r},
\label{ExcGGA}
\end{equation}
where $e_\text{xc}^\text{GGA}$ is some function. The GGAs are often called {\it semilocal} approximations in the sense that $e_\text{xc}^\text{GGA}$ does not only use the local value of the spin densities $\rho_\uparrow(\b{r})$ and $\rho_\downarrow(\b{r})$ but also ``semilocal information'' through its gradients\footnote{For generality and simplicity, we consider here that the GGAs depend on the spin density gradients $\nabla \rho_\uparrow$ and $\nabla \rho_\downarrow$, but due to rotational invariance GGAs actually depend only on the scalar quantities $(\nabla \rho_\uparrow)^2$, $(\nabla \rho_\downarrow)^2$, and $\nabla \rho_\uparrow \cdot \nabla \rho_\downarrow$.} $\nabla \rho_\uparrow(\b{r})$ and $\nabla \rho_\downarrow(\b{r})$.

Many GGA functionals have been proposed. They generally provide a big improvement over LDA for molecular systems. However, their accuracy is still limited, in particular by self-interaction and static-correlation errors. We review here some of the most widely used GGA functionals.

\vspace{0.4cm}
\noindent
{\bf B88 exchange functional}

In the Becke 88 (B88 or B) exchange functional~\cite{Bec-PRA-88}, the exchange energy density is written as
\begin{eqnarray}
e_\x^\text{B88}(\rho_\uparrow,\rho_\downarrow,\nabla \rho_\uparrow,\nabla \rho_\downarrow) = e_\x^\text{UEG}(\rho_\uparrow,\rho_\downarrow) - \sum_{\sigma \in \{\uparrow,\downarrow\}} \rho_\sigma^{4/3} \frac{\beta x_\sigma^2}{1+6\beta x_\sigma \sinh^{-1}(x_\sigma)},
\label{exB88}
\end{eqnarray}
where $x_\sigma=|\nabla \rho_\sigma|/\rho_\sigma^{4/3}$. The fact that $e_\x^\text{B88}$ depends linearly on $\rho_\sigma^{4/3}$ and nonlinearly only on the dimensionless reduced density gradient $x_\sigma(\b{r})$ guarantees the scaling relation of Eq.~(\ref{Exrhog}). Using the exponential decay of the ground-state spin densities of Coulombic systems, $\rho_\sigma(\b{r}) \isPropTo{|\b{r}|\to\infty}  e^{-\alpha_\sigma |\b{r}|}$, it can be verified that the chosen form for $e_\x^\text{B88}$ satisfies the exact asymptotic behavior of the exchange energy density per particle [Eq.~(\ref{epsxinf})], although the corresponding exchange potential does not satisfy the exact $-1/r$ asymptotic behavior [Eq.~(\ref{vxrinf})]~\cite{EngCheMacVos-ZPD-92}. For small $x_\sigma$, $e_\x^\text{B88}$ is correctly quadratic in $x_\sigma$. The parameter $\beta=0.0042$ was found by fitting to HF exchange energies of rare-gas atoms. A very similar value of $\beta$ can also be found by imposing the coefficient $B_\x$ of the approximate large-$Z$ asymptotic expansion of the exchange energy of neutral atoms [Eq.~(\ref{ExrhoZZinf})]~\cite{EllBur-CJC-09}. It turns out that imposing the coefficient of the second-order gradient expansion [Eq.~(\ref{ExGEA})] would lead to a value of $\beta$ about two times smaller and would greatly deteriorate the accuracy of the functional for atoms and molecules. 

\vspace{0.4cm}
\noindent
{\bf LYP correlation functional}

The Lee--Yang--Parr (LYP)~\cite{LeeYanPar-PRB-88} correlation functional is one of the rare functionals which have not been constructed starting from LDA. It originates from the Colle--Salvetti~\cite{ColSal-TCA-75} correlation-energy approximation depending on the curvature of the HF hole. By using a gradient-expansion approximation of the curvature of the HF hole, LYP turned the Colle-Salvetti expression into a density functional depending on the density, the density gradient, and the Laplacian of the density. The dependence on the Laplacian of the density can be exactly eliminated by an integration by parts~\cite{MieSavStoPre-CPL-89}, giving the following correlation energy density
\begin{eqnarray}
e_\c^\text{LYP}(\rho_\uparrow,\rho_\downarrow,\nabla \rho_\uparrow,\nabla \rho_\downarrow) = 
-a \frac{4}{1+d\rho^{-1/3}} \frac{\rho_\uparrow \rho_\downarrow}{\rho} 
-a \, b \, \omega(\rho) 
\Biggl\{ \rho_\uparrow \rho_\downarrow \phantom{xxxxxxxxxxxxxxxxxxxxxx}
\nonumber\\
\times \Bigg[ \sum_{\sigma \in \{\uparrow,\downarrow\}} \left( 2^{11/3}C_\text{F} \rho_\sigma^{8/3} -\left(\frac{5}{2}-\frac{\delta(\rho)}{18}\right) |\nabla \rho_\sigma|^2 -\frac{\delta(\rho) -11}{9} \frac{\rho_\sigma}{\rho} |\nabla \rho_\sigma|^2 \right) +\left( \frac{47}{18} - \frac{7\delta(\rho)}{18} \right) |\nabla \rho|^2  \Bigg] 
\nonumber\\
-\frac{2}{3} \rho^2 |\nabla \rho|^2 
+ \left( \frac{2}{3} \rho^2 - \rho_\uparrow^2 \right) |\nabla \rho_\downarrow|^2
+ \left( \frac{2}{3} \rho^2 - \rho_\downarrow^2 \right) |\nabla \rho_\uparrow|^2 \Biggl\},
\phantom{xxxxx}
\end{eqnarray}
where $\omega(\rho) = \rho^{-11/3} \exp(-c \rho^{-1/3})/(1+d\rho^{-1/3})$, $\delta(\rho) = c \rho^{-1/3} + d \rho^{-1/3}/(1+d\rho^{-1/3})$, and $C_\text{F}=(3/10)(3\pi^2)^{2/3}$. The parameters $a=0.04918$, $b=0.132$, $c=0.2533$, and $d=0.349$ were obtained in the original Colle-Salvetti expression by a fit to Helium data. Note that the LYP correlation energy vanishes for fully spin-polarized densities ($\rho_\uparrow=0$ or $\rho_\downarrow=0$) and therefore correctly vanishes for one-electron systems [Eq.~(\ref{Ec1espin})].

\vspace{0.4cm}
\noindent
{\bf PW91 exchange-correlation functional}

The Perdew--Wang 91 (PW91) (see Refs.~\cite{Per-INC-91,PerCheVosJacPedSinFio-PRB-92,BurPerWan-INC-98}) exchange-correlation functional is based on a model of the exchange hole $h_\x(\b{r}_1,\b{r}_2)$ in Eq.~(\ref{epsx}) and of the coupling-constant-integrated correlation hole $\bar{h}_\c(\b{r}_1,\b{r}_2)$ in Eq.~(\ref{epsc}). The idea is to start from the GEA model of these holes given as gradient expansions and remove the unrealistic long-range parts of these holes to restore important constraints satisfied by the LDA. Specifically, the spurious positive parts of the GEA exchange hole are removed to enforce the negativity condition of Eq.~(\ref{nxneg}) and a cutoff in $|\b{r}_1 - \b{r}_2|$ is applied to enforce the normalization condition of Eq.~(\ref{intnx}). Similarly, a cutoff is applied on the GEA correlation hole to enforce the condition that the hole integrates to zero [Eq.~(\ref{inthcl})]. The exchange and correlation energies per particle calculated from these numerical holes are then fitted to functions of the density and density gradient chosen to satisfy a number of exact constraints.

The spin-independent PW91 exchange energy density is written as
\begin{eqnarray}
e_\x^\text{PW91}(\rho,\nabla \rho) = e_\x^\text{UEG}(\rho) F_\x^\text{PW91}(s),
\label{epsxPW91}
\end{eqnarray}
where the so-called enhancement factor is
\begin{eqnarray}
F_\x^\text{PW91}(s)= \frac{1+0.19645 s \sinh^{-1}(7.7956 s) + [0.2743 - 0.1508 \exp(-100 s^2)] s^2}{1+0.19645 s \sinh^{-1}(7.7956 s) + 0.004 s^4},
\label{FxPW91}
\end{eqnarray}
with the reduced density gradient $s=|\nabla \rho|/(2 k_\text{F} \rho) = x/[2(3 \pi^2)^{1/3}]$ where $k_\text{F} = (3 \pi^2 \rho)^{1/3}$ is the Fermi wave vector. The spin-dependent PW91 exchange energy density is simply obtained from the spin-scaling relation [Eq.~(\ref{Exspinscaling})]: $e_\x^\text{PW91}(\rho_\uparrow,\rho_\downarrow,\nabla \rho_\uparrow,\nabla \rho_\downarrow) = [e_\x^\text{PW91}(2\rho_\uparrow,2\nabla \rho_\uparrow) + e_\x^\text{PW91}(2\rho_\downarrow,2\nabla \rho_\downarrow)]/2$. The enhancement factor $F_\x^\text{PW91}(s)$ satisfies the second-order gradient expansion [Eq.~(\ref{ExGEA})], $F_\x^\text{PW91}(s)= 1 + \mu \; s^2 + O(s^4)$ with $\mu=-16\pi (\pi/3)^{2/3} C_\x^{(2)} = 10/81 \approx 0.1235$, the local Lieb--Oxford bound, $F_\x^\text{PW91}(s) \leq -C_\text{LO}/(C_\x 2^{1/3}) \approx 1.804$,
which is a sufficient and necessary condition for a spin-dependent GGA exchange functional to satisfy the Lieb--Oxford lower bound [Eq.~(\ref{LObound})] for all densities~\cite{PerRuzSunBur-JCP-14} (note however that $1.804$ is not an optimal bound), and the condition $\lim_{s\to\infty} s^{1/2} F_\x^\text{PW91}(s) < \infty$ which guarantees the non-uniform scaling finiteness conditions of Eqs.~(\ref{Exgamma1inf}) and~(\ref{Exgamma20})~\cite{LevPer-PRB-93,PerRuzSunBur-JCP-14}.

The PW91 correlation energy density is written as
\begin{eqnarray}
e_\c^\text{PW91}(\rho_\uparrow,\rho_\downarrow,\nabla \rho_\uparrow, \nabla \rho_\downarrow) = \rho \left[ \varepsilon_\c^\text{UEG}(\rho_\uparrow,\rho_\downarrow)  + H^\text{PW91}(\rho_\uparrow,\rho_\downarrow,t) \right],
\label{epscPW91}
\end{eqnarray}
where the gradient correction $H^\text{PW91}(\rho_\uparrow,\rho_\downarrow,t) = H_0(\rho_\uparrow,\rho_\downarrow,t) + H_1(\rho_\uparrow,\rho_\downarrow,t)$ depends on another reduced density gradient (adapted to correlation) $t=|\nabla \rho|/(2 \phi_2(\zeta) k_\s \rho) = y/[4\phi_2(\zeta)(3/\pi)^{1/6}]$ where $k_\s=\sqrt{4 k_\text{F}/\pi}$ is the Thomas--Fermi screening wave vector and the spin-scaling function $\phi_2(\zeta)$ is defined by Eq.~(\ref{phinzeta}), with
\begin{eqnarray}
H_0(\rho_\uparrow,\rho_\downarrow,t) = \phi_2(\zeta)^3 \frac{\beta^2}{2\alpha} \ln \left[ 1 + \frac{2\alpha}{\beta} t^2 \frac{1+{\cal A} t^2}{1+{\cal A} t^2 + {\cal A}^2 t^4} \right],
\label{}
\end{eqnarray}
\begin{eqnarray}
{\cal A} = \frac{2\alpha}{\beta} \left[ \exp(-2 \alpha \varepsilon_\c^\text{UEG}(\rho_\uparrow,\rho_\downarrow)/(\phi_2(\zeta)^3\beta^2)) -1 \right]^{-1},
\label{}
\end{eqnarray}
\begin{eqnarray}
H_1(\rho_\uparrow,\rho_\downarrow,t) = 16 \left(\frac{3}{\pi} \right)^{1/3} [ C_\xc(r_\s) - C_{\c,\text{MB}}^{(2)}(r_\s\to 0) -C_\x^{(2)}] \phi_2(\zeta)^3 t^2 e^{-100 \phi_2(\zeta)^4 k_\s^2 t^2/k_\text{F}^2},
\label{}
\end{eqnarray}
and $C_\xc(r_\s)$ is taken from Ref.~\cite{RasGel-PRB-86}. The function $H_0(\rho_\uparrow,\rho_\downarrow,t)$ was chosen so that it fulfills the second-order gradient expansion [Eq.~(\ref{EcGEAspin})], $H_0(\rho_\uparrow,\rho_\downarrow,t) = \beta \phi_2(\zeta)^3 t^2 + O(t^4)$, using an approximate $\zeta$ dependence~\cite{WanPer-PRB-91} and the Ma-Brueckner high-density-limit second-order coefficient~\cite{MaBru-PR-68} $\beta = 16(3/\pi)^{1/3} C_{\c,\text{MB}}^{(2)}(r_\s\to 0) \approx 0.06673$, and so that it cancels the LDA correlation in the large-$t$ limit, $\lim_{t\to\infty} H_0(\rho_\uparrow,\rho_\downarrow,t) = - \varepsilon_\c^\text{UEG}(\rho_\uparrow,\rho_\downarrow)$. The only fitted parameter is $\alpha=0.09$. The function $H_1(\rho_\uparrow,\rho_\downarrow,t)$ only serves to restore the correct second-order gradient expansion, such that $H^\text{PW91}(\rho_\uparrow,\rho_\downarrow,t) = 16 (3/\pi)^{1/3} C_\xc(r_\s) \phi_2(\zeta)^3 t^2 + O(t^4)$, while keeping the large-$t$ limit unchanged. 

\vspace{0.4cm}
\noindent
{\bf PBE exchange-correlation functional}

The Perdew--Burke--Ernzerhof (PBE)~\cite{PerBurErn-PRL-96} exchange-correlation functional is a simplification of the PW91 functional with no fitted parameters which gives almost the same energies. The spin-independent PBE exchange energy density is written as
\begin{eqnarray}
e_\x^\text{PBE}(\rho,\nabla \rho) = e_\x^\text{UEG}(\rho) F_\x^\text{PBE}(s),
\label{epsxPBE}
\end{eqnarray}
where the enhancement factor is
\begin{eqnarray}
F_\x^\text{PBE}(s)= 1 + \kappa - \frac{\kappa}{1 + \mu s^2/\kappa}.
\label{FxPBE}
\end{eqnarray}
The function $F_\x^\text{PBE}(s)$ has the second-order gradient expansion $F_\x^\text{PBE}(s)= 1 + \mu \; s^2 + O(s^4)$, and the parameter is chosen as $\mu = 16\pi (\pi/3)^{2/3} C_{\c,\text{MB}}^{(2)}(r_\s\to 0) \approx 0.21951$ so as to cancel the correlation second-order gradient expansion. The second parameter $\kappa$ is chosen so as to saturate the local Lieb--Oxford bound, i.e.  $\lim_{s\to\infty} F_\x^\text{PBE}(s) = 1 + \kappa = -C_\text{LO}/(C_\x 2^{1/3})\approx 1.804$, leading to $\kappa = 0.804$. The same exchange functional form was in fact proposed earlier in the Becke 86 (B86) functional~\cite{Bec-JCP-86} with empirical parameters ($\mu=0.235$, $\kappa=0.967$). 

A revised version of the PBE exchange functional, called revPBE, was proposed where the local Lieb--Oxford bound constraint is relaxed and the parameter $\kappa = 1.245$ is found instead by fitting to exchange-only total atomic energies for He and Ar, resulting in more accurate atomic total energies and molecular atomization energies~\cite{ZhaYan-PRL-98}. Another revised version of the PBE exchange functional, called RPBE, was also proposed to achieve a similar improvement, while still enforcing the local Lieb--Oxford bound, by changing the form of the enhancement factor to $F_\x^\text{RPBE}(s) = 1+\kappa (1-\exp(-\mu s^2/\kappa))$ with the same parameters as in the original PBE~\cite{HamHanNor-PRB-99}.

The PBE correlation energy density is written as
\begin{eqnarray}
e_\c^\text{PBE}(\rho_\uparrow,\rho_\downarrow,\nabla \rho_\uparrow, \nabla \rho_\downarrow) = \rho \left[ \varepsilon_\c^\text{UEG}(\rho_\uparrow,\rho_\downarrow)  + H^\text{PBE}(\rho_\uparrow,\rho_\downarrow,t) \right],
\label{epscPBE}
\end{eqnarray}
with the gradient correction
\begin{eqnarray}
H^\text{PBE}(\rho_\uparrow,\rho_\downarrow,t) = A(0) \phi_2(\zeta)^3 \ln \left[ 1 + \frac{\beta}{A(0)} t^2 \frac{1+ {\cal A} t^2}{1+{\cal A} t^2 + {\cal A}^2 t^4} \right]
\label{}
\end{eqnarray}
and
\begin{eqnarray}
{\cal A} = \frac{\beta}{A(0)} \left[ \exp(-\varepsilon_\c^\text{UEG}(\rho_\uparrow,\rho_\downarrow)/(A(0) \phi_2(\zeta)^3)) -1 \right]^{-1}.
\label{}
\end{eqnarray}
As in the PW91 correlation functional, the function $H^\text{PBE}(\rho_\uparrow,\rho_\downarrow,t)$ has the second-order gradient expansion $H^\text{PBE}(\rho_\uparrow,\rho_\downarrow,t) = \beta \phi_2(\zeta)^3 t^2 + O(t^4)$ where $\beta = 16(3/\pi)^{1/3} C_{\c,\text{MB}}^{(2)}(r_\s\to 0) \approx 0.06673$, and it cancels the LDA correlation in the large-$t$ limit, $\lim_{t\to\infty} H^\text{PBE}(\rho_\uparrow,\rho_\downarrow,t) = - \varepsilon_\c^\text{UEG}(\rho_\uparrow,\rho_\downarrow)$. In contrast with the PW91 correlation functional, under uniform coordinate scaling to the high-density limit, the PBE correlation functional correctly cancels out the logarithm divergence of the LDA correlation functional [Eq.~(\ref{epscUEGgammainf})], i.e. $H^\text{PBE}(\gamma^3\rho_{\uparrow},\gamma^3\rho_\downarrow,\gamma^{1/2}t) \isEquivTo{\g \to \infty} A(0) \phi_2(\zeta)^3 \; \ln \g$ where $A(0) \phi_2(\zeta)^3$ is a good approximation to the coefficient $A(\zeta)$~\cite{WanPer-PRB-91}. 

A variant of the PBE exchange-correlation functional, called PBEsol~\cite{PerRuzCsoVydScuConZhoBur-PRL-08}, targeted for solid-state systems, was proposed where the correct second-order exchange gradient-expansion coefficient is restored, i.e. $\mu_\text{PBEsol} = 16\pi (\pi/3)^{2/3} C_{\x}^{(2)} = 10/81 \approx 0.1235$, and the second-order correlation gradient-expansion coefficient $\beta_\text{PBEsol}=0.046$ is found by fitting to jellium surface exchange-correlation energies.

\vspace{0.4cm}
\noindent
{\bf B97-GGA exchange-correlation functional}

The Becke 97 GGA (B97-GGA) exchange-correlation functional is the GGA part of the B97 hybrid functional~\cite{Bec-JCP-97} (see Section~\ref{sec:hybrid}). The B97-GGA exchange energy density is
\begin{eqnarray}
e_\x^\text{B97-GGA}(\rho_\uparrow,\rho_\downarrow,\nabla \rho_\uparrow, \nabla \rho_\downarrow) = \sum_{\sigma \in \{\uparrow,\downarrow\}} e_{\x,\sigma}^\text{UEG}(\rho_\sigma) \; g_{\x}(x_\sigma),
\label{exB97-GGA}
\end{eqnarray}
where $e_{\x,\sigma}^\text{UEG}(\rho_\sigma) = e_\x^\text{UEG}(\rho_\sigma,0)$ is the spin-$\sigma$ contribution to the UEG exchange energy density and the gradient correction $g_{\x}(x_\sigma)$ is a function of $x_\sigma = |\nabla \rho_\sigma |/\rho_\sigma^{4/3}$,
\begin{eqnarray}
g_{\x}(x_\sigma) = \sum_{i=0}^m c_{\x,i} \; u_{\x}(x_\sigma)^i,
\label{gx}
\end{eqnarray}
with $u_{\x}(x_\sigma) = \gamma_{\x} x_\sigma^2/(1+ \gamma_{\x} x_\sigma^2)$. The B97-GGA correlation energy density is written as the sum of opposite- and same-spin contributions
\begin{eqnarray}
e_\c^\text{B97-GGA}(\rho_\uparrow,\rho_\downarrow,\nabla \rho_\uparrow, \nabla \rho_\downarrow) = e_{\c,\uparrow\downarrow}^\text{B97-GGA}(\rho_\uparrow,\rho_\downarrow,\nabla \rho_\uparrow, \nabla \rho_\downarrow) + \sum_{\sigma \in \{\uparrow,\downarrow\}} e_{\c,\sigma\sigma}^\text{B97-GGA}(\rho_\sigma,\nabla \rho_\sigma),
\label{ecB97-GGA}
\end{eqnarray}
where 
\begin{eqnarray}
e_{\c,\uparrow\downarrow}^\text{B97-GGA}(\rho_\uparrow,\rho_\downarrow,\nabla \rho_\uparrow, \nabla \rho_\downarrow) = e_{\c,\uparrow\downarrow}^\text{UEG}(\rho_\uparrow,\rho_\downarrow)\; g_{\c,\uparrow\downarrow}(x_{\uparrow\downarrow})
\end{eqnarray}
and
\begin{eqnarray}
e_{\c,\sigma\sigma}^\text{B97-GGA}(\rho_\sigma,\nabla \rho_\sigma) = e_{\c,\sigma\sigma}^\text{UEG}(\rho_\sigma)\; g_{\c,\sigma\sigma}(x_\sigma).
\label{ecssB97-GGA}
\end{eqnarray}
In these expressions, $e_{\c,\uparrow\downarrow}^\text{UEG}(\rho_\uparrow,\rho_\downarrow)=e_{\c}^\text{UEG}(\rho_\uparrow,\rho_\downarrow) - e_{\c}^\text{UEG}(\rho_\uparrow,0) - e_{\c}^\text{UEG}(\rho_\downarrow,0)$ and $e_{\c,\sigma\sigma}^\text{UEG}(\rho_\alpha) = e_{\c}^\text{UEG}(\rho_\sigma,0)$ are estimations of the opposite- and same-spin contributions to the UEG correlation energy density~\cite{StoPavPre-TCA-78,StoGolPre-TCA-80}. The opposite-spin gradient correction is taken as a function of $x_{\uparrow\downarrow} = \sqrt{(x_\uparrow^2 + x_\downarrow^2)/2}$,
\begin{eqnarray}
g_{\c,\uparrow\downarrow}(x_{\uparrow\downarrow}) = \sum_{i=0}^m c_{\c,i}^{\uparrow\downarrow} \; u_{\c}^{\uparrow\downarrow} (x_{\uparrow\downarrow})^i,
\label{gcud}
\end{eqnarray}
with $u_{\c}^{\uparrow\downarrow}(x_{\uparrow\downarrow}) = \gamma_{\c}^{\uparrow\downarrow} x_{\uparrow\downarrow}^2/(1+ \gamma_{\c}^{\uparrow\downarrow} x_{\uparrow\downarrow}^2)$, and the same-spin gradient correction is
\begin{eqnarray}
g_{\c,\sigma\sigma}(x_\sigma) = \sum_{i=0}^m c_{\c,i}^{\sigma\sigma} \; u_{\c}^{\sigma\sigma} (x_\sigma)^i,
\label{gcss}
\end{eqnarray}
with $u_{\c}^{\sigma\sigma}(x_\sigma) = \gamma_{\c}^{\sigma\sigma} x_\sigma^2/(1+ \gamma_{\c}^{\sigma\sigma} x_\sigma^2)$. The parameters $\gamma_\x = 0.004$, $\gamma_{\c}^{\uparrow\downarrow}=0.006$, and $\gamma_{\c}^{\sigma\sigma}=0.2$, were roughly optimized on atomic exchange and correlation energies. The other parameters $c_{\x,i}$, $c_{\c,i}^{\uparrow\downarrow}$, $c_{\c,i}^{\sigma\sigma}$ for a polynomial degree $m=2$ in Eqs.~(\ref{gx}),~(\ref{gcud}), and~(\ref{gcss}) were optimized in the B97 hybrid functional in the presence of a fraction of HF exchange energy (see Section~\ref{sec:hybrid}). 

The Hamprecht--Cohen--Tozer--Handy (HCTC)~\cite{HamCohTozHan-JCP-98} exchange-correlation functional uses the same form as the B97-GGA exchange-correlation functional but with a polynomial degree $m=4$ and the parameters $c_{\x,i}$, $c_{\c,i}^{\uparrow\downarrow}$, $c_{\c,i}^{\sigma\sigma}$ were optimized without HF exchange a set of energetic properties (atomic total energies, ionization energies, atomization energies), nuclear gradients, and accurate exchange-correlation potentials.

\subsection{Meta-generalized-gradient approximations}

The {\it meta-generalized-gradient approximations} (meta-GGAs or mGGAs) are of the generic form, in their spin-independent version,
\begin{equation}
E_\xc^\mGGA[\rho,\tau] = \int_{\mathbb{R}^3} e_\xc^\text{mGGA}(\rho(\b{r}),\nabla \rho(\b{r}),\nabla^2 \rho(\b{r}), \tau(\b{r})) \d\b{r},
\label{ExcmGGA}
\end{equation}
i.e., they use more ingredients than the GGAs, namely the Laplacian of the density $\nabla^2 \rho(\b{r})$ and/or the non-interacting positive kinetic energy density $\tau(\b{r})$ associated with a single-determinant wave function $\Phi$,
\begin{eqnarray}
\tau(\b{r}) = \tau_{\Phi}(\b{r}) &=& \frac{N}{2} \int_{ \{\uparrow,\downarrow\} \times (\mathbb{R}^3 \times \{\uparrow,\downarrow\})^{N-1}} \left| \nabla_\b{r} \Phi(\b{x},\b{x}_2,...,\b{x}_N) \right|^2 \d\sigma \d \b{x}_2 ... \d \b{x}_N 
\nonumber\\
&=& \frac{1}{2} \sum_{i=1}^{N} \left| \nabla \varphi_{i} (\b{r}) \right|^2,
\label{tau}
\end{eqnarray}
where $\{\varphi_{i}\}_{i=1,...,N}$ are the orbitals occupied in $\Phi$. The meta-GGAs are considered as part of the family of semilocal approximations, in the sense that $\tau(\b{r})$ contains semilocal information with respect to the orbitals. 

Meta-GGAs can be viewed as implicit functionals of the density only, i.e. $E_\xc^\mGGA[\rho,\tau_{\Phi[\rho]}]$, since $\tau(\b{r})$ can be considered itself as an implicit functional of the density via the KS single-determinant wave function $\Phi[\rho]$. This view in which $E_\xc^\mGGA[\rho]$ is a proper approximation to the exchange-correlation density functional $E_\xc[\rho]$ of the KS scheme is normally adopted when constructing meta-GGAs approximations. However, the calculation of the functional derivative of $E_\xc^\mGGA[\rho]$ with respect to the density then requires the use of the complicated optimized-effective-potential method (see Section~\ref{sec:orbdep}). Therefore, in practical calculations, meta-GGAs are usually reinterpreted as explicit functionals of a single-determinant wave function $\Phi$, i.e. $E_\xc^\mGGA[\rho_{\Phi},\tau_{\Phi}]$,~\cite{NeuNobHan-MP-96,NeuHan-CPL-96,AdaErnScu-JCP-00,ArbKauMalRevMal-PCCP-02,FurPer-JCP-06,SunMarCsoRuzHaoKimKrePer-PRB-11,ZahLeaGor-JCP-13,SouShaTou-JCP-14} or, in other words, approximations to an exact GKS exchange-correlation functional (see Section~\ref{GKS}).

In the latter approach, which we will here refer to as the meta-Kohn--Sham (mKS) scheme, we introduce a functional $E_\xc^\mKS[\rho,\tau]$ (to which meta-GGAs are approximations) defined for $\rho$ and $\tau$ simultaneously representable by a single-determinant wave function $\Phi \in {\cal S}^N$ and which defines the GKS functional $E_\xc^{S}[\Phi]=E_\xc^\mKS[\rho_{\Phi},\tau_{\Phi}]$ [see Eq.~(\ref{ExcGKS})] giving the exact ground-state energy via Eq.~(\ref{E0GKS}),
\begin{eqnarray}
E_0 = \inf_{\Phi \in {\cal S}^N} \left\{ \bra{\Phi} \hat{T} + \hat{V}_\tne \ket{\Phi} + E_\H[\rho_\Phi] + E_\xc^\mKS[\rho_{\Phi},\tau_{\Phi}] \right\},
\end{eqnarray}
which, by taking variations with respect to the orbitals, gives the mKS equations:
\begin{equation}
\left( -\frac{1}{2}\nabla^2 +v_{\tne}(\b{r}) + v_\H(\b{r}) + v_\xc^\mKS(\b{r}) \right) \varphi_i(\b{r}) = \varepsilon_i \varphi_i(\b{r}).
\label{KSeqsmGGA}
\end{equation}
Here, $v_\xc^\mKS(\b{r})=v_{\xc,1}^\mKS(\b{r})+v_{\xc,2}^\mKS(\b{r})$ contains a usual local potential
\begin{equation}
v_{\xc,1}^\mKS(\b{r}) = \frac{\delta E_\xc^\mKS[\rho,\tau]}{\delta \rho(\b{r})},
\label{VxcmGGA1}
\end{equation}
and a non-multiplicative operator~\cite{AdaErnScu-JCP-00,ArbKauMalRevMal-PCCP-02,FurPer-JCP-06,SunMarCsoRuzHaoKimKrePer-PRB-11,ZahLeaGor-JCP-13}
\begin{equation}
v_{\xc,2}^\mKS(\b{r}) = 
- \frac{1}{2} \nabla \cdot \left( \frac{\delta E_\xc^\mKS[\rho,\tau]}{\delta \tau(\b{r})} \nabla \right),
\label{VxcmGGA2}
\end{equation}
evaluated with $\rho(\b{r}) =\sum_{i=1}^N |\varphi_i(\b{r})|^2$ and $\tau(\b{r})=(1/2) \sum_{i=1}^{N} | \nabla \varphi_{i} (\b{r}) |^2$.
Interestingly, the mKS equations can be rewritten as a Schr\"odinger-like equation with a position-dependent mass $m(\b{r})$~\cite{EicHel-JCP-14},
\begin{equation}
\left( -\frac{1}{2}\nabla \cdot \frac{1}{m(\b{r})} \nabla  +v_{\tne}(\b{r}) + v_\H(\b{r}) + v_{\xc,1}^\mKS(\b{r}) \right) \varphi_i(\b{r}) = \varepsilon_i \varphi_i(\b{r}),
\label{KSeqsmKSlocalm}
\end{equation}
where $m(\b{r}) = \left(1+\delta  E_\xc^\mKS[\rho,\tau]/\delta \tau(\b{r})\right)^{-1}$. As in the KS scheme, the functional $E_\xc^\mKS[\rho,\tau]$ is decomposed into exchange and correlation contributions: $E_\xc^\mKS[\rho,\tau] = E_\x^\mKS[\rho,\tau] + E_\c^\mKS[\rho,\tau]$. In the spin-dependent version of the mKS scheme, we consider a similar functional of the spin-resolved densities and non-interacting positive kinetic energy densities $E_\xc^\mKS[\rho_\uparrow,\rho_\downarrow,\tau_\uparrow,\tau_\downarrow]$ and the spin-scaling relation of Eq.~(\ref{Exspinscaling}) is generalized to
\begin{equation}
E_\x^\mKS[\rho_\uparrow,\rho_\downarrow,\tau_\uparrow,\tau_\downarrow]  = \frac{1}{2} \left( E_\x^\mKS[2\rho_\uparrow,2\tau_\uparrow] + E_\x^\mKS[2\rho_\downarrow,2\tau_\downarrow] \right).
\label{ExspinscalingmGGA}
\end{equation}
Correspondingly, the spin-dependent versions of the meta-GGAs are formulated in terms of the spin-resolved quantities $\rho_\uparrow$, $\rho_\downarrow$, $\nabla \rho_\uparrow$, $\nabla \rho_\downarrow$, $\nabla^2 \rho_\uparrow$, $\nabla^2 \rho_\downarrow$, $\tau_\uparrow$, and $\tau_\downarrow$. 

One motivation for the introduction of the variable $\tau(\b{r})$ is that it appears in the expansion of the spherically averaged exchange hole [entering in Eq.~(\ref{epsx})] for small interelectronic distances $r_{12}$~\cite{Bec-IJQC-83}, which for the case of a closed-shell system is
\begin{eqnarray}
\frac{1}{4\pi r_{12}^2} \int_{S(\b{0},r_{12})} h_\x(\b{r}_1,\b{r}_1+\b{r}_{12}) \d \b{r}_{12} = -\frac{\rho(\b{r}_1)}{2} \phantom{xxxxxxxxxxxxxxxxxxx}
\nonumber\\
- \frac{1}{3} \left( \frac{1}{4} \nabla^2 \rho(\b{r}_1) - \tau(\b{r}_1) + \frac{|\nabla \rho(\b{r}_1)|^2}{8 \rho(\b{r}_1)}\right) r_{12}^2 + O(r_{12}^4),
\label{hxsmallr12}
\end{eqnarray}
where $S(\b{0},r_{12})$ designates the sphere centered at $\b{0}$ and of radius $r_{12}=|\b{r}_{12}|$. Thus $\tau(\b{r})$ is needed to describe the curvature of the exchange hole.

Another important motivation is that $\tau(\b{r})$ is useful for identifying different types of spatial regions of electronic systems~\cite{SunXiaFanHauHaoRuzCsoScuPer-PRL-13}. This is done by comparing $\tau(\b{r})$ with the von Weizs\"acker kinetic energy density,
\begin{equation}
\tau^\text{W}(\b{r}) = \frac{|\nabla \rho(\b{r}) |^2}{8 \rho(\b{r})},
\label{tauW}
\end{equation}
which is the exact non-interacting kinetic energy density for one-electron systems and two-electron spin-unpolarized systems, and, more generally, for one-orbital regions as introduced in Section~\ref{sec:onetwoelectron}. For example, the indicator
\begin{equation}
z(\b{r}) = \frac{\tau^\text{W}(\b{r})}{\tau(\b{r})},
\label{z}
\end{equation}
which takes its values in the range $[0,1]$~\cite{KurPerBla-IJQC-99}, identifies one-orbital regions ($z=1$). A better indicator is
\begin{equation}
\alpha(\b{r}) = \frac{\tau(\b{r}) - \tau^\text{W}(\b{r})}{\tau^\text{UEG}(\b{r})},
\label{alpha}
\end{equation}
where $\tau^\text{UEG}(\b{r}) = (3/10) (3\pi^2)^{2/3} \rho(\b{r})^{5/3}$ is the non-interacting kinetic energy density of the UEG. This indicator $\alpha(\b{r})$ distinguishes one-orbital regions ($\alpha =0$), slowly varying density regions ($\alpha \approx 1$), and regions of density overlap between closed shells that characterize noncovalent bonds ($\alpha \gg 1$).

Nowadays, $\nabla^2 \rho(\b{r})$ is rarely used to construct meta-GGAs because it contains similar information to $\tau(\b{r})$, which can be seen by the second-order gradient expansion of $\tau(\b{r})$~\cite{BraJenChu-PL-76}:
\begin{equation}
\tau^\text{GEA2}(\b{r}) = \tau^\text{UEG}(\b{r}) + \frac{1}{72} \frac{|\nabla \rho(\b{r}) |^2}{\rho(\b{r})} + \frac{1}{6} \nabla^2 \rho(\b{r}).
\label{tauGEA}
\end{equation}

In comparison to GGAs, meta-GGAs are more versatile and generally constitute an improvement. Significantly, thanks to the use of $\tau$, self-interaction errors in the correlation functional can be essentially eliminated with meta-GGAs. They still suffer however from self-interaction errors in the exchange functional. We now describe some of the most used meta-GGA functionals.

\vspace{0.4cm}
\noindent
{\bf TPSS exchange-correlation functional}

In the Tao--Perdew--Staroverov--Scuseria (TPSS)~\cite{TaoPerStaScu-PRL-03,PerTaoStaScu-JCP-04} functional, the exchange energy density is written as
\begin{eqnarray}
e_\x^\text{TPSS}(\rho,\nabla \rho,\tau) = e_\x^\text{UEG}(\rho) F_\x^\text{TPSS}(s,z),
\label{epsxTPSS}
\end{eqnarray}
where the enhancement factor is a function of $s=|\nabla \rho|/(2 k_\text{F} \rho)$ and $z=\tau^\text{W}/\tau$,
\begin{eqnarray}
F_\x^\text{TPSS}(s,z)= 1 + \kappa - \frac{\kappa}{1 + x^\text{TPSS}(s,z)/\kappa},
\label{FxTPSS}
\end{eqnarray}
with $\kappa=0.804$ so as to saturate the local Lieb--Oxford bound (just like in the PBE exchange functional) and
\begin{eqnarray}
x^\text{TPSS}(s,z) &=& \Biggl[ \left( \frac{10}{81} + c \frac{z^2}{(1+z^2)^2} \right) s^2 + \frac{146}{2025} \tilde{q}_b^2
-\frac{73}{405} \tilde{q}_b \sqrt{\frac{1}{2} \left( \frac{3}{5} z \right)^2 + \frac{1}{2} s^4} 
\nonumber\\
&&+ \frac{1}{\kappa} \left(\frac{10}{81} \right)^2 s^4 + 2 \sqrt{e} \frac{10}{81} \left( \frac{3}{5} z \right)^2 + e \mu s^6 \Biggl]/\left(1+\sqrt{e} s^2\right)^2,
\label{xTPSS}
\end{eqnarray}
and $\tilde{q}_b = (9/20) (\alpha -1)/[1 + b \alpha (\alpha -1)]^{1/2} + 2s^2/3$ (where $\alpha = (\tau - \tau^\text{W})/\tau^\text{UEG} = (5s^2/3) (z^{-1} -1)$) is a quantity that tends to the reduced density Laplacian $q=\nabla^2 \rho/(4 k_\text{F}^2 \rho)$ in the slowly varying density limit [using Eq.~(\ref{tauGEA})]. The function $x^\text{TPSS}(s,z)$ is chosen so as to satisfy the fourth-order gradient expansion [Eq.~(\ref{ExGEA})] which can be written in the form of the enhancement factor $F_\x^\text{GEA4}(s,z) = 1+ (10/81) s^2 +(146/2025) q^2 -(73/405) s^2 q$. The constant $\mu=0.21951$ is chosen to retain the same large-$s$ behavior of the PBE exchange functional, i.e. $F_\x^\text{TPSS}(s,z) \isEquivTo{s\to \infty} F_\x^\text{PBE}(s)$. The constants $c=1.59096$ and $e=1.537$ are chosen so as to eliminate the divergence of the potential at the nucleus for a two-electron exponential density and to yield the correct exchange energy ($-0.3125$ hartree) for the exact ground-state density of the hydrogen atom. Finally, the constant $b=0.40$ is chosen, quite arbitrarily, as the smallest value that makes $F_\x^\text{TPSS}(s,z)$ a monotonically increasing function of $s$.

The TPSS correlation functional is constructed by making minor refinements to the previously developed Perdew--Kurth--Zupan--Blaha (PKZB)~\cite{PerKurZupBla-PRL-99} meta-GGA correlation functional,
\begin{eqnarray}
e_\c^\text{TPSS}(\rho_\uparrow,\rho_\downarrow,\nabla \rho_\uparrow, \nabla \rho_\downarrow, \tau_\uparrow,\tau_\downarrow) &=& \rho \; \varepsilon_\c^\text{revPKZB}(\rho_\uparrow,\rho_\downarrow,\nabla \rho_\uparrow, \nabla \rho_\downarrow, \tau_\uparrow,\tau_\downarrow) 
\nonumber\\
&& \times \left[ 1 + d \; \varepsilon_\c^\text{revPKZB}(\rho_\uparrow,\rho_\downarrow,\nabla \rho_\uparrow, \nabla \rho_\downarrow, \tau_\uparrow,\tau_\downarrow) z^3 \right],
\label{epscTPSS}
\end{eqnarray}
where the revised PKZB correlation energy per particle is
\begin{eqnarray}
\varepsilon_\c^\text{revPKZB}(\rho_\uparrow,\rho_\downarrow,\nabla \rho_\uparrow, \nabla \rho_\downarrow, \tau_\uparrow,\tau_\downarrow) = \varepsilon_\c^\text{PBE}(\rho_\uparrow,\rho_\downarrow,\nabla \rho_\uparrow, \nabla \rho_\downarrow) \left[ 1 + C(\zeta,\xi)z^2\right]
\nonumber\\
 - \left[1 + C(\zeta,\xi) \right] z^2 \sum_{\sigma \in \{\uparrow,\downarrow\}} \frac{\rho_\sigma}{\rho} \tilde{\varepsilon}_{\c,\sigma}^\text{PBE}(\rho_\uparrow,\rho_\downarrow,\nabla \rho_\uparrow, \nabla \rho_\downarrow),
\label{epscrevPKZB}
\end{eqnarray}
with $\tilde{\varepsilon}_{\c,\sigma}^\text{PBE}(\rho_\uparrow,\rho_\downarrow,\nabla \rho_\uparrow, \nabla \rho_\downarrow)=\max[\varepsilon_{\c}^\text{PBE}(\rho_\sigma,0,\nabla \rho_\sigma, 0),\varepsilon_{\c}^\text{PBE}(\rho_\uparrow,\rho_\downarrow,\nabla \rho_\uparrow, \nabla \rho_\downarrow)]$\\ where $\varepsilon_{\c}^\text{PBE}(\rho_\uparrow,\rho_\downarrow,\nabla \rho_\uparrow, \nabla \rho_\downarrow)$ is the PBE correlation energy per particle. Equation~(\ref{epscrevPKZB}) constitutes a one-electron self-interaction correction on the PBE correlation functional. Indeed, for one-electron densities we have $z=1$ and $\zeta=\pm 1$, and the TPSS correlation energy correctly vanishes [Eqs.~(\ref{Ec1espin})]. The TPSS correlation functional preserves many properties of the PBE correlation functional: it has correct uniform coordinate scaling in the high- and low-density limits, vanishing correlation energy in the large density-gradient limit, and the same second-order gradient expansion (since the additional terms beyond PBE are at least in $z^2$ and thus only change the fourth-order terms of the gradient expansion). The parameters $d=2.8$ hartree$^{-1}$ and $C(0,0)=0.53$ are chosen so as to recover the PBE surface correlation energy of jellium~\cite{LanKoh-PRB-70} over the range of valence-electron bulk densities. The rest of the function is taken as
\begin{eqnarray}
C(\zeta,\xi) = \frac{0.53 + 0.87 \zeta^2 + 0.50 \zeta^4 + 2.26 \zeta^6}{(1+\xi^2 [ (1+\zeta)^{-4/3} + (1-\zeta)^{-4/3}]/2)^4},
\label{}
\end{eqnarray}
where $\xi = |\nabla \zeta|/(2 k_\text{F})$ is a reduced spin-polarization gradient. The function $C(\zeta,\xi)$ is chosen so as to make the exchange-correlation energy independent of the spin polarization $\zeta$ in the low-density limit [Eq.~(\ref{Ecrhog0})] and to avoid that the self-interaction correction introduces additional correlation energy density in the core-valence overlap region of monovalent atoms such as Li. 

\vspace{0.4cm}
\noindent
{\bf M06-L exchange-correlation functional}

In the Minnesota 06 local (M06-L) exchange-correlation functional~\cite{ZhaTru-JCP-06}, the exchange energy density is written as
\begin{eqnarray}
e_\x^\text{M06-L}(\rho_\uparrow,\rho_\downarrow,\nabla \rho_\uparrow, \nabla \rho_\downarrow,\tau_\uparrow,\tau_\downarrow) = \sum_{\sigma \in \{\uparrow,\downarrow\}} e_{\x,\sigma}^\text{PBE}(\rho_\sigma,\nabla \rho_\sigma)f(w_\sigma) + e_{\x,\sigma}^\text{UEG}(\rho_\sigma)h_\x(x_\sigma, Z_\sigma).
\label{exM06L}
\end{eqnarray}
The first term in Eq.~(\ref{exM06L}), which has the same form as in the previously developed M05 exchange functional~\cite{ZhaSchTru-JCP-05}, contains the spin-$\sigma$ PBE exchange energy density $e_{\x,\sigma}^\text{PBE}(\rho_\sigma,\nabla \rho_\sigma) = e_{\x}^\text{PBE}(\rho_\sigma,0,\nabla \rho_\sigma,0)$ and the kinetic-energy density correction factor
\begin{eqnarray}
f(w_\sigma) = \sum_{i=0}^{11} a_i w_\sigma^i,
\label{fws}
\end{eqnarray}
where $w_\sigma=(\tau_\sigma^\text{UEG}/\tau_\sigma -1)/(\tau_\sigma^\text{UEG}/\tau_\sigma+1)$ with $\tau_\sigma^\text{UEG}=(3/10) (6\pi^2)^{2/3} \rho_\sigma^{5/3}$ is an indicator of the delocalization of the exchange hole~\cite{Bec-JCP-00}. The second term in Eq.~(\ref{exM06L}), which has the same form as in the VS98 exchange functional~\cite{VooScu-JCP-98}, contains the spin-$\sigma$ UEG exchange energy density $e_{\x,\sigma}^\text{UEG}(\rho_\sigma) = e_\x^\text{UEG}(\rho_\sigma,0)$ and the correction factor 
\begin{eqnarray}
h_\x(x_\sigma, Z_\sigma) = h(x_\sigma, Z_\sigma,d_{\x,0},d_{\x,1},d_{\x,2},d_{\x,3},d_{\x,4},\alpha_\x), 
\label{hx}
\end{eqnarray}
where $x_\sigma=|\nabla \rho_\sigma|/\rho_\sigma^{4/3}$ and $Z_\sigma = 2(\tau_\sigma-\tau_\sigma^\text{UEG})/\rho_\sigma^{5/3}$ and $h$ is the parametrized function
\begin{eqnarray}
h(x,Z,d_0,d_1,d_2,d_3,d_4,\alpha) = \frac{d_{0}}{\gamma(x,Z,\alpha)} + \frac{d_{1} x^2 + d_{2} Z}{\gamma(x,Z,\alpha)^2} + \frac{d_{3} x^4 + d_{4} x^2 Z}{\gamma(x,Z,\alpha)^3},
\end{eqnarray}
with $\gamma(x,Z,\alpha)=1+\alpha(x^2+Z)$.

The M06-L correlation energy is written as the sum of opposite- and same-spin contributions, similarly to the B97-GGA correlation functional [Eq.~(\ref{ecB97-GGA})],
\begin{eqnarray}
e_\c^\text{M06-L}(\rho_\uparrow,\rho_\downarrow,\nabla \rho_\uparrow, \nabla \rho_\downarrow,\tau_\uparrow,\tau_\downarrow) &=& e_{\c,\uparrow\downarrow}^\text{M06-L}(\rho_\uparrow,\rho_\downarrow,\nabla \rho_\uparrow, \nabla \rho_\downarrow,\tau_\uparrow,\tau_\downarrow) 
\nonumber\\
&&+ \sum_{\sigma \in \{\uparrow,\downarrow\}} e_{\c,\sigma\sigma}^\text{M06-L}(\rho_\sigma,\nabla \rho_\sigma,\tau_\sigma),
\end{eqnarray}
where 
\begin{eqnarray}
e_{\c,\uparrow\downarrow}^\text{M06}(\rho_\uparrow,\rho_\downarrow,\nabla \rho_\uparrow, \nabla \rho_\downarrow,\tau_\uparrow,\tau_\downarrow) = e_{\c,\uparrow\downarrow}^\text{UEG}(\rho_\uparrow,\rho_\downarrow)\; \left[ g_{\c,\uparrow\downarrow}(x_{\uparrow\downarrow}) + h_{\c,\uparrow\downarrow}(x_{\uparrow\downarrow},Z_{\uparrow\downarrow}) \right],
\end{eqnarray}
and
\begin{eqnarray}
e_{\c,\sigma\sigma}^\text{M06-L}(\rho_\sigma,\nabla \rho_\sigma) = e_{\c,\sigma\sigma}^\text{UEG}(\rho_\sigma)\; \left[ g_{\c,\sigma\sigma}(x_\sigma) + h_{\c,\sigma\sigma}(x_\sigma,Z_\sigma) \right] D_\sigma(z_\sigma),
\end{eqnarray}
where the spin-decomposed UEG correlation energies $e_{\c,\uparrow\downarrow}^\text{UEG}(\rho_\uparrow,\rho_\downarrow)$ and $e_{\c,\sigma\sigma}^\text{UEG}(\rho_\sigma)$ were already defined after Eq.~(\ref{ecssB97-GGA}), and the gradient corrections $g_{\c,\uparrow\downarrow}(x_{\uparrow\downarrow})$ and $g_{\c,\sigma\sigma}(x_\sigma)$ are given in Eqs.~(\ref{gcud}) and~(\ref{gcss}). The additional correction factors are 
\begin{eqnarray}
h_\c^{\uparrow\downarrow}(x_{\uparrow\downarrow}, Z_{\uparrow\downarrow}) = h(x_{\uparrow\downarrow}, Z_{\uparrow\downarrow},d_{\c,0}^{\uparrow\downarrow},d_{\c,1}^{\uparrow\downarrow},d_{\c,2}^{\uparrow\downarrow},d_{\c,3}^{\uparrow\downarrow},d_{\c,4}^{\uparrow\downarrow},\alpha_\c^{\uparrow\downarrow}),
\label{hcud}
\end{eqnarray}
where $x_{\uparrow\downarrow} = \sqrt{(x_\uparrow^2 + x_\downarrow^2)/2}$, $Z_{\uparrow\downarrow} = Z_\uparrow + Z_\downarrow$, and 
\begin{eqnarray}
h_\c^{\sigma\sigma}(x_{\sigma}, Z_{\sigma}) = h(x_{\sigma}, Z_{\sigma},d_{\c,0}^{\sigma\sigma},d_{\c,1}^{\sigma\sigma},d_{\c,2}^{\sigma\sigma},d_{\c,3}^{\sigma\sigma},d_{\c,4}^{\sigma\sigma},\alpha_\c^{\sigma\sigma}).
\label{hcss}
\end{eqnarray}
The factor $D_\sigma(z_\sigma)=1-z_\sigma$, where $z_\sigma=\tau^\text{W}_\sigma/\tau_\sigma$ and $\tau^\text{W}_\sigma=|\nabla \rho_\sigma|^2/(8\rho_\sigma)$, ensures that the correlation energy correctly vanishes for one-electron systems~\cite{Bec-JCP-98}.

The parameters $\gamma_{\c}^{\uparrow\downarrow}=0.0031$, and $\gamma_{\c}^{\sigma\sigma}=0.06$, were optimized on the correlation energies of He and Ne. The parameters $\alpha_\x=0.001867$, $\alpha_{\c}^{\uparrow\downarrow}=0.003050$, and $\alpha_{\c}^{\sigma\sigma}=0.005151$ were taken from Ref.~\cite{VooScu-JCP-98}. The constraints $a_0+d_{\x,0}=1$, $c_{\c,0}^{\uparrow\downarrow}+d_{\c,0}^{\uparrow\downarrow}=1$, and $c_{\c,0}^{\sigma\sigma}+d_{\c,0}^{\sigma\sigma}=1$ are enforced to obtain the correct UEG limit. The remaining 34 free parameters $a_i$, $c_{\c,i}^{\uparrow\downarrow}$, $c_{\c,i}^{\sigma\sigma}$ for a polynomial degree $m=4$ in Eqs.~(\ref{fws}),~(\ref{gcud}), and~(\ref{gcss}), and $d_{\x,i}$, $d_{\c,i}^{\uparrow\downarrow}$, $d_{\c,i}^{\sigma\sigma}$ in Eqs.~(\ref{hx}),~(\ref{hcud}), and~(\ref{hcss}) were optimized on a large set of diverse physicochemical properties concerning main-group thermochemistry, reaction barrier heights, noncovalent interactions, electronic spectroscopy, and transition metal bonding.

\vspace{0.4cm}
\noindent
{\bf SCAN exchange-correlation functional}

In the SCAN (strongly constrained and appropriately normed)~\cite{SunRuzPer-PRL-15} exchange-correlation functional, the exchange energy density is written as
\begin{eqnarray}
e_\x^\text{SCAN}(\rho,\nabla \rho,\tau) = e_\x^\text{UEG}(\rho) F_\x^\text{SCAN}(s,\alpha),
\label{exSCAN}
\end{eqnarray}
where the enhancement factor is a function of $s=|\nabla \rho|/(2 k_\text{F} \rho)$ and $\alpha = (\tau - \tau^\text{W})/\tau^\text{UEG}$,
\begin{eqnarray}
F_\x^\text{SCAN}(s,\alpha) = [h_\x^1(s,\alpha) +f_\x(\alpha) (h_\x^0 - h_\x^1(s,\alpha))] g_\x(s),
\label{FxSCAN}
\end{eqnarray}
which interpolates between $\alpha=0$ and $\alpha\approx 1$, and extrapolates to $\alpha\to\infty$ using the function  
\begin{eqnarray}
f_\x(\alpha) = \exp[-c_{1\x}\alpha/(1-\alpha)] \theta(1-\alpha) - d_\x \exp[\c_{2\x}/(1-\alpha)] \theta(\alpha-1),
\label{}
\end{eqnarray}
where $\theta$ is the Heaviside step function. The function $g_\x(s) = 1 - \exp(-a_1 s^{-1/2})$ is chosen to make $F_\x^\text{SCAN}(s,\alpha)$ vanish like $s^{-1/2}$ as $s\to\infty$, which guarantees the non-uniform scaling finiteness conditions [Eqs.~(\ref{Exgamma1inf}) and~(\ref{Exgamma20})]~\cite{LevPer-PRB-93,PerRuzSunBur-JCP-14}, and $a_1=4.9479$ is taken to recover the exact exchange energy of the hydrogen atom. For $\alpha \approx 1$ (slowly varying density regions), $F_\x^\text{SCAN}(s,\alpha) \approx h_\x^1(s,\alpha) g_\x(s)$ where $h_\x^1(s,\alpha)$ is a PBE-like resummation of the fourth-order gradient expansion [Eq.~(\ref{ExGEA})],
\begin{eqnarray}
h_\x^1(s,\alpha)=1 + k_1 - \frac{k_1}{1+x^\text{SCAN}(s,\alpha)/k_1},
\label{}
\end{eqnarray}
where
\begin{eqnarray}
x^\text{SCAN}(s,\alpha) = \mu s^2 [ 1 + (b_4 s^2/\mu) e^{-|b_4| s^2 /\mu}] + [b_1 s^2 + b_2 (1-\alpha) e^{ -b_3 (1-\alpha)^2}]^2,
\label{}
\end{eqnarray}
with $\mu = 10/81$, $b_2=(5913/405000)^{1/2}$, $b_1=(511/13500)/(2b_2)$, $b_3=0.5$, and $b_4=\mu^2/k_1-1606/18225-b_1^2$. For $\alpha=0$ (one-orbital regions), $F_\x^\text{SCAN}(s,\alpha=0) = h_\x^0 g_\x(s)$ where $h_\x^0=1.174$ is chosen to saturate the local two-electron tight bound $F_\x^\text{SCAN}(s,\alpha=0) \leq 1.174$, which is a sufficient and necessary condition for a meta-GGA exchange functional to satisfy the global tight bound of Eq.~(\ref{LObound2eC1}) for all two-electron spin-unpolarized densities~\cite{PerRuzSunBur-JCP-14}. 

The SCAN correlation energy density is written as
\begin{eqnarray}
e_\c^\text{SCAN}(\rho_\uparrow,\rho_\downarrow,\nabla \rho_\uparrow, \nabla \rho_\downarrow, \tau_\uparrow,\tau_\downarrow) = \rho  \; [\varepsilon_\c^{1}(\rho_\uparrow,\rho_\downarrow,t) + f_\c(\alpha) (\varepsilon_\c^{0}(\rho_\uparrow,\rho_\downarrow,s)-\varepsilon_\c^{1}(\rho_\uparrow,\rho_\downarrow,t)) ],
\label{ecSCAN}
\end{eqnarray}
which is again an interpolation between $\alpha=0$ and $\alpha=1$, and an extrapolation to $\alpha\to\infty$ using the function
\begin{eqnarray}
f_\c(\alpha) = \exp[-c_{1\c}\alpha/(1-\alpha)] \theta(1-\alpha) - d_\c \exp[\c_{2\c}/(1-\alpha)] \theta(\alpha-1).
\label{}
\end{eqnarray}
For $\alpha=1$, the correlation energy par particle is taken as a revised version of the PBE correlation energy per particle,
\begin{eqnarray}
\varepsilon_\c^{1}(\rho_\uparrow,\rho_\downarrow,t) = \varepsilon_\c^{\text{UEG}}(\rho_\uparrow,\rho_\downarrow) + H_1^\text{SCAN}(\rho_\uparrow,\rho_\downarrow,t)
\label{epscSCAN1}
\end{eqnarray}
where 
\begin{eqnarray}
H_1^\text{SCAN}(\rho_\uparrow,\rho_\downarrow,t) = A(0) \phi_2(\zeta)^3 \ln \left[ 1 + w_1 (1- g({\cal A} t^2)) \right],
\end{eqnarray}
with $t=|\nabla \rho|/(2 \phi_2(\zeta) k_\s \rho)$, $w_1 = \exp[-\varepsilon_\c^{\text{UEG}}(\rho_\uparrow,\rho_\downarrow)/(A(0)\phi_2(\zeta)^3)]-1$, ${\cal A} = \beta(r_\s)/(A(0)w_1)$, and $g({\cal A} t^2)=1/(1+4{\cal A} t^2)^{1/4}$. The function has a second-order gradient expansion $H_1^\text{SCAN}(\rho_\uparrow,\rho_\downarrow,t) = \beta(r_\s) \phi_2(\zeta) t^2 + O(t^4)$ where the coefficient $\beta(r_\s) = 0.066725(1+0.1 r_\s)/(1+ 0.1778 r_\s)$ is a rough fit of the density dependence of the second-order gradient expansion correlation coefficient beyond the Ma-Brueckner high-density-limit value and designed so that for $r_\s \to \infty$ the second-order gradient expansion terms for exchange and correlation cancel each other~\cite{PerRuzCsoConSun-PRL-09}. For $\alpha=0$, the correlation energy par particle is constructed to be accurate for one- and two-electron systems and is written as
\begin{eqnarray}
\varepsilon_\c^{0}(\rho_\uparrow,\rho_\downarrow,s) = [ \varepsilon_\c^{\text{LDA0}}(\rho) + H_0^\text{SCAN}(\rho,s) ] G_\c(\zeta).
\label{epscSCAN0}
\end{eqnarray}
The spin function $G_\c(\zeta) = [1-2.3631(\phi_4(\zeta)-1)](1-\zeta^{12})$ is designed to make the correlation energy vanish for one-electron densities ($\alpha=0$ and $\zeta=\pm 1$) and to make the exchange-correlation energy independent of $\zeta$ in the low-density limit [Eq.~(\ref{Ecrhog0})]. Equation~(\ref{epscSCAN0}) includes a LDA-type term~\cite{SunPerYanPen-JCP-16}
\begin{eqnarray}
\varepsilon_\c^{\text{LDA0}}(\rho) = - \frac{b_{1\c}}{1+b_{2\c} r_\s^{1/2} + b_{3\c} r_\s},
\label{}
\end{eqnarray}
and a gradient correction
\begin{eqnarray}
H_0^\text{SCAN}(\rho,s) = b_{1\c} \ln \left[ 1+ w_0 (1-g_\infty(\zeta=0,s))\right],
\label{}
\end{eqnarray}
with $w_0 = \exp(-\varepsilon_\c^{\text{LDA0}}(\rho)/b_{1\c}) -1$ and $g_\infty(\zeta=0,s)= \lim_{\zeta \to 0} \lim_{r_\s \to \infty} g({\cal A} t) = 1/(1+0.512104 s^2)^{1/4}$. The parameter $b_{1\c}=0.0285764$ is determined so that the high-density limit of $\varepsilon_\c^{0}(\rho_\uparrow,\rho_\downarrow,s)$ reproduces the exact correlation energy of the Helium isoelectronic series in the large-nuclear charge limit, i.e. $\lim_{Z\to\infty} E_\c[\rho_{N=2,Z}] = E_\c^{\GL}[\rho_{N=2,Z=1}^\text{H}]=-0.0467$ hartree [Eq.~(\ref{EcZinf})]. The parameter $b_{3\c}=0.125541$ is determined to saturate the lower bound on the exchange-correlation energies of two-electron densities [Eq.~(\ref{LObound2eC2})]. The parameter $b_{2\c}=0.0889$ is determined to reproduce the exact exchange-correlation energy of the He atom.

The remaining seven parameters ($k_1 = 0.065$, $c_{1\x} = 0.667$, $c_{2\x} = 0.8$, $d_\x = 1.24$, $c_{1\c} = 0.64$, $c_{2\c} =1.5$, and $d_\c = 0.7$) are determined by fitting to the approximate asymptotic expansions of the exchange and correlation energies of neutral atoms in large nuclear charge limit [Eqs.~(\ref{ExrhoZZinf}) and~(\ref{EcrhoZZinf})], the binding energy curve of compressed Ar$_2$, and jellium surface exchange-correlation energies.

\section{Single-determinant hybrid approximations}
\stepcounter{myequation}
\label{sec:sdha}

\subsection{Hybrid approximations}
\label{sec:hybrid}

Based on arguments relying on the adiabatic-connection formalism, in 1993 Becke~\cite{Bec-JCP-93a} proposed to mix a fraction of the exact or Hartree-Fock (HF) exchange energy $E_\x^{\HF}$ with GGA functionals. In particular, he proposed a \textit{three-parameter hybrid (3H) approximation}~\cite{Bec-JCP-93} of the form, written here in its spin-independent version,
\begin{equation}
E_\xc^{\text{3H}}[\Phi] = a \; E_\x^{\HF}[\Phi] + b \; E_\x^\GGA[\rho_\Phi] + (1-a-b) \; E_\x^\LDA[\rho_\Phi] + c  \; E_\c^\GGA[\rho_\Phi] + (1-c) \; E_\c^\LDA[\rho_\Phi],
\label{Exc3H}
\end{equation}
with empirical parameters $a$, $b$, and $c$. The functional $E_\xc^{\text{3H}}[\Phi]$ is thought of as a functional of a single-determinant wave function $\Phi \in {\cal S}^N$ since $E_\x^{\HF}[\Phi]$ is itself a functional of $\Phi$,
\begin{eqnarray}
E_{\x}^\HF[\Phi] &=& \bra{\Phi} \hat{W}_\text{ee} \ket{\Phi} - E_\H[\rho_\Phi]
\nonumber\\
&=&- \frac{1}{2} \sum_{\sigma\in\{\uparrow,\downarrow\}} \sum_{i=1}^{N_\sigma} \sum_{j=1}^{N_\sigma} \int_{\mathbb{R}^3\times\mathbb{R}^3} \frac{\varphi_{i \sigma}^*(\b{r}_1)\varphi_{j \sigma}(\b{r}_1) \varphi_{j \sigma}^*(\b{r}_2)\varphi_{i \sigma}(\b{r}_2)}{|\b{r}_1-\b{r}_2|}\d\b{r}_1 \d\b{r}_2,
\label{ExHF}
\end{eqnarray}
where $\{\varphi_{i\sigma}\}_{i=1,...,N_\sigma}$ are the orbitals occupied in $\Phi$. In 1996, Becke proposed a simpler \textit{one-parameter hybrid (1H) approximation}~\cite{Bec-JCP-96},
\begin{eqnarray}
E_\xc^{\text{1H}}[\Phi] = a \; E_\x^{\HF}[\Phi] +  (1-a) \; E_\x^\text{GGA}[\rho_\Phi] + E_\c^\text{GGA}[\rho_\Phi],
\label{Exc1H}
\end{eqnarray}
where the fraction $a$ of HF exchange has to be determined. For simplicity, we considered GGA functionals $E_\x^\text{GGA}[\rho_\Phi]$ and $E_\c^\text{GGA}[\rho_\Phi]$ in Eq.~(\ref{Exc1H}) but we can more generally use meta-GGA functionals $E_\x^\text{mGGA}[\rho_\Phi,\tau_\Phi]$ and $E_\c^\text{mGGA}[\rho_\Phi,\tau_\Phi]$.

These hybrid approximations should be considered as approximations of the GKS exchange-correlation functional $E_\xc^{S}[\Phi]$ in Eq.~(\ref{ExcGKS}) with $S[\Phi]=a \; E_\x^{\HF}[\Phi]$. The corresponding GKS equations [Eq.~(\ref{GKS})] then include the term
\begin{equation}
\frac{\delta S[\Phi]}{\delta \varphi_{i\sigma}^*(\b{r})} = a \int_{\mathbb{R}^3} v_{\x,\sigma}^\HF (\b{r},\b{r}') \varphi_{i\sigma} (\b{r}') \d \b{r}',
\label{dSdphihybrid}
\end{equation}
where $v_{\x,\sigma}^\HF (\b{r},\b{r}')$ is the nonlocal HF exchange potential
\footnote{The possibility of combining a nonlocal HF potential with a local correlation potential was mentioned already in 1965 in the paper by Kohn and Sham~\cite{KohSha-PR-65}.}
\begin{equation}
v_{\x,\sigma}^\HF (\b{r},\b{r}') = - \sum_{j=1}^{N_\sigma} \frac{\varphi_{j\sigma}(\b{r})\varphi_{j\sigma}^*(\b{r}')}{|\b{r}-\b{r}'|}.
\end{equation}

The main benefit of adding a fraction of HF exchange is to decrease the self-interaction error (see Section~\ref{sec:onetwoelectron}) introduced by semilocal exchange functionals which tends to favor too much delocalized electron densities over localized electron densities. The fraction of HF exchange should however be small enough to keep the compensation of errors usually occurring between the approximate semilocal exchange and correlation functionals. First, Becke used the value $a=0.5$ in the so-called Becke Half-and-Half functional~\cite{Bec-JCP-93a}, but then fits to various experimental data often repeatedly gave an optimal parameter $a$ around $0.20$-$0.25$. A rationale has been proposed in favor of the value $0.25$~\cite{PerErnBur-JCP-96}. By decreasing self-interaction errors in the exchange energy, hybrid approximations are often a big improvement over semilocal approximations for molecular systems with sufficiently large electronic gaps. However, for systems with small HOMO-LUMO gaps, such as systems with stretched chemical bonds or with transition metal elements, they tend to increase static-correlation errors. 

An interesting extension of the hybrid approximations are the so-called local hybrids, which use a position-dependent fraction $a(\b{r})$ of a (non-uniquely defined) HF exchange energy density $e_\x^{\HF}(\b{r})$~\cite{JarScuErn-JCP-03} (see, Ref.~\cite{MaiArbKau-WIRES-18} for a recent review), and which belong to the wider family of hyper-GGA functionals in which the correlation energy can also be expressed as a function of $e_\x^{\HF}(\b{r})$~\cite{PerSch-AIPCP-01}. The local-hybrid approximations are much more flexible than the global hybrid approach exposed in this Section but require more complicated and computationally expensive implementations. For this reason, they have not been often used and we will not consider them any further here.

We now describe some of the most used hybrid approximations.

\vspace{0.4cm}
\noindent
{\bf B3LYP exchange-correlation functional}

The B3LYP exchange-correlation functional~\cite{SteDevChaFri-JPC-94} is the most famous and widely used three-parameter hybrid approximation [Eq.~(\ref{Exc3H})]. It uses the B88 exchange functional and the LYP correlation functional,
\begin{eqnarray}
E_\xc^{\text{B3LYP}}[\Phi] &=& a \; E_\x^{\HF}[\Phi] + b \; E_\x^\text{B88}[\rho_{\uparrow,\Phi},\rho_{\downarrow,\Phi}] + (1-a-b) \; E_\x^\LSDA[\rho_{\uparrow,\Phi},\rho_{\downarrow,\Phi}] 
\nonumber\\
&& + c  \; E_\c^\text{LYP}[\rho_{\uparrow,\Phi},\rho_{\downarrow,\Phi}] + (1-c) \; E_\c^\LSDA[\rho_{\uparrow,\Phi},\rho_{\downarrow,\Phi}],
\label{B3LYP}
\end{eqnarray}
and the parameters $a=0.20$, $b=0.72$, and $c=0.81$ were found by optimizing on a set of atomization energies, ionization energies, proton affinities of small molecules and first-row total atomic energies~\cite{Bec-JCP-93}. A caveat is that the VWN parametrization of the RPA correlation energy (sometimes referred to as VWN3) of the UEG was actually used for $E_\c^\LSDA[\rho_{\uparrow},\rho_{\downarrow}]$ instead of the VWN parametrization of the accurate correlation energy (sometimes referred to as VWN5) of the UEG~\cite{VosWilNus-CJP-80}.

\vspace{0.4cm}
\noindent
{\bf B97 exchange-correlation functional}

The Becke 97 (B97) exchange-correlation functional~\cite{Bec-JCP-97} is a GGA hybrid of the form
\begin{eqnarray}
E_\xc^{\text{B97}}[\Phi] = a \; E_\x^{\HF}[\Phi] +  (1-a) \; E_\x^\text{B97-GGA}[\rho_{\uparrow,\Phi},\rho_{\downarrow,\Phi}] + E_\c^\text{B97-GGA}[\rho_{\uparrow,\Phi},\rho_{\downarrow,\Phi}],
\label{ExcB97}
\end{eqnarray}
where the form of the B97-GGA exchange and correlation functionals were given in Eqs.~(\ref{exB97-GGA}) and~(\ref{ecB97-GGA}). The fraction of HF exchange $a=0.1943$ and the remaining parameters $c_{\x,0}=1.00459$, $c_{\x,1}=0.629639$, $c_{\x,2}=0.928509$, $c_{\c,0}^{\uparrow\downarrow}=0.9454$, $c_{\c,1}^{\uparrow\downarrow}=0.7471$, $c_{\c,2}^{\uparrow\downarrow}=-4.5961$, $c_{\c,0}^{\sigma\sigma}=0.1737$, $c_{\c,1}^{\sigma\sigma}=2.3487$, and $c_{\c,2}^{\sigma\sigma}=-2.4868$ for a polynomial degree $m=2$ in Eqs.~(\ref{gx}),~(\ref{gcud}), and~(\ref{gcss}) were optimized on a set of total energies, atomization energies, ionization energies, and proton affinities. Note that, for $x_\sigma=0$, the UEG limit is not imposed, which would require the parameters $c_{\x,0}$, $c_{\c,0}^{\uparrow\downarrow}$, and $c_{\c,0}^{\sigma\sigma}$ to be all strictly equal to $1$. With the above optimized parameters, we see that it is nearly satisfied for the exchange energy and the opposite-spin correlation energy, but very far from it for the same-spin correlation energy which is drastically reduced compared to the LDA.

\vspace{0.4cm}
\noindent
{\bf PBE0 exchange-correlation functional}

The PBE0 exchange-correlation functional~\cite{AdaBar-JCP-99,ErnScu-JCP-99a} is a GGA hybrid using the PBE exchange and correlation functionals,
\begin{eqnarray}
E_\xc^{\text{PBE0}}[\Phi] = a \; E_\x^{\HF}[\Phi] +  (1-a) \; E_\x^\text{PBE}[\rho_{\uparrow,\Phi},\rho_{\downarrow,\Phi}] + E_\c^\text{PBE}[\rho_{\uparrow,\Phi},\rho_{\downarrow,\Phi}],
\label{ExcPBE0}
\end{eqnarray}
and the fraction of the HF exchange is fixed at $a=0.25$ according to the rationale of Ref.~\cite{PerErnBur-JCP-96}. This functional is also known under the name PBE1PBE. The ``1'' in the latter name emphasizes that there is one parameter, $a$, while the ``0'' in the more common name PBE0 emphasizes that this parameter is not found by fitting.

\vspace{0.4cm}
\noindent
{\bf TPSSh exchange-correlation functional}

The TPSSh exchange-correlation functional~\cite{StaScuTaoPer-JCP-03} is a meta-GGA hybrid using the TPSS exchange and correlation functionals,
\begin{eqnarray}
E_\xc^{\text{TPSSh}}[\Phi] = a \; E_\x^{\HF}[\Phi] +  (1-a) \; E_\x^\text{TPSS}[\rho_{\uparrow,\Phi},\rho_{\downarrow,\Phi},\tau_{\uparrow,\Phi},\tau_{\downarrow,\Phi}] + E_\c^\text{TPSS}[\rho_{\uparrow,\Phi},\rho_{\downarrow,\Phi},\tau_{\uparrow,\Phi},\tau_{\downarrow,\Phi}], \;
\label{ExcTPSSh}
\end{eqnarray}
and the fraction of the HF exchange $a=0.10$ was determined by optimizing on a large set of atomization energies.

\vspace{0.4cm}
\noindent
{\bf M06 and M06-2X exchange-correlation functionals}

The M06 exchange-correlation functional~\cite{ZhaTru-TCA-08} is a meta-GGA hybrid using the M06-L exchange and correlation functionals,
\begin{eqnarray}
E_\xc^{\text{M06}}[\Phi] = a \; E_\x^{\HF}[\Phi] +  (1-a) \; E_\x^\text{M06-L}[\rho_{\uparrow,\Phi},\rho_{\downarrow,\Phi},\tau_{\uparrow,\Phi},\tau_{\downarrow,\Phi}] + E_\c^\text{M06-L}[\rho_{\uparrow,\Phi},\rho_{\downarrow,\Phi},\tau_{\uparrow,\Phi},\tau_{\downarrow,\Phi}], \;
\label{ExcM06}
\end{eqnarray}
and the parameters in the M06-L exchange and correlation functionals were reoptimized together with the fraction of HF exchange $a=0.27$ on the same large set of diverse physicochemical properties used for the M06-L functional. In the M06-2X exchange-correlation functional the fraction of HF exchange is doubled, i.e. $a=0.54$, and the parameters were reoptimized with the function $h_\x(x_\sigma, Z_\sigma)$ in Eq.~(\ref{hx}) set to zero and excluding transition metal properties in the training set. With this large fraction of HF exchange, the M06-2X functional is designed for systems without transition metal elements.

\subsection{Range-separated hybrid approximations}
\label{sec:rsh}

Based on earlier ideas of Savin~\cite{Sav-INC-96} (exposed in details in Section~\ref{sec:rsdh}), in 2001, Iikura, Tsuneda, Yanai, and Hirao~\cite{IikTsuYanHir-JCP-01} proposed a \textit{long-range correction (LC) scheme} in which the exchange-correlation energy is written as, in its spin-independent version,
\begin{eqnarray}
E_\xc^{\text{LC}}[\Phi] =  E_\x^{\lr,\mu,\HF}[\Phi] +  E_\x^{\sr,\mu,\GGA}[\rho_\Phi] + E_\c^{\GGA}[\rho_\Phi].
\label{ExcLC}
\end{eqnarray}
This scheme has also been referred to as the range-separated hybrid exchange (RSHX) scheme~\cite{GerAng-CPL-05a}.
In Eq.~(\ref{ExcLC}), $E_\x^{\lr,\mu,\HF}[\Phi]$ is the HF exchange energy for a long-range electron-electron interaction $w_{\ee}^{\lr,\mu}(r_{12}) = \erf(\mu r_{12})/r_{12}$ (where $\erf$ is the error function and the parameter $\mu \in [0,+\infty)$ controls the range of the interaction),
\begin{eqnarray}
E_{\x}^{\lr,\mu,\HF}[\Phi] = 
- \frac{1}{2} \sum_{\sigma\in\{\uparrow,\downarrow\}} \sum_{i=1}^{N_\sigma} \sum_{j=1}^{N_\sigma} \int_{\mathbb{R}^3\times\mathbb{R}^3} \varphi_{i\sigma}^*(\b{r}_1)\varphi_{j\sigma}(\b{r}_1) \varphi_{j\sigma}^*(\b{r}_2)\varphi_{i\sigma}(\b{r}_2) w_{\ee}^{\lr,\mu}(r_{12}) \d\b{r}_1 \d\b{r}_2, \;\;\;
\label{ExlrHF}
\end{eqnarray}
and $E_\x^{\sr,\mu,\GGA}[\rho]$ is a GGA exchange energy functional for the complementary short-range interaction $w_{\ee}^{\sr,\mu}(r_{12}) = 1/r_{12} - w_{\ee}^\lr(r_{12})$. This latter functional can be thought of as an approximation to the short-range exchange functional
\begin{equation}
E_\x^{\sr,\mu}[\rho]  = \frac{1}{2} \int_{\mathbb{R}^3\times\mathbb{R}^3} \rho(\b{r}_1) h_{\x}(\b{r}_1,\b{r}_2) w_{\ee}^{\sr,\mu}(r_{12}) \d\b{r}_1 \d\b{r}_2,
\label{Exsrint}
\end{equation}
where $h_{\x}(\b{r}_1,\b{r}_2)$ is the KS exchange hole of Section~\ref{exchangecorrelationholes}. For $\mu=0$, the long-range HF exchange energy vanishes, i.e. $E_{\x}^{\lr,\mu=0,\HF}[\Phi]=0$, and the short-range exchange functional reduces to the standard KS exchange functional, i.e. $E_\x^{\sr,\mu=0}[\rho] = E_\x[\rho]$. Reversely, for $\mu\to\infty$, the long-range HF exchange energy reduces to the full-range HF exchange energy, i.e. $E_{\x}^{\lr,\mu\to\infty,\HF}[\Phi]=E_{\x}^{\HF}[\Phi]$, and the short-range exchange functional vanishes, i.e. $E_\x^{\sr,\mu\to\infty}[\rho] = 0$. Significantly, for large $\mu$, the short-range exchange functional becomes a local functional of the density~\cite{GilAdaPop-MP-96,TouColSav-PRA-04}:
\begin{equation}
E_\x^{\sr,\mu}[\rho] \isEquivTo{\mu \to \infty} -\frac{\pi}{4\mu^2} \int_{\mathbb{R}^3}  \rho(\b{r})^2 \d\b{r}.
\label{Exsrmuinf}
\end{equation}

Like the hybrid approximations of Section~\ref{sec:hybrid}, Eq.~(\ref{ExcLC}) should be considered as an approximation of the GKS exchange-correlation functional $E_\xc^{S}[\Phi]$ in Eq.~(\ref{ExcGKS}) with $S[\Phi]= E_\x^{\lr,\mu,\HF}[\Phi]$, and the corresponding GKS equations [Eq.~(\ref{GKS})] then includes a long-range nonlocal HF exchange potential $v_{\x,\sigma}^{\lr,\mu,\HF}(\b{r}_1,\b{r}_2) = - \sum_{j=1}^{N_\sigma} \varphi_{j\sigma}(\b{r}_1)\varphi_{j\sigma}^*(\b{r}_2) w_{\ee}^{\lr,\mu}(r_{12})$. Similarly to the hybrid approximations, the introduction of a fraction of long-range HF exchange reduces the self-interaction error (see, e.g., Ref.~\cite{MusTou-MP-17}). In addition, the short-range exchange part is easier to approximate with semilocal density-functional approximations, as Eq.~(\ref{Exsrmuinf}) strongly suggests. In particular, the $-1/r$ asymptotic behavior of the exchange potential [Eq.~(\ref{vxrinf})], which is difficult to satisfy with semilocal approximations, does not apply anymore to the short-range exchange potential. 

In 2004, Yanai, Tew, and Handy~\cite{YanTewHan-CPL-04}, introduced a more flexible scheme called the Coulomb-attenuating method (CAM)~\cite{YanTewHan-CPL-04} in which fractions of HF exchange are added at both short range and long range,
\begin{equation}
E_\xc^{\text{CAM}}[\Phi] =  a \; E_\x^{\sr,\mu,\HF}[\Phi] +  b \; E_\x^{\lr,\mu,\HF}[\Phi] + (1-a) \; E_\x^{\sr,\mu,\GGA}[\rho_\Phi] + (1-b) \; E_\x^{\lr,\mu,\GGA}[\rho_\Phi] + E_\c^{\GGA}[\rho_\Phi],
\label{ExcCAM}
\end{equation}
where $E_\x^{\sr,\mu,\HF}[\Phi]=E_\x^{\HF}[\Phi] - E_\x^{\lr,\mu,\HF}[\Phi]$ is the short-range HF exchange energy and $E_\x^{\lr,\mu,\GGA}=E_\x^{\GGA} - E_\x^{\sr,\mu,\GGA}$ is a long-range GGA exchange energy. The reintroduction of HF exchange at short range further reduces the self-interaction error and improves thermodynamic properties such as atomization energies. Again, Eq.~(\ref{ExcCAM}) should be considered as an approximation of the GKS exchange-correlation functional $E_\xc^{S}[\Phi]$ in Eq.~(\ref{ExcGKS}) with $S[\Phi]= a \; E_\x^{\sr,\mu,\HF}[\Phi] +  b \; E_\x^{\lr,\mu,\HF}[\Phi]$. Other forms of modified electron-electron interactions are also possible (see, e.g., Refs.~\cite{SavFla-IJQC-95,TouColSav-PRA-04,HenIzmScuSav-JCP-07}).

The approximations in Eqs.~(\ref{ExcLC}) and (\ref{ExcCAM}) are usually collectively referred to as \textit{range-separated hybrid approximations}. Range-separated hybrids in the form of Eq.~(\ref{ExcCAM}) are more flexible than the hybrid approximations of Section~\ref{sec:hybrid}, and consequently are potentially more accurate, in particular for long-range electronic excitations. However, like the hybrid approximations, the presence of HF exchange tends to induce static-correlation errors for systems with small HOMO-LUMO gaps. 

The range-separation parameter $\mu$ (also sometimes denoted as $\omega$) is generally chosen empirically, e.g. by fitting to experimental data. In practice, a value around $\mu \approx 0.3 - 0.5$ bohr$^{-1}$, fixed for all systems, is often found to be optimal. It has also been proposed to adjust the value of $\mu$ in each system, e.g. by requiring that the opposite of the HOMO energy be equal to the ionization energy calculated by total energy differences~\cite{SteKroBae-JACS-09,SteKroBae-JCP-09,BaeLivSal-ARPC-10}. These so-called optimally tuned range-separated hybrids are well suited for the calculation of charge-transfer electronic excitations but have the disadvantage of not being size consistent~\cite{KarKroKum-JCP-13}. 

A natural idea is to use a position-dependent range-separation parameter $\mu(\b{r})$ which allows the range of the modified interaction to adapt to the local average electron-electron distance in the diverse spatial regions of the system. These locally range-separated hybrids~\cite{KruScuPerSav-JCP-08,AscKum-JCP-19,KlaBah-JCTC-20} are promising but they induced computational complications and are still in the early stages of development. We will thus not consider them any further here.

We now describe some of the most used approximations in the context of the range-separated hybrids.

\vspace{0.4cm}
\noindent
{\bf Short-range LDA exchange functional}

The short-range LDA exchange functional~\cite{Sav-INC-96,GilAdaPop-MP-96} can be obtained by using in Eq.~(\ref{Exsrint}) the LDA exchange hole [Eq.~(\ref{hxLDA})], which leads to
\begin{equation}
E_\x^{\sr,\mu,\LDA}[\rho]  = \int_{\mathbb{R}^3} e_\x^{\sr,\mu,\UEG}(\rho(\b{r})) \d\b{r},
\label{ExsrLDA}
\end{equation}
with the short-range UEG exchange energy density
\begin{equation}
e_\x^{\sr,\mu,\UEG}(\rho) = e_\x^{\UEG}(\rho) \left[ 1 - \frac{8\mut}{3} \left( \sqrt{\pi} \erf\left( \frac{1}{2\mut}\right) + (2\mut-4\mut^3) e^{-1/(4\mut^2)} -3\mut+4\mut^3\right) \right],
\label{exsrLDA}
\end{equation}
where $\mut=\mu/(2k_\text{F})$ is a dimensionless range-separation parameter. The spin-dependent version is obtained from the same spin-scaling relation as in the standard case [Eq.~(\ref{Exspinscaling})]. The short-range LDA exchange functional becomes exact for large $\mu$ [Eq.~(\ref{Exsrmuinf})] and is the first building block for constructing short-range exchange GGA functionals. 

\vspace{0.4cm}
\noindent
{\bf CAM-B3LYP exchange-correlation functional}

The CAM-B3LYP exchange-correlation functional~\cite{YanTewHan-CPL-04} uses Eq.~(\ref{ExcCAM}) with short- and long-range versions of the B88 exchange functional and the same correlation functional used in B3LYP (i.e., $0.81 \; E_\c^\text{LYP} + 0.19 \; E_\c^\text{LSDA}$),
\begin{eqnarray}
E_\xc^{\text{CAM-B3LYP}}[\Phi] =  a \; E_\x^{\sr,\mu,\HF}[\Phi] +  b \; E_\x^{\lr,\mu,\HF}[\Phi] 
+ (1-a) \; E_\x^{\sr,\mu,\text{B88}}[\rho_{\uparrow,\Phi},\rho_{\downarrow,\Phi}] 
\phantom{xxxxxxxx}
\nonumber\\
\phantom{xxxxxxxx} + (1-b) \; E_\x^{\lr,\mu,\text{B88}}[\rho_{\uparrow,\Phi},\rho_{\downarrow,\Phi}]
+ 0.81 E_\c^\text{LYP}[\rho_{\uparrow,\Phi},\rho_{\downarrow,\Phi}] + 0.19 E_\c^\text{LSDA}[\rho_{\uparrow,\Phi},\rho_{\downarrow,\Phi}],
\label{ExcCAM-B3LYP}
\end{eqnarray}
where the parameters $a=0.19$ and $b=0.65$ were optimized on atomization energies and the range-separation parameter $\mu=0.33$ bohr$^{-1}$ was taken from Ref.~\cite{TawTsuYanYanHir-JCP-04} where it was optimized on equilibrium distances of diatomics molecules. In this expression, the short-range B88 exchange functional $E_\x^{\sr,\mu,\text{B88}}$ is defined by using in Eq.~(\ref{Exsrint}) the following generic GGA model for the exchange hole~\cite{IikTsuYanHir-JCP-01} (given here in its spin-independent version)
\begin{equation}
h_\x^\text{GGA}(\rho,\nabla \rho,r_{12})=- \rho \; \frac{9}{2} \left( \frac{j_1(k_\text{GGA} r_{12})}{k_\text{GGA} r_{12}} \right)^2,
\label{hxGGA}
\end{equation}
with $k_\text{GGA} = k_\text{F}/\sqrt{e_\x^\text{GGA}(\rho,\nabla \rho)/e_\x^\text{UEG}(\rho)}$. The exchange-hole model of Eq.~(\ref{hxGGA}) properly yields the GGA exchange energy density $e_\x^\text{GGA}(\rho,\nabla \rho)$ for $\mu=0$ and thus allows one to extend any standard GGA exchange functional to a short-range GGA exchange functional. Note however that it does not fulfill the sum rule [Eq.~(\ref{intnx})]. The long-range B88 exchange functional is then simply $E_\x^{\lr,\mu,\text{B88}} = E_\x^{\text{B88}} - E_\x^{\sr,\mu,\text{B88}}$.

\vspace{0.4cm}
\noindent
{\bf LC-$\omega$PBE exchange-correlation functional}

The LC-$\omega$PBE exchange-correlation functional~\cite{VydHeyKruScu-JCP-06,VydScu-JCP-06} uses a short-range version of the PBE exchange functional as well as the standard PBE correlation functional,
\begin{eqnarray}
E_\xc^{\text{LC-$\omega$PBE}}[\Phi] =  E_\x^{\lr,\mu,\HF}[\Phi] +  E_\x^{\sr,\mu,\text{PBE}}[\rho_{\uparrow,\Phi},\rho_{\downarrow,\Phi}] + E_\c^\text{PBE}[\rho_{\uparrow,\Phi},\rho_{\downarrow,\Phi}].
\label{ExcLCwPBE}
\end{eqnarray}
The short-range PBE exchange functional is obtained by using in Eq.~(\ref{Exsrint}) the following GGA exchange hole model constructed to yield the PBE exchange energy~\cite{ErnPer-JCP-98},
\begin{eqnarray}
h_\x^\text{PBE}(\rho,\nabla \rho,r_{12}) = \rho J^\text{PBE}(s, k_\text{F} r_{12}),
\label{hxPBE}
\end{eqnarray}
where $s=|\nabla \rho|/(2k_\text{F} \rho)$ and
\begin{eqnarray}
J^\text{PBE}(s,u) &=&  \Biggl[ -\frac{{\cal A}}{u^2} \frac{1}{1+(4/9) {\cal A} u^2} + 
\nonumber\\
&&\left( \frac{{\cal A}}{u^2} + {\cal B} + {\cal C} [ 1+ s^2 {\cal F}(s)] u^2 + {\cal E} [ 1+s^2 {\cal G}(s)]u^4 \right) 
e^{-{\cal D} u^2} \Biggl]  e^{-s^2 {\cal H}(s) u^2}.
\label{JPBE}
\end{eqnarray}
Here, ${\cal A}$, ${\cal B}$, ${\cal C}$, ${\cal D}$, and ${\cal E}$ are constants chosen to obtain an oscillation-averaged UEG exchange hole for $s=0$, and ${\cal F}(s)$, ${\cal G}(s)$ and ${\cal H}(s)$ are functions determined so that the hole yields the PBE exchange density for $\mu=0$, and satisfies the sum rule [Eq.~(\ref{intnx})] and the small-$r_{12}$ expansion [Eq.~(\ref{hxsmallr12})] using the gradient expansion of $\tau$ of Eq.~(\ref{tauGEA}). The range-separation parameter is fixed at $\mu=\omega=0.4$ bohr$^{-1}$ which has been found to be close to optimal for atomization energies, reaction barrier heights, and ionization energies~\cite{VydHeyKruScu-JCP-06}.

\vspace{0.4cm}
\noindent
{\bf $\omega$B97X exchange-correlation functional}

The $\omega$B97X exchange-correlation functional~\cite{ChaHea-JCP-08} has the form of Eq.~(\ref{ExcCAM}) with $b=1$:
\begin{eqnarray}
E_\xc^\text{$\omega$B97X}[\Phi] &=&  a \; E_\x^{\sr,\mu,\HF}[\Phi] +  E_\x^{\lr,\mu,\HF}[\Phi] 
+ (1-a) \; E_\x^{\sr,\mu,\text{B97-GGA}}[\rho_{\uparrow,\Phi},\rho_{\downarrow,\Phi}] 
\nonumber\\
&&+ E_\c^\text{B97-GGA}[\rho_{\uparrow,\Phi},\rho_{\downarrow,\Phi}].
\label{ExcwB97X}
\end{eqnarray}
The short-range B97-GGA exchange density is defined as 
\begin{eqnarray}
e_\x^{\sr,\mu,\text{B97-GGA}}(\rho_\uparrow,\rho_\downarrow,\nabla \rho_\uparrow, \nabla \rho_\downarrow) = \sum_{\sigma \in \{\uparrow,\downarrow\}} e_{\x,\sigma}^{\sr,\mu,\text{UEG}}(\rho_\sigma) \; g_{\x}(x_\sigma),
\label{srexB97-GGA}
\end{eqnarray}
where $e_{\x,\sigma}^{\sr,\mu,\text{UEG}}(\rho_\sigma) = e_\x^{\sr,\mu,\text{UEG}}(\rho_\sigma,0)$ is the spin-$\sigma$ contribution to the short-range UEG exchange energy density [Eq.~(\ref{exsrLDA})] and the gradient correction $g_{\x}(x_\sigma)$ where $x_\sigma = |\nabla \rho_\sigma |/\rho_\sigma^{4/3}$ has the same form as in Eq.~(\ref{gx}) with polynomial degree $m=4$. In Eq.~(\ref{ExcwB97X}), the correlation functional has the same form as the B97-GGA correlation functional but again with polynomial degree $m=4$ in Eqs.~(\ref{gcud}) and~(\ref{gcss}). The fraction of short-range HF exchange $a\approx 0.16$, the range-separation parameter $\mu=\omega=0.3$ bohr$^{-1}$, and the linear coefficients in Eqs.~(\ref{gx}),~(\ref{gcud}), and~(\ref{gcss}) were optimized on sets of atomic energies, atomization energies, ionization energies, electron and proton affinities, reaction barrier heights, and noncovalent interactions, with the constraints $a+c_{\x,0}=1$, $c_{\c,0}^{\uparrow\downarrow}=1$, and $c_{\c,0}^{\sigma\sigma}=1$ to enforce the correct UEG limit.

\vspace{0.4cm}
\noindent
{\bf HSE exchange-correlation functional}

The Heyd--Scuseria--Ernzerhof (HSE) exchange-correlation functional~\cite{HeyScuErn-JCP-03} is of the form of Eq.~(\ref{ExcCAM}) with $b=0$ (i.e., no long-range HF exchange),
\begin{eqnarray}
E_\xc^{\text{HSE}}[\Phi] &=&  a E_\x^{\sr,\mu,\HF}[\Phi] +  (1-a)E_\x^{\sr,\mu,\text{PBE}}[\rho_{\uparrow,\Phi},\rho_{\downarrow,\Phi}] + E_\x^{\lr,\mu,\text{PBE}}[\rho_{\uparrow,\Phi},\rho_{\downarrow,\Phi}] 
\nonumber\\
&&+ E_\c^\text{PBE}[\rho_{\uparrow,\Phi},\rho_{\downarrow,\Phi}],
\label{ExcHSE}
\end{eqnarray}
and involves the long-range PBE exchange functional $E_\x^{\lr,\mu,\text{PBE}}= E_\x^{\text{PBE}}- E_\x^{\sr,\mu,\text{PBE}}$ complementary to the short-range PBE exchange functional constructed from the PBE exchange hole model [Eqs.~(\ref{hxPBE}) and~(\ref{JPBE})]. In order to reproduce reliable values for the band gap in semiconducting solids, the range-separation parameter is fixed at $\mu=0.15$ bohr$^{-1}$, which is a very small value compared to the other range-separated hybrids. It means that the range of electron-electron distances covered by HF exchange is large, and the HSE functional could be thought of as a regular hybrid approximation but with the very long-range contribution of the HF exchange removed. This is particularly appropriate for solids since in these systems the very long-range HF exchange is effectively balanced by the correlation effects (a phenomenon known as screening). The fraction of (short-range) HF exchange is fixed at $a=0.25$ like in the PBE0 hybrid functional.

\section{Multideterminant hybrid approximations}
\stepcounter{myequation}
\label{sec:mdha}

\subsection{Double-hybrid approximations}
\label{sec:dh}

In 2006, Grimme~\cite{Gri-JCP-06} introduced a two-parameter {\it double-hybrid} (2DH) approximation, written here in its spin-independent version,
\begin{equation}
E_\xc^{\text{2DH}} = a_\x \; E_\x^{\HF}[\Phi] +  (1-a_\x) \; E_\x^{\GGA}[\rho_\Phi] + (1-a_\c) E_\c^{\GGA}[\rho_\Phi] + a_\c E_\c^{\MP},
\label{2DH}
\end{equation}
mixing a fraction $a_\x$ of the HF exchange energy with a GGA exchange functional, and a fraction $a_\c$ of the second-order M{\o}ller--Plesset (MP2) correlation energy $E_\c^{\MP}$ with a GGA correlation functional. In Eq.~(\ref{2DH}), the first three terms are first calculated in a self-consistent manner, and then the last term $E_\c^{\MP}$ is added perturbatively using the orbitals determined in the first step. The expression of $E_\c^{\MP}$ is~\cite{SzaOst-BOOK-96}
\begin{equation}
E_\text{c}^{\text{MP2}}  = - \frac{1}{4} \sum_{i=1}^{N}\sum_{j=1}^{N} \; \sum_{a \geq N+1} \sum_{b \geq N+1} \frac{|\bra{\phi_i \phi_j}\ket{\phi_a \phi_b}|^2}{\varepsilon_a+\varepsilon_b-\varepsilon_i-\varepsilon_j},
\label{EcMP2}
\end{equation}
where $i,j$ and $a,b$ run over occupied and virtual spin orbitals, respectively, $\varepsilon_k$ are spin orbital energies, and $\bra{\phi_i \phi_j}\ket{\phi_a \phi_b} = \braket{\phi_i\phi_j}{\phi_a\phi_b} - \braket{\phi_i\phi_j}{\phi_b\phi_a}$ are antisymmetrized two-electron integrals with (in physicists' notation)
\begin{eqnarray}
\braket{\phi_p \phi_q}{\phi_r\phi_s} = \int_{(\mathbb{R}^3\times\{\uparrow,\downarrow\})^2} \frac{\phi_p^*(\b{x}_1) \phi_q^*(\b{x}_2) \phi_r(\b{x}_1) \phi_s(\b{x}_2)}{|\b{r}_1-\b{r}_2|} \d\b{x}_1 \d\b{x}_2.
\label{2eintspinorb}
\end{eqnarray}
Note that the notation in Eq.~(\ref{EcMP2}) assumes that the one-electron wave-function space is spanned by a discrete set of spin orbitals. In the exact theory, the continuum limit of the set of virtual spin orbitals is implied.

The rigorous framework underlying these double-hybrid approximations was established by Sharkas, Toulouse, and Savin~\cite{ShaTouSav-JCP-11}. The idea is to decompose the universal density functional of Eq.~(\ref{FnLevy}) as
\begin{eqnarray}
F[\rho] = \min_{\Psi \in {\cal W}^{N}_{\rho}} \bra{\Psi} \hat{T} + \l \hat{W}_\ee \ket{\Psi} + \bar{E}_\Hxc^\l[\rho],
\label{Fmultidethybrid}
\end{eqnarray}
where $\l \in [0,1]$ is a coupling constant and $\bar{E}_\Hxc^\l[\rho]$ is a complementary density functional defined to make Eq.~(\ref{Fmultidethybrid}) exact. From Eqs.~(\ref{FKS}) and~(\ref{Flndecomp}), we see that $\bar{E}_\Hxc^\l[\rho] = E_\Hxc[\rho] - E_\Hxc^\l[\rho]$ where $E_\Hxc[\rho]$ is the standard Hartree-exchange-correlation functional of the KS scheme and $E_\Hxc^\l[\rho]$ is the Hartree-exchange-correlation functional along the adiabatic connection. The Hartree and exchange contributions are simply linear in $\l$,
\begin{eqnarray}
\bar{E}_\H^\l[\rho] = (1-\l) E_\H[\rho], 
\end{eqnarray}
\begin{eqnarray}
\bar{E}_\x^\l[\rho] = (1-\l) E_\x[\rho],
\end{eqnarray}
where $E_\H[\rho]$ and $E_\x[\rho]$ are the standard Hartree and exchange functionals of the KS scheme. Moreover, from the uniform coordinate scaling relation of Eq.~(\ref{Ecrhog}), we have 
\begin{eqnarray}
\bar{E}_\c^\l[\rho] = E_\c[\rho] - \l^2 E_\c[\rho_{1/\l}], 
\label{barEcl}
\end{eqnarray}
where $E_\c[\rho]$ is the standard correlation functional of the KS scheme and $\rho_{1/\l}(\b{r})= (1/\l)^3 \rho(\b{r}/\l)$ is the scaled density. The decomposition in Eq.~(\ref{Fmultidethybrid}) leads to the following expression of the exact ground-state energy
\begin{eqnarray}
E_0 = \inf_{\Psi \in {\cal W}^{N}} \left\{ \bra{\Psi} \hat{T} + \hat{V}_\text{ne} + \l \hat{W}_\ee \ket{\Psi} + \bar{E}_\Hxc^\l[\rho_\Psi] \right\},
\label{E0multidethybrid}
\end{eqnarray}
where the infimum is over general multideterminant wave functions $\Psi \in {\cal W}^{N}$. This constitutes a \textit{multideterminant extension of the KS scheme}. Note that this multideterminant KS scheme can trivially be extended to spin-dependent density functionals and functionals depending on the kinetic-energy density~\cite{SouShaTou-JCP-14}.

The double-hybrid ansatz can be seen as a particular approximation within this multideterminant KS scheme~\cite{ShaTouSav-JCP-11}. To see this, we define a density-scaled one-parameter hybrid (DS1H) approximation by restricting the minimization in Eq.~(\ref{E0multidethybrid}) to single-determinant wave functions $\Phi\in {\cal S}^{N}$,
\begin{eqnarray}
E^{\text{DS1H},\l}_0 &=& \inf_{\Phi \in {\cal S}^{N}} \Bigl\{\bra{\Phi}\hat{T}+\hat{V}_{\text{ne}}+\l\hat{W}_{\ee} \ket{\Phi}+\bar{E}_{\Hxc}^{\l}[\rho_{\Phi}]\Bigl\}, 
\end{eqnarray} 
obtaining an energy which necessarily depends on $\l$. A minimizing single-determinant wave function $\Phi^\l$ must satisfy the self-consistent eigenvalue equation
\begin{eqnarray}
\left( \hat{T}+\hat{V}_{\text{ne}}+\l \hat{V}_{\Hx}^{\HF}[\Phi^\l] + \hat{\bar{V}}_{\Hxc}^{\l}[\rho_{\Phi^\l}] \right) \ket{\Phi^\l} = {\cal E}_0^{\l} \ket{\Phi^\l},
\label{DS1Heigenval}
\end{eqnarray} 
where $\hat{V}_{\Hx}^{\HF}[\Phi^\l]$ is the nonlocal HF potential operator evaluated with the DS1H wave function $\Phi^\l$ and $\hat{\bar{V}}_{\Hxc}^{\l}[\rho_{\Phi^\l}]$ is the local Hartree-exchange-correlation potential operator generated by the energy functional $\bar{E}_{\Hxc}^{\l}[\rho]$ and evaluated at the DS1H density $\rho_{\Phi^\l}$. If written explicitly in terms of spin orbitals, Eq.~(\ref{DS1Heigenval}) would have the form of the GKS equations [Eq.~(\ref{GKSeqs})]. The DS1H ground-state energy can be finally written as
\begin{eqnarray}
E^{\text{DS1H},\l}_0 &=& \bra{\Phi^\l} \hat{T}+\hat{V}_{\text{ne}} \ket{\Phi^\l} + E_{\H}[\rho_{\Phi^\l}] + \l E_\x^{\HF}[\Phi^\l]
+ (1-\l) E_\x[\rho_{\Phi^\l}] + \bar{E}^\l_\c[\rho_{\Phi^\l}],
\label{DS1H}
\end{eqnarray} 
where the full Hartree energy $E_{\H}[\rho]$ has been recomposed. The exchange-correlation energy in Eq.~(\ref{DS1H}) is of the form of a hybrid approximation [Eq.~(\ref{Exc1H})].

All that is missing in Eq.~(\ref{DS1H}) is the correlation energy associated with the scaled interaction $\l\hat{W}_{\ee}$. It can be calculated by a nonlinear Rayleigh-Schr\"odinger perturbation theory~\cite{AngGerSavTou-PRA-05,FroJen-PRA-08,Ang-PRA-08} starting from the DS1H reference. Consider the following energy expression with the perturbation parameter $\alpha \in [0,1]$,
\begin{eqnarray}
E^{\l,\alpha}_0 &=& \inf_{\Psi \in {\cal W}^N}\Bigl\{\bra{\Psi}\hat{T}+\hat{V}_{\text{ne}}+\l \hat{V}_{\Hx}^\HF[\Phi^\l]  + \alpha \l \hat{W} \ket{\Psi} +\bar{E}_{\Hxc}^{\l}[\rho_{\Psi}]\Bigl\},
\label{Ela}
\end{eqnarray}
where $\l\hat{W}=\l \left( \hat{W}_{\ee} - \hat{V}_{\Hx}^\HF[\Phi^\l] \right)$ is the scaled M{\o}ller--Plesset perturbation operator. For $\alpha=0$, the stationary equation associated with Eq.~(\ref{Ela}) reduces to the DS1H eigenvalue equation [Eq.~(\ref{DS1Heigenval})]. For $\alpha=1$, Eq.~(\ref{Ela}) reduces to Eq.~(\ref{E0multidethybrid}), so $E^{\l,\alpha=1}_0$ is the exact energy, independently of $\l$. The sum of the zeroth-order energy and first-order energy correction gives simply the DS1H energy, $E^{\text{DS1H},\l}_0=E^{\l,(0)}_0+E^{\l,(1)}_0$. Thanks to the existence of a Brillouin theorem just like in standard M{\o}ller--Plesset perturbation theory (see Refs.~\cite{AngGerSavTou-PRA-05,FroJen-PRA-08,Ang-PRA-08}), only double excitations contribute to the first-order wave-function correction $\Psi^{\l,(1)}$ and the second-order energy correction has a standard MP2 form
\begin{eqnarray}
E^{\l,(2)}_0 = \l^2 \bra{\Phi^\l} \hat{W} \ket{\Psi^{\l,(1)}}= \l^2 E_\c^{\text{MP2}},
\label{}
\end{eqnarray}
where $E_\c^{\text{MP2}}$ has the expression in Eq.~(\ref{EcMP2}) with DS1H spin orbitals and associated orbital eigenvalues (which implicitly depend on $\l$). This second-order perturbation theory defines a density-scaled one-parameter double-hybrid (DS1DH) approximation
\begin{eqnarray}
E^{\text{DS1DH},\l}_0 = E^{\text{DS1H},\l}_0 + E^{\l,(2)}_0,
\label{DS1DH}
\end{eqnarray}
which contains the exchange-correlation energy contribution
\begin{eqnarray}
E^{\text{DS1DH},\l}_{\xc} &=& \l E_\x^{\HF}[\Phi^\l] + (1-\l) E_\x[\rho_{\Phi^\l}] + \bar{E}^\l_\c[\rho_{\Phi^\l}] + \l^2 E_\c^{\text{MP2}}.
\label{ExcDS1DH}
\end{eqnarray} 
To make connection with the double-hybrid ansatz of Eq.~(\ref{2DH}), we can also define a one-parameter double-hybrid (1DH) approximation, obtained by neglecting the density scaling in the correlation functional, i.e. $E_\c[\rho_{1/\l}]\approx E_\c[\rho]$ in Eq.~(\ref{barEcl}),
\begin{eqnarray}
E^{\text{1DH},\l}_{\xc} &=& \l E_\x^{\HF}[\Phi^\l] + (1-\l) E_\x[\rho_{\Phi^\l}] + (1-\l^2) E_\c[\rho_{\Phi^\l}] + \l^2 E_\c^{\text{MP2}},
\label{Exc1DH}
\end{eqnarray} 
which, after using semilocal approximations for $E_\x[\rho]$ and $E_\c[\rho]$, has the form of Eq.~(\ref{2DH}) with parameters $a_\x=\l$ and $a_\c=\l^2$. In this rigorous formulation of the double-hybrid approximations, the fraction of HF exchange is thus connected to the fraction of MP2 correlation. Taking into account approximately the scaling of the density in $E_\c[\rho_{1/\l}]$, it has also been proposed to use $a_\c=\l^3$~\cite{TouShaBreAda-JCP-11}. Fromager~\cite{Fro-JCP-11} also proposed an extension of this rigorous formulation in order to justify the use of double-hybrid approximations with two parameters such that $a_\c \leq a_\x^2=\l^2$.

An essential advantage of double-hybrid approximations is that the presence of nonlocal MP2 correlation allows one to use a larger fraction of nonlocal HF exchange, which helps decreasing the self-interaction error. This usually provides an improvement over hybrid approximations for molecular systems with sufficiently large electronic gaps. However, a large fraction of HF exchange and a fraction of MP2 correlation also generally means large static-correlation errors in systems with small HOMO-LUMO gaps. 

The first and still best known double-hybrid approximation is B2PLYP~\cite{Gri-JCP-06} which is based on the B88 exchange functional and the LYP correlation functional,
\begin{equation}
E_\xc^{\text{B2PLYP}} = a_\x \; E_\x^{\HF}[\Phi] +  (1-a_\x) \; E_\x^\text{B88}[\rho_{\uparrow,\Phi},\rho_{\downarrow,\Phi}] + (1-a_\c) E_\c^\text{LYP}[\rho_{\uparrow,\Phi},\rho_{\downarrow,\Phi}] + a_\c E_\c^{\MP},
\label{}
\end{equation}
and the parameters $a_\x=0.53$ and $a_\c=0.27$ have been optimized on a set of atomization energies. Interestingly, even though the two parameters have been optimized without any constraint, we have $a_\c \approx a_\x^2 = 0.28$ as predicted by Eq.~(\ref{Exc1DH}). 

It has also been proposed to use the spin-component-scaled (SCS) version of MP2~\cite{Gri-JCP-03} to construct spin-component-scaled double-hybrid approximations of the form~\cite{KozGruMar-JPCC-10,KozMar-PCCP-11} 
\begin{eqnarray}
E_\xc^{\text{SCS-DH}} &=& a_\x \; E_\x^{\HF}[\Phi] +  (1-a_\x) \; E_\x^{\GGA}[\rho_\Phi] + (1-a_\c) E_\c^{\GGA}[\rho_\Phi] + c_\text{OS} E^{\text{MP2}}_{\c,\text{OS}} + c_\text{SS} E^{\text{MP2}}_{\c,\text{SS}},
\nonumber\\
\label{2DHSCS}
\end{eqnarray} 
which contains four empirical parameters $a_\x$, $a_\c$, $c_\text{OS}$, and $c_\text{SS}$. In this expression, $E^{\text{MP2}}_{\c,\text{OS}}$ and $E^{\text{MP2}}_{\c,\text{SS}}$ are the opposite-spin (OS) and same-spin (SS) contributions to the MP2 correlation energy obtained by restricting the sums over $i$ and $j$ in Eq.~(\ref{EcMP2}) to spin orbitals of opposite and same spins, respectively. Since in MP2 the same-spin component is usually overestimated relative to the opposite-spin component, this SCS variant is a simple way to achieve higher accuracy without increasing computational cost. 

For reviews on different flavors of double hybrids and their assessments, the reader may consult Refs.~\cite{SanAda-PCCP-13,GoeGri-WIRE-14,SuXu-WIRES-16,MehCasGor-PCCP-18}. It has also been proposed to construct double-hybrid approximations where the MP2 correlation term is extended to a higher-order correlation method such as RPA~\cite{RuzPerCso-JCTC-10,AhnHehVogTraLeuKlo-CP-14,MezCsoRuzKal-JCTC-15,GriSte-PCCP-16,MezCsoRuzKal-JCTC-17} or coupled-cluster~\cite{GarBulHenScu-JCP-15,ChaGoeRad-JCC-16}. More generally, the multideterminant extension of the KS scheme of Eq.~(\ref{E0multidethybrid}) allows one to define hybrids combining any wave-function method with density functionals. For example, a multiconfiguration hybrid approximation based on Eq.~(\ref{E0multidethybrid}) which combines a multiconfiguration self-consistent-field (MCSCF) wave function with density functionals has been proposed in the goal of tackling strongly correlated systems~\cite{ShaSavJenTou-JCP-12}. This approach has also been used to combine valence-bond (VB) theory~\cite{YinZhoZheLuaSuWu-FC-19} or variational two-electron reduced-density-matrix theory~\cite{MosLieDep-JCTC-20} with DFT.

\subsection{Range-separated double-hybrid approximations}
\label{sec:rsdh}

\subsubsection{Range-separated one-parameter double-hybrid approximations}
\label{sec:rs1dh}

In 2005, \'Angy\'an, Gerber, Savin, and Toulouse~\cite{AngGerSavTou-PRA-05} introduced what could be called the first \textit{range-separated one-parameter double-hybrid approximation}, i.e. combining HF exchange and MP2 correlation with density functionals using a one-parameter decomposition of the electron-electron interaction. This is based on the \textit{range-separated multideterminant extension of the KS scheme} introduced earlier by Savin~\cite{Sav-INC-96} (see, also, Refs.~\cite{SavFla-IJQC-95,Sav-INC-96a,TouColSav-PRA-04}) and which actually predates and inspired the multideterminant extension of the KS scheme of Eq.~(\ref{E0multidethybrid}).

The idea is to decompose the universal density functional of Eq.~(\ref{FnLevy}) as
\begin{eqnarray}
F[\rho] = \min_{\Psi \in {\cal W}^{N}_{\rho}} \bra{\Psi} \hat{T} + \hat{W}_\ee^{\lr,\mu} \ket{\Psi} + \bar{E}_\Hxc^{\sr,\mu}[\rho],
\label{FRSDFT}
\end{eqnarray}
where $\hat{W}_\ee^{\lr,\mu}$ is the long-range electron-electron interaction operator (associated with the pair potential $w_\ee^{\lr,\mu}(r_{12}) = \erf(\mu r_{12})/r_{12}$ as already used in the range-separated hybrids of Section~\ref{sec:rsh}) and $\bar{E}_\Hxc^{\sr,\mu}[\rho]$ is the complementary short-range density functional defined to make Eq.~(\ref{FRSDFT}) exact. As before, the parameter $\mu \in [0,+\infty)$ controls the range of the separation. The complementary short-range functional can be written as $\bar{E}_\Hxc^{\sr,\mu}[\rho] = E_\Hxc[\rho] - E_\Hxc^{\lr,\mu}[\rho]$, where $E_\Hxc[\rho]$ is the standard Hartree-exchange-correlation functional of the KS scheme and $E_\Hxc^{\lr,\mu}[\rho]$ is the Hartree-exchange-correlation functional associated with the long-range interaction $w_\ee^{\lr,\mu}(r_{12})$. It is often convenient to decompose the short-range functional as (see Refs.~\cite{Tou-THESIS-05,TouGorSav-TCA-05,StoTeaTouHelFro-JCP-13} for an alternative decomposition)
\begin{eqnarray}
\bar{E}_\Hxc^{\sr,\mu}[\rho] = E_\H^{\sr,\mu}[\rho] + E_\x^{\sr,\mu}[\rho] + \bar{E}_\c^{\sr,\mu}[\rho],
\label{}
\end{eqnarray}
where $E_\H^{\sr,\mu}[\rho]$ is the short-range Hartree functional,
\begin{equation}
E_\H^{\sr,\mu}[\rho] = \frac{1}{2} \int_{\mathbb{R}^3\times\mathbb{R}^3} \rho(\b{r}_1) \rho(\b{r}_2) w_\ee^{\sr,\mu}(r_{12}) \d \b{r}_1 \d \b{r}_2,
\label{}
\end{equation}
with the short-range interaction $w_\ee^{\sr,\mu}(r_{12})=1/r_{12} - w_\ee^{\lr,\mu}(r_{12})$, $E_\x^{\sr,\mu}[\rho]$ is the short-range exchange functional [Eq.~(\ref{Exsrint})] which can also be written as
\begin{eqnarray}
E_\x^\sr[\rho] &=& \bra{\Phi[\rho]} \hat{W}_\ee^{\sr,\mu} \ket{\Phi[\rho]} - E_\H^{\sr,\mu}[\rho],
\label{}
\end{eqnarray}
with the KS single-determinant wave function $\Phi[\rho]$, and $\bar{E}_\c^{\sr,\mu}[\rho]$ is the complementary short-range correlation functional. Just like for Eq.~(\ref{E0multidethybrid}), the decomposition in Eq.~(\ref{FRSDFT}) leads to the following expression of the exact ground-state energy
\begin{eqnarray}
E_0 = \inf_{\Psi \in {\cal W}^{N}} \left\{ \bra{\Psi} \hat{T} + \hat{V}_\text{ne} + \hat{W}_\ee^{\lr,\mu} \ket{\Psi} + \bar{E}_\Hxc^{\sr,\mu}[\rho_\Psi] \right\},
\label{E0RSDFT}
\end{eqnarray}
where the infimum is over general multideterminant wave functions $\Psi \in {\cal W}^{N}$. 

To obtain a MP2/DFT hybrid scheme, we proceed analogously to Section~\ref{sec:dh}. First, we define the following range-separated hybrid (RSH) approximation by restricting the minimization in Eq.~(\ref{E0RSDFT}) to single-determinant wave functions $\Phi\in {\cal S}^{N}$,
\begin{eqnarray}
E^{\text{RSH},\mu}_0 &=& \inf_{\Phi \in {\cal S}^{N}} \Bigl\{\bra{\Phi}\hat{T}+\hat{V}_{\text{ne}}+\hat{W}_{\ee}^{\lr,\mu} \ket{\Phi}+\bar{E}_{\Hxc}^{\sr,\mu}[\rho_{\Phi}]\Bigl\}, 
\end{eqnarray} 
obtaining an energy which necessarily depends on $\mu$. A minimizing single-determinant wave function $\Phi^\mu$ must satisfy the self-consistent eigenvalue equation
\begin{eqnarray}
\left( \hat{T}+\hat{V}_{\text{ne}}+\hat{V}_{\Hx}^{\lr,\mu,\HF}[\Phi^\mu] + \hat{\bar{V}}_{\Hxc}^{\sr,\mu}[\rho_{\Phi^\mu}] \right) \ket{\Phi^\mu} = {\cal E}_0^{\mu} \ket{\Phi^\mu},
\label{RSHeigenval}
\end{eqnarray} 
where $\hat{V}_{\Hx}^{\lr,\mu,\HF}[\Phi^\mu]$ is the nonlocal long-range HF potential operator evaluated with the RSH wave function $\Phi^\mu$ and $\hat{\bar{V}}_{\Hxc}^{\sr,\mu}[\rho_{\Phi^\mu}]$ is the local short-range Hartree-exchange-correlation potential operator generated by the energy functional $\bar{E}_{\Hxc}^{\sr,\mu}[\rho]$ and evaluated at the RSH density $\rho_{\Phi^\mu}$. The RSH ground-state energy can be finally written as
\begin{eqnarray}
E^{\text{RSH},\mu}_0 &=& \bra{\Phi^\mu} \hat{T}+\hat{V}_{\text{ne}} \ket{\Phi^\mu} + E_{\H}[\rho_{\Phi^\mu}] + E_\x^{\lr,\mu,\HF}[\Phi^\mu]
+ E_\x^{\sr,\mu}[\rho_{\Phi^\mu}] + \bar{E}^{\sr,\mu}_\c[\rho_{\Phi^\mu}],
\label{RSH}
\end{eqnarray} 
where the full Hartree energy $E_{\H}[\rho]$ has been recomposed. The exchange-correlation energy in Eq.~(\ref{RSH}) has a similar form as in the LC scheme of Eq.~(\ref{ExcLC}).

To calculate the missing long-range correlation energy in Eq.~(\ref{RSH}), we can define a nonlinear Rayleigh-Schr\"odinger perturbation theory~\cite{AngGerSavTou-PRA-05,FroJen-PRA-08,Ang-PRA-08} starting from the RSH reference. We start from the following energy expression with the perturbation parameter $\alpha \in [0,1]$,
\begin{eqnarray}
E^{\mu,\alpha}_0 &=& \inf_{\Psi \in {\cal W}^N}\Bigl\{\bra{\Psi}\hat{T}+\hat{V}_{\text{ne}}+\hat{V}_{\Hx}^{\lr,\mu,\HF}[\Phi^\mu]  + \alpha \hat{W}^{\lr,\mu} \ket{\Psi} +\bar{E}_{\Hxc}^{\lr,\mu}[\rho_{\Psi}]\Bigl\},
\label{Emua}
\end{eqnarray}
where $\hat{W}^{\lr,\mu}=\left( \hat{W}_{\ee}^{\lr,\mu} - \hat{V}_{\Hx}^{\lr,\mu,\HF}[\Phi^\mu] \right)$ is the long-range M{\o}ller--Plesset perturbation operator. For $\alpha=0$, the stationary equation associated with Eq.~(\ref{Emua}) reduces to the RSH eigenvalue equation [Eq.~(\ref{RSHeigenval})]. For $\alpha=1$, Eq.~(\ref{Emua}) reduces to Eq.~(\ref{E0RSDFT}), so $E^{\mu,\alpha=1}_0$ is the exact energy, independently of $\mu$. The sum of the zeroth-order energy and first-order energy correction gives simply the RSH energy, $E^{\text{RSH},\mu}_0=E^{\mu,(0)}_0+E^{\mu,(1)}_0$. As in Section~\ref{sec:dh}, only double excitations contribute to the first-order wave-function correction $\Psi^{\mu,(1)}$ and the second-order energy correction has a standard MP2 form
\begin{eqnarray}
E^{\mu,(2)}_0 = \bra{\Phi^\mu} \hat{W}^{\lr,\mu} \ket{\Psi^{\mu,(1)}}= E_\c^{\lr,\mu,\text{MP2}},
\label{}
\end{eqnarray}
where $E_\c^{\lr,\mu,\text{MP2}}$ has the same expression as in Eq.~(\ref{EcMP2}) with RSH spin orbitals and associated orbital eigenvalues (which implicitly depend on $\mu$) but using the long-range two-electron integrals
\begin{eqnarray}
\braket{\phi_p \phi_q}{\phi_r\phi_s}^{\lr,\mu} = \int_{(\mathbb{R}^3\times\{\uparrow,\downarrow\})^2} \phi_p^*(\b{x}_1) \phi_q^*(\b{x}_2) \phi_r(\b{x}_1) \phi_s(\b{x}_2) w_\ee^{\lr,\mu}(r_{12}) \d\b{x}_1 \d\b{x}_2,
\label{2eintspinorblr}
\end{eqnarray}
instead of the standard two-electron integrals of Eq.~(\ref{2eintspinorb}). This second-order perturbation theory defines a RSH+MP2 approximation,
\begin{eqnarray}
E^{\text{RSH+MP2},\mu}_0 = E^{\text{RSH},\mu}_0 + E_\c^{\lr,\mu,\text{MP2}},
\label{RSH+MP2}
\end{eqnarray}
which contains the exchange-correlation energy contribution
\begin{eqnarray}
E^{\text{RSH+MP2},\mu}_{\xc} &=& E_\x^{\lr,\mu,\HF}[\Phi^\mu] + E_\x^{\sr,\mu}[\rho_{\Phi^\mu}] + \bar{E}^{\sr,\mu}_\c[\rho_{\Phi^\mu}] + E_\c^{\lr,\mu,\text{MP2}}.
\label{ExcRSH+MP2}
\end{eqnarray} 
When using semilocal density-functional approximations for the short-range functionals $E_\x^{\sr,\mu}[\rho]$ and $\bar{E}^{\sr,\mu}_\c[\rho]$, the RSH+MP2 exchange-correlation energy expression of Eq.~(\ref{ExcRSH+MP2}) thus constitutes range-separated double-hybrid approximations similar to the double hybrids of Section~\ref{sec:dh}. The optimal value for the range-separation parameter is often around $\mu \approx 0.5$ bohr$^{-1}$~\cite{GerAng-CPL-05a,MusReiAngTou-JCP-15}. This scheme has the advantage of dropping the long-range part of both the exchange and correlation density functionals which are usually not well described by semilocal density-functional approximations. Moreover, using a long-range MP2 correlation energy has the advantage of leading to a correct qualitative description of London dispersion interaction energies~\cite{AngGerSavTou-PRA-05,GerAng-CPL-05b,GerAng-JCP-07,TayAngGalZhaGygHirSonRahLilPodBulHenScuTouPevTruSza-JCP-16}, while displaying a fast convergence with the one-electron basis size~\cite{FraMusLupTou-JCP-15}. Similar to the SCS double hybrids [Eq.~(\ref{2DHSCS})], a SCS variant of the RSH+MP2 scheme has also been proposed~\cite{SanCivUsvTouShaMas-JCP-15}.

The range-separated multideterminant extension of the KS scheme of Eq.~(\ref{E0RSDFT}) allows one to define various hybrid schemes combining any wave-function method with density functionals. For example, one can go beyond second order by using long-range coupled-cluster~\cite{GolWerSto-PCCP-05,GolWerStoLeiGorSav-CP-06,TouZhuSavJanAng-JCP-11,GarBulHenScu-PCCP-15} or random-phase approximations~\cite{TouGerJanSavAng-PRL-09,JanHenScu-JCP-09,TouZhuAngSav-PRA-10,PaiJanHenScuGruKre-JCP-10,TouZhuSavJanAng-JCP-11}. To describe strongly correlated systems, one can also use for the long-range part wave-function methods such as configuration interaction (CI)~\cite{LeiStoWerSav-CPL-97,PolSavLeiSto-JCP-02,Cas-JCP-18,FerGinTou-JCP-19}, MCSCF~\cite{FroTouJen-JCP-07,FroReaWahWahJen-JCP-09,HedTouJen-JCP-18}, density-matrix renormalization group (DMRG)~\cite{HedKneKieJenRei-JCP-15}, or multireference perturbation theory~\cite{FroCimJen-PRA-10}. Density-matrix functional theory (DMFT)~\cite{Per-PRA-10,RohTouPer-PRA-10,RohPer-JCP-11} and Green-function methods~\cite{RebTou-JCP-16,KanZgi-JCTC-17} have also been used for the long-range part.

We now consider the approximations used for $E_\x^{\sr,\mu}[\rho]$ and $\bar{E}^{\sr,\mu}_\c[\rho]$. In Section~\ref{sec:rsh}, we have already described the short-range exchange LDA [Eq.~(\ref{ExsrLDA})] and some short-range exchange GGAs for $E_\x^{\sr,\mu}[\rho]$. Here, we describe the short-range LDA correlation functional and another short-range GGA exchange-correlation functional.

\vspace{0.4cm}
\noindent
{\bf Short-range LDA correlation functional}

The complementary short-range LDA (or LSDA) correlation functional is
\begin{equation}
\bar{E}_\c^{\sr,\mu,\LSDA}[\rho_\uparrow,\rho_\downarrow]  = \int_{\mathbb{R}^3} \bar{e}_\c^{\sr,\mu,\UEG}(\rho_\uparrow(\b{r}),\rho_\downarrow(\b{r})) \d\b{r},
\label{EcsrLDA}
\end{equation}
where $\bar{e}_\c^{\sr,\mu,\UEG}(\rho_\uparrow,\rho_\downarrow) = \rho \; \bar{\varepsilon}_\c^{\sr,\mu,\UEG}(\rho_\uparrow,\rho_\downarrow)$ is the complementary short-range UEG correlation energy density. In this expression, $\bar{\varepsilon}_\c^{\sr,\mu,\UEG}(\rho_\uparrow,\rho_\downarrow)$ is defined by
\begin{equation}
\bar{\varepsilon}_\c^{\sr,\mu,\UEG}(\rho_\uparrow,\rho_\downarrow) = \varepsilon_\c^{\UEG}(\rho_\uparrow,\rho_\downarrow) - \varepsilon_\c^{\lr,\mu,\UEG}(\rho_\uparrow,\rho_\downarrow),
\label{epscsrUEG}
\end{equation}
where $\varepsilon_\c^{\UEG}(\rho_\uparrow,\rho_\downarrow)$ and $\varepsilon_\c^{\lr,\mu,\UEG}(\rho_\uparrow,\rho_\downarrow)$ are the correlation energies per particle of the UEG with the standard Coulomb and long-range electron-electron interactions, respectively. A simple spin-independent parametrization of $\bar{\varepsilon}_\c^{\sr,\mu,\UEG}$ was given in Ref.~\cite{TouSavFla-IJQC-04}. A better spin-dependent parametrization was constructed in Ref.~\cite{PazMorGorBac-PRB-06} which uses the PW92 parametrization for $\varepsilon_\c^{\UEG}(\rho_\uparrow,\rho_\downarrow)$ [Eq.~(\ref{PW92})] and the following parametrization for $\varepsilon_\c^{\lr,\mu,\UEG}(\rho_\uparrow,\rho_\downarrow)$ in terms of $r_\s=(3/(4\pi \rho))^{1/3}$ and $\zeta=(\rho_\uparrow-\rho_\downarrow)/\rho$:
\begin{eqnarray}
\varepsilon_\c^{\lr,\mu,\UEG}(\rho_\uparrow,\rho_\downarrow) = \phantom{xxxxxxxxxxxxxxxxxxxxxxxxxxxxxxxxxxxxxxxxxxxxxxxxxxx}
\nonumber\\
 \frac{\phi_2(\zeta)^3 Q\left(\frac{\mu \sqrt{r_\s}}{\phi_2(\zeta)}\right) + a_1(r_\s,\zeta) \mu^3 + a_2(r_\s,\zeta) \mu^4 + a_3(r_\s,\zeta) \mu^5 + a_4(r_\s,\zeta) \mu^6 + a_5(r_\s,\zeta) \mu^8}{(1+b_0(r_\s)^2 \mu^2)^4}.
\label{}
\end{eqnarray}
In this expression, $\phi_2(\zeta)$ is a spin-scaling function defined by Eq.~(\ref{phinzeta}), $Q(x)$ is a function determined from the small-$\mu$ and/or small-$r_\s$ limit,
\begin{equation}
Q(x)=\frac{2\ln(2) -2}{\pi^2} \ln \left( \frac{1 + a x + b x^2 + c  x^3}{1 + a x + d x^2} \right),
\label{}
\end{equation}
with $a=5.84605$, $c=3.91744$, $d=3.44851$, $b=d-3 \pi \alpha/[4 \ln(2) - 4]$, $\alpha=4/(9\pi)^{1/3}$, and the functions $a_i(r_\s,\zeta)$ are 
\begin{subequations}
\begin{equation}
a_1(r_\s,\zeta) = 4 b_0(r_\s)^6 C_3(r_\s,\zeta) + b_0(r_\s)^8 C_5(r_\s,\zeta),
\label{}
\end{equation}
\begin{equation}
a_2(r_\s,\zeta) = 4 b_0(r_\s)^6 C_2(r_\s,\zeta) + b_0(r_\s)^8 C_4(r_\s,\zeta) + 6 b_0(r_\s)^4 \varepsilon_\c^\text{PW92}(r_\s,\zeta),
\label{}
\end{equation}
\begin{equation}
a_3(r_\s,\zeta) = b_0(r_\s)^8 C_3(r_\s,\zeta),
\label{}
\end{equation}
\begin{equation}
a_4(r_\s,\zeta) = b_0(r_\s)^8 C_2(r_\s,\zeta) + 4 b_0(r_\s)^6 \varepsilon_\c^\text{PW92}(r_\s,\zeta),
\label{}
\end{equation}
\begin{equation}
a_5(r_\s,\zeta) = b_0(r_\s)^8 \varepsilon_\c^\text{PW92}(r_\s,\zeta),
\label{}
\end{equation}
\end{subequations}
where $\varepsilon_\c^\text{PW92}(r_\s,\zeta)$ is the PW92 parametrization of the UEG correlation energy per particle. The functions $C_i(r_\s,\zeta)$ are determined from the large-$\mu$ limit,
\begin{subequations}
\begin{equation}
C_2(r_\s,\zeta) = - \frac{3 g_\c(0,r_\s,\zeta)}{8 r_\s^3},
\label{}
\end{equation}
\begin{equation}
C_3(r_\s,\zeta) = - \frac{ g(0,r_\s,\zeta)}{\sqrt{2\pi} r_\s^3},
\label{}
\end{equation}
\begin{equation}
C_4(r_\s,\zeta) = - \frac{9 [g_\c''(0,r_\s,\zeta)+(1-\zeta^2) D_2(r_\s)]}{64 r_\s^3}, 
\label{}
\end{equation}
\begin{equation}
C_5(r_\s,\zeta) = - \frac{9 [g''(0,r_\s,\zeta)+(1-\zeta^2) D_3(r_\s)]}{40\sqrt{2\pi} r_\s^3}, 
\label{}
\end{equation}
\end{subequations}
where $g(0,r_\s,\zeta)$ is the on-top pair-distribution function\footnote{For a general system, the pair-distribution function $g(\b{r}_1,\b{r}_2)$ is defined from the pair density $\rho_2(\b{r}_1,\b{r}_2)$ [Eq.~(\ref{n2def})] as $\rho_2(\b{r}_1,\b{r}_2) = \rho(\b{r}_1) \rho(\b{r}_2) g(\b{r}_1,\b{r}_2)$. The on-top value is the value at electron coalescence, i.e. for $\b{r}_1=\b{r}_2$.} of the Coulombic UEG and $g''(0,r_\s,\zeta)$ is its second-order derivative with respect to $r_{12}$ at $r_{12}=0$, and similarly for their correlation parts $g_\c(0,r_\s,\zeta)=g(0,r_\s,\zeta)-(1-\zeta^2)/2$  and $g_\c''(0,r_\s,\zeta)=g''(0,r_\s,\zeta)-\phi_8(\zeta)/(5 \alpha^2 r_\s^2)$ with $\phi_8(\zeta)$ defined by Eq.~(\ref{phinzeta}). The $\zeta$-dependence of the latter quantities is assumed to be exchange-like, i.e. $g(0,r_\s,\zeta) \approx (1-\zeta^2) g(0,r_\s,\zeta=0)$ and $g''(0,r_\s,\zeta) \approx \zeta_+^2 g''(0,r_\s/\zeta_+^{1/3},\zeta=1) + \zeta_-^2 g''(0,r_\s/\zeta_-^{1/3},\zeta=1)$ where $\zeta_{\pm} = (1\pm\zeta)/2$. The on-top pair-distribution function has been parametrized in Ref.~\cite{GorPer-PRB-01} as
\begin{equation}
g(0,r_\s,\zeta=0) = (1-B r_\s + C r_\s^2 + D r_\s^3 + E r_\s^4) e^{-F r_\s},
\label{}
\end{equation}
with $B=0.7317 -d$, $C=0.08193$, $D=-0.01277$, $E=0.001859$, and $F=0.7524$. The remaining functions were determined by fitting to QMC data:
\begin{subequations}
\begin{equation}
 b_0(r_\s) = 0.784949 r_\s,
\label{}
\end{equation}
\begin{equation}
g''(0,r_\s,\zeta=1) = \frac{2^{5/3}}{5\alpha^2 r_\s^2} \frac{1-0.02267 r_\s}{1+0.4319 r_\s +0.04 r_\s^2},
\label{}
\end{equation}
\begin{equation}
D_2(r_\s) = \frac{e^{-0.547 r_\s}}{r_s^2} (-0.388 r_\s + 0.676 r_\s^2),
\label{}
\end{equation}
\begin{equation}
D_3(r_\s) = \frac{e^{-0.31 r_\s}}{r_s^3} (-4.95 r_\s + r_\s^2).
\label{}
\end{equation}
\end{subequations}

\vspace{0.4cm}
\noindent
{\bf Short-range PBE(GWS) exchange-correlation functional}

The Goll-Werner-Stoll (GWS) variant of the short-range PBE exchange-correlation functional~\cite{GolWerSto-PCCP-05,GolWerStoLeiGorSav-CP-06} is a slight modification of the short-range PBE functional developed in Ref.~\cite{TouColSav-JCP-05}. The exchange energy density is
\begin{eqnarray}
e_\x^{\sr,\mu,\text{PBE(GWS)}}(\rho,\nabla \rho) &=&  e_\x^{\sr,\mu,\text{UEG}} (\rho) F_\x(s,\mut),
\end{eqnarray}
with an enhancement factor of the same form as in the standard PBE exchange functional,
\begin{eqnarray}
F_\x(s,\mut) = 1 + \kappa - \frac{\kappa}{1+b(\mut) s^2/\kappa},
\end{eqnarray}
with $s=|\nabla \rho|/(2 k_\text{F} \rho)$ and $\mut=\mu/(2k_\text{F})$. In this expression, $\kappa=0.840$, as in the standard PBE exchange functional, to saturate the local Lieb--Oxford bound (for $\mu=0$) and $b(\mut)=b^\text{PBE} [b^\text{T}(\mut)/b^\text{T}(0)] e^{-\alpha_\text{x} \mut^2}$ where $b^\text{PBE}=0.21951$ is the second-order gradient-expansion coefficient of the standard PBE exchange functional, and $b^\text{T}(\mut)$ is a function coming from the second-order GEA of the short-range exchange energy~\cite{Tou-THESIS-05,TouColSav-JCP-05},
\begin{equation}
b^\text{T}(\mut) = \frac{-c_1(\mut) +c_2(\mut) e^{1/(4\mut^2)}}{c_3(\mut) +54 c_4(\mut) e^{1/(4\mut^2)}},
\label{berf}
\end{equation}
with $c_1(\mut)=1 +22\mut^2 +144\mut^4$, $c_2(\mut)=2\mut^2(-7+72\mut^2)$, $c_3(\mut)=-864\mut^4(-1+2\mut^2)$, and $c_4(\mut)=\mut^2[-3 -24\mut^2 +32\mut^4 +8\mut\sqrt{\pi}\erf(1/(2\mut))]$. Finally, $\alpha_\text{x}=19.0$ is a damping parameter optimized for the He atom. 

Similarly, the correlation energy density has the same form as the standard PBE correlation functional,
\begin{equation}
\bar{e}_\c^{\sr,\mu,\text{PBE(GWS)}}(\rho_\uparrow,\rho_\downarrow,\nabla \rho_\uparrow, \nabla \rho_\downarrow) =  \rho \left[ \bar{\varepsilon}_\c^{\sr,\mu,\text{UEG}} (\rho_\uparrow,\rho_\downarrow) + H^\mu(\rho_\uparrow,\rho_\downarrow,t) \right],
\label{}
\end{equation}
with $t=|\nabla \rho|/(2 \phi_2(\zeta) k_\s \rho)$ and the gradient correction 
\begin{eqnarray}
H^\mu(\rho_\uparrow,\rho_\downarrow,t) = A(0) \phi_2(\zeta)^3 \ln \left[ 1 + \frac{\beta(\mu)}{A(0)} t^2 \frac{1+ {\cal A}(\mu) t^2}{1+{\cal A}(\mu) t^2 + {\cal A}(\mu)^2 t^4} \right],
\label{}
\end{eqnarray}
where
\begin{eqnarray}
{\cal A}(\mu) = \frac{\beta(\mu)}{A(0)} \left[ \exp(-\bar{\varepsilon}_\c^{\sr,\mu,\text{UEG}}(\rho_\uparrow,\rho_\downarrow)/(A(0) \phi_2(\zeta)^3)) -1 \right]^{-1},
\label{}
\end{eqnarray}
and
\begin{eqnarray}
\beta(\mu) = \beta^\text{PBE} \left( \frac{\bar{\varepsilon}_\c^{\sr,\mu,\text{UEG}}(\rho_\uparrow,\rho_\downarrow)}{\bar{\varepsilon}_\c^{\sr,\mu=0,\text{UEG}}(\rho_\uparrow,\rho_\downarrow)} \right)^{\alpha_\c},
\label{betamu}
\end{eqnarray}
and the value of $A(0)$ is given after Eq.~(\ref{epscunifsmallrs}).
In Eq.~(\ref{betamu}), $\beta = 0.066725$ is the second-order gradient coefficient of the standard PBE correlation functional and $\alpha_\c = 2.78$ is a damping parameter optimized for the He atom. 

For $\mu=0$, this short-range PBE exchange-correlation functional reduces to the standard PBE exchange-correlation functional and for large $\mu$ it reduces to the short-range LDA exchange-correlation functional.

\subsubsection{Range-separated two-parameter double-hybrid approximations}

In 2018, Kalai and Toulouse~\cite{KalTou-JCP-18} introduced what we will call \textit{range-separated two-parameter double-hybrid approximations}, combining HF exchange and MP2 correlation with density functionals using a two-parameter decomposition of the electron-electron in a way reminiscent of the CAM decomposition [Eq.~(\ref{ExcCAM})] (see, also, Refs.~\cite{CorFro-IJQC-14,GarBulHenScu-PCCP-15}). This is based on a multideterminant extension of the KS scheme which generalizes the schemes of Sections~\ref{sec:dh} and ~\ref{sec:rs1dh}.

We first decompose the universal density functional of Eq.~(\ref{FnLevy}) as
\begin{eqnarray}
F[\rho] = \min_{\Psi \in {\cal W}^{N}_{\rho}} \bra{\Psi} \hat{T} + \hat{W}_\ee^{\lr,\mu} + \l \hat{W}_\ee^{\sr,\mu} \ket{\Psi} + \bar{E}_\Hxc^{\sr,\mu,\l}[\rho],
\label{FRS2DFT}
\end{eqnarray}
where the parameter $\mu \in [0,+\infty)$ controls the range of the separation as always, the parameter $\l \in [0,1]$ corresponds to the fraction of the short-range electron-electron interaction in the wave-function part, and $\bar{E}_\Hxc^{\sr,\mu,\l}[\rho]$ is the complementary short-range density functional defined to make this decomposition exact. As before, the latter functional can be decomposed as
\begin{eqnarray}
\bar{E}_\Hxc^{\sr,\mu,\l}[\rho] = E_\H^{\sr,\mu,\l}[\rho] + E_\x^{\sr,\mu,\l}[\rho] + \bar{E}_\c^{\sr,\mu,\l}[\rho].
\label{}
\end{eqnarray}
The Hartree and exchange contributions are linear in $\l$,
\begin{eqnarray}
E_\H^{\sr,\mu,\l}[\rho] = (1-\l) E_\H^{\sr,\mu}[\rho],
\label{}
\end{eqnarray}
\begin{eqnarray}
E_\x^{\sr,\mu,\l}[\rho] = (1-\l) E_\x^{\sr,\mu}[\rho],
\label{Exsrmul}
\end{eqnarray}
where $E_\H^{\sr,\mu}[\rho]$ and $E_\x^{\sr,\mu}[\rho]$ are the short-range Hartree and exchange functionals introduced in Section~\ref{sec:rs1dh}, and the correlation contribution can be written as
\begin{eqnarray}
\bar{E}_\c^{\sr,\mu,\l}[\rho] = E_\c[\rho] - E_\c^{\mu,\l}[\rho],
\label{}
\end{eqnarray}
where $E_\c[\rho]$ is the standard KS correlation functional and $E_\c^{\mu,\l}[\rho]$ is the correlation functional associated with the interaction $w_\text{ee}^{\lr,\mu}(r_{12}) + \l w_\text{ee}^{\sr,\mu}(r_{12})$. The exact ground-state energy can then be expressed as
\begin{eqnarray}
E_0 = \inf_{\Psi \in {\cal W}^{N}} \left\{ \bra{\Psi} \hat{T} + \hat{V}_\text{ne} + \hat{W}_\ee^{\lr,\mu} + \l \hat{W}_\ee^{\sr,\mu} \ket{\Psi} + \bar{E}_\Hxc^{\sr,\mu,\l}[\rho_\Psi] \right\},
\label{E0RS2DFT}
\end{eqnarray}
which constitutes a generalization of Eqs.~(\ref{E0multidethybrid}) and~(\ref{E0RSDFT}).

To obtain a MP2/DFT hybrid scheme, we proceed in full analogy to Sections~\ref{sec:dh} and~\ref{sec:rs1dh}. First, we define the following single-determinant range-separated two-parameter hybrid (RS2H) approximation,
\begin{eqnarray}
E^{\text{RS2H},\mu,\l}_0 &=& \inf_{\Phi \in {\cal S}^{N}} \Bigl\{\bra{\Phi}\hat{T}+\hat{V}_{\text{ne}}+\hat{W}_{\ee}^{\lr,\mu} + \l \hat{W}_\ee^{\sr,\mu} \ket{\Phi}+\bar{E}_{\Hxc}^{\sr,\mu,\l}[\rho_{\Phi}]\Bigl\}, 
\label{RS2H}
\end{eqnarray} 
and use it as a reference for defining a perturbation theory similarly to Eqs.~(\ref{Ela}) and~(\ref{Emua}). At second order, we obtain
\begin{eqnarray}
E^{\text{RS2H+MP2},\mu,\l}_0 = E^{\text{RS2H},\mu,\l}_0 + E^{\mu,\l,\MP}_\c,
\label{RS2H+MP2}
\end{eqnarray}
where $E^{\mu,\l,\MP}_\c$ is the MP2 correlation energy expression evaluated with RS2H spin orbitals and orbital eigenvalues, and the two-electron integrals associated with the interaction $w_\text{ee}^{\lr,\mu}(r_{12}) + \l w_\text{ee}^{\sr,\mu}(r_{12})$. This RS2H+MP2 scheme thus contains the exchange-correlation energy contribution
\begin{eqnarray}
E^{\text{RS2H+MP2},\mu,\l}_{\xc} &=& E_\x^{\lr,\mu,\HF}[\Phi^{\mu,\l}] + \l E_\x^{\sr,\mu,\HF}[\Phi^{\mu,\l}] + (1-\l) E_\x^{\sr,\mu}[\rho_{\Phi^{\mu,\l}}] 
\nonumber\\
&& + \bar{E}^{\sr,\mu,\l}_\c[\rho_{\Phi^{\mu,\l}}] + E_\c^{\mu,\l,\text{MP2}},
\label{ExcRS2H+MP2}
\end{eqnarray} 
where $\Phi^{\mu,\l}$ is a minimizing single-determinant wave function in Eq.~(\ref{RS2H}).

A good approximation for the $\l$-dependence of the complementary correlation functional $\bar{E}^{\sr,\mu,\l}_\c[\rho]$ is~\cite{KalTou-JCP-18}
\begin{eqnarray}
\bar{E}^{\sr,\mu,\l}_\c[\rho] \approx \bar{E}^{\sr,\mu}_\c[\rho] - \l^2 \bar{E}^{\sr,\mu \sqrt{\l}}_\c[\rho],
\label{Ecsrmul}
\end{eqnarray} 
where $\bar{E}_\c^{\sr,\mu}[\rho]$ is the short-range correlation functional introduced in Section~\ref{sec:rs1dh}. In particular, the $\l$-dependence in Eq.~(\ref{Ecsrmul}) is correct both in the high-density limit, for a non-degenerate KS system, and in the low-density limit. Thanks to Eqs.~(\ref{Exsrmul}) and~(\ref{Ecsrmul}), the semilocal density-functional approximations for $E^{\sr,\mu}_\x[\rho]$ and $\bar{E}^{\sr,\mu}_\c[\rho]$ of Section~\ref{sec:rs1dh} can be reused here without developing new ones. In Ref.~\cite{KalTou-JCP-18}, the short-range PBE(GWS) exchange and correlation functionals were used, and the optimal parameters $\mu=0.46$ bohr$^{-1}$ and $\l=0.58$ were found on small sets of atomization energies and reaction barrier heights, i.e. values similar to the ones usually used separately in range-separated hybrids and double hybrids.

The RS2H+MP2 scheme improves a bit over the RSH+MP2 scheme of Section~\ref{sec:rs1dh}, particularly for interaction energies of hydrogen-bonded systems. Even if the presence of short-range MP2 correlation deteriorates in principle the convergence rate with the one-electron basis size, in practice the fraction of pure short-range MP2 correlation ($\l^2 \approx 0.34$) is small enough to keep a fast basis convergence. Accuracy can be improved, particularly for dispersion interactions, by supplanting the MP2 term by coupled-cluster or random-phase approximations~\cite{KalMusTou-JCP-19}. Like for the approach of Section~\ref{sec:rs1dh}, many wave-function methods could be used in the general scheme of Eq.~(\ref{E0RS2DFT}).

\section{Semiempirical dispersion corrections and nonlocal van der Waals density functionals}

Among the previously considered exchange-correlation approximations, only the range-separated double hybrids of Section~\ref{sec:rsdh}, thanks to their long-range nonlocal correlation component, are capable of fully describing London dispersion interactions, crucial for describing weakly bonded systems. To improve the other approximations (semilocal functionals, single-determinant hybrids, double hybrids without range separation) for weakly bonded systems, it has been proposed to add to them a semiempirical dispersion correction or a nonlocal van der Waals density functional. We now describe these approaches.

\subsection{Semiempirical dispersion corrections}

To explicitly account for London dispersion interactions, it has been proposed in the 2000s to add to the standard approximate functionals a \textit{semiempirical dispersion correction} of the form~\cite{ElsHobFraSuhKax-JCP-01,WuYan-JCP-02,Gri-JCC-04,Gri-JCC-06}
\begin{eqnarray}
E_\text{disp} = - s \sum_{\alpha=1}^{N_\text{n}} \sum_{\substack{\beta=1\\ \beta>\alpha}}^{N_\text{n}} f(R_{\alpha \beta}) \frac{C_{6}^{\alpha\beta}}{R_{\alpha \beta}^6},
\label{Edisp}
\end{eqnarray} 
where $R_{\alpha \beta}$ is the distance between each pair of atoms and $C_{6}^{\alpha\beta}$ is the London dispersion coefficient between these atoms. Here, $f(R_{\alpha \beta})$ is a damping function which tends to 1 at large $R_{\alpha \beta}$ and tends to zero at small $R_{\alpha \beta}$, e.g.
\begin{eqnarray}
f(R_{\alpha \beta})=\frac{1}{1+e^{-d (R_{\alpha \beta} / R_{\alpha \beta}^\text{vdW} -1)}},
\label{fdamp}
\end{eqnarray} 
with the sum of tabulated atomic van der Waals radii $R_{\alpha \beta}^\text{vdW} = R_{\alpha}^\text{vdW} + R_{\beta}^\text{vdW}$ and a constant $d$, and $s$ is a scaling parameter that can be adjusted for each approximate functional. The dispersion coefficient $C_{6}^{\alpha\beta}$ for any pair of atoms is empirically calculated from tabulated same-atom dispersion coefficients $C_{6}^{\alpha\alpha}$ and/or atomic polarizabilities. This approach was named ``DFT-D'' by Grimme~\cite{Gri-JCC-04}. 

The last version of DFT-D (referred to as DFT-D3) also includes $C_8^{\alpha \beta}$ two-body terms and $C_9^{\alpha \beta \gamma}$ three-body terms~\cite{GriAntEhrKri-JCP-10}. There have also been various proposals to make the determination of dispersion coefficients less empirical, such as the scheme of Becke and Johnson~\cite{BecJoh-JCP-07} based on the exchange-hole dipole moment, the scheme of Tkatchenko and Scheffler~\cite{TkaSch-PRL-09} based on a Hirshfeld atomic partitioning, or the scheme of Sato and Nakai~\cite{SatNak-JCP-10} based on the local-response approximation~\cite{DobDin-PRL-96}.

The ``DFT-D'' approach provides a big and inexpensive improvement for the description of weakly bonded systems. One limitation is that the semiempirical dispersion correction, being just a force field in its simplest variant, affects only the molecular geometry of the system but not directly its electronic structure. Some of the most used DFT-D functionals are:
\begin{itemize}
\item The PBE-D exchange-correlation functional~\cite{Gri-JCC-06}, based on the PBE functional with a scaling parameter $s=0.75$;
\item The B97-D exchange-correlation functional~\cite{Gri-JCC-06}, based on the B97-GGA functional with a scaling parameter $s=1.25$ and reoptimized linear coefficients in Eqs.~(\ref{gx}),~(\ref{gcud}), and~(\ref{gcss}) in the presence of the semiempirical dispersion correction;
\item The B3LYP-D exchange-correlation functional~\cite{Gri-JCC-06}, based on the B3LYP hybrid functional with a scaling parameter $s=1.05$;
\item The $\omega$B97X-D exchange-correlation functional~\cite{ChaHea-PCCP-08}, based on the $\omega$B97X range-separated hybrid functional with a scaling parameter $s=1$, a modified damping function, and reoptimized parameters in $\omega$B97X in the presence of the semiempirical dispersion correction.
\end{itemize}

The semiempirical dispersion correction can also be added to double-hybrid approximations. For example, B2PLYP-D~\cite{SchGri-PCCP-07} is based on the B2PLYP double hybrid with a scaling parameter $s=0.55$. The scaling parameter is small since the fraction of MP2 correlation in B2PLYP already partially takes into account dispersion interactions. It has also been proposed to add a semiempirical dispersion correction to the SCS version of the double hybrids [Eq.~(\ref{2DHSCS})], resulting in a family of dispersion-corrected spin-component-scaled double-hybrid (DSD) approximations~\cite{KozGruMar-JPCC-10,KozMar-PCCP-11,KozMar-JCC-13}. An example of double hybrid is this latter family is DSD-BLYP~\cite{KozGruMar-JPCC-10} which uses the B88 exchange functional and the LYP correlation functional.

\subsection{Nonlocal van der Waals density functionals}

Another approach to describe dispersion interactions is to add to the standard approximate functionals a so-called \textit{nonlocal van der Waals density functional} of the form~\cite{DioRydSchLan-PRL-04,VydVoo-JCP-09,VydVoo-PRL-09,LeeMurKonLunLan-PRB-10,VydVoo-JCP-10b}
\begin{eqnarray}
E^\text{nl}_\c[\rho] = \frac{1}{2} \int_{\mathbb{R}^3\times\mathbb{R}^3} \rho(\b{r}_1) \rho(\b{r}_2)  \phi(\b{r}_1,\b{r}_2) \d \b{r}_1 \d \b{r}_2,
\label{Ecnl}
\end{eqnarray} 
where $\phi(\b{r}_1,\b{r}_2)$ is a correlation kernel. Two main families of such nonlocal correlation functionals exist: the ``van der Waals density functionals'' (vdW-DF)~\cite{DioRydSchLan-PRL-04,LeeMurKonLunLan-PRB-10} and the Vydrov-Van Voorhis (VV) functionals~\cite{VydVoo-JCP-09,VydVoo-PRL-09,VydVoo-JCP-10b}.

We will only describe the last version of the VV functionals, i.e. the VV10 nonlocal correlation functional~\cite{VydVoo-JCP-10b}. In this functional, the correlation kernel is taken as
\begin{eqnarray}
\phi^\text{VV10}(\b{r}_1,\b{r}_2) = -\frac{3}{2g(\b{r}_1,r_{12}) g(\b{r}_2,r_{12})(g(\b{r}_1,r_{12}) + g(\b{r}_2,r_{12}))} + \beta \delta(\b{r}_1-\b{r}_2),
\label{}
\end{eqnarray} 
where $r_{12} = |\b{r}_2 - \b{r}_1|$ is the interelectronic distance, $\beta$ is a constant determining the local (delta-distribution) part of the kernel, and the function $g$ is defined as
\begin{eqnarray}
g(\b{r},r_{12}) = \omega_0(\b{r}) r_{12}^2 + \kappa(\b{r}).
\label{grR}
\end{eqnarray} 
In Eq.~(\ref{grR}), $\omega_0(\b{r}) = \sqrt{\omega_\text{g}(\b{r})^2 + \frac{\omega_\text{p}(\b{r})^2}{3}}$ involves the square of the local plasma frequency $\omega_\text{p}(\b{r})^2 = 4 \pi \rho(\b{r})$ and the square of the local band gap $\omega_\text{g}(\b{r})^2 =  C |\nabla \rho(\b{r}) |^4/\rho(\b{r})^4$ where $C$ is an adjustable parameter controlling the large-$r_{12}$ asymptotic dispersion coefficients, and $\kappa(\b{r})=b \; k_\text{F}(\b{r})^2/\omega_\text{p}(\b{r})$ where $k_\text{F}(\b{r}) = (3\pi^2 \rho(\b{r}))^{1/3}$ is the local Fermi wave vector and $b$ is an adjustable parameter controlling the short-range damping of the large-$r_{12}$ asymptote. As expected for dispersion interactions, in the large-$r_{12}$ limit, $\phi^\text{VV10}(\b{r}_1,\b{r}_2)$ behaves as $1/r_{12}^6$:
\begin{eqnarray}
\phi^\text{VV10}(\b{r}_1,\b{r}_2)  \isEquivTo{r_{12} \to \infty} -\frac{3}{2 \omega_0(\b{r}_1) \omega_0(\b{r}_2)(\omega_0(\b{r}_1) + \omega_0(\b{r}_2)) r_{12}^6}.
\label{}
\end{eqnarray} 
The constant $\beta = (3/b^2)^{3/4}/16$ is chosen to make $E^\text{nl}_\c[\rho]$ vanish in the uniform density limit, leaving thus this limit unchanged when $E^\text{nl}_\c[\rho]$ is added to another density functional. The adjustable parameters $C \approx 0.009$ and $b \approx 6$ are found by optimization of $C_6$ dispersion coefficients and of weak intermolecular interaction energies, respectively, the precise values depending on which exchange-correlation functional the VV10 correction is added to.

Nonlocal van der Waals density functionals are necessarily more computationally expensive than semiempirical dispersion corrections. However, they have the advantage of being less empirical and, since they are functionals of the density, of impacting directly on the electronic structure of the system. The VV10 nonlocal functional has been incorporated in a number of recently developed exchange-correlation functionals, for example:
\begin{itemize}
\item The $\omega$B97X-V exchange-correlation functional~\cite{MarHea-PCCP-14}, based on the $\omega$B97X range-separated hybrid [Eq.~(\ref{ExcwB97X})] with reoptimized linear coefficients in Eq.~(\ref{gx}) with polynomial degree $m=2$ and in Eqs.~(\ref{gcud}) and~(\ref{gcss}) with polynomial degree $m=1$, as well as reoptimized VV10 parameters $C=0.01$ and $b=6.0$;

\item The $\omega$B97M-V exchange-correlation functional~\cite{MarHea-JCP-16}, based on the $\omega$B97X range-separated hybrid [Eq.~(\ref{ExcwB97X})] but with more general and combinatorially optimized meta-GGA exchange and correlation enhancement factors and the same VV10 parameters $C=0.01$ and $b=6.0$ as in $\omega$B97X-V.
\end{itemize}

\section{Orbital-dependent exchange-correlation density functionals}
\stepcounter{myequation}
\label{sec:orbdep}

We discuss here some exchange-correlation density functionals explicitly depending on the KS orbitals (for a review, see Ref.~\cite{KumKro-RMP-08}). Since the KS orbitals are themselves functionals of the density, these exchange-correlation expressions are thus {\it implicit} functionals of the density (for notational simplicity, this dependence on the density of the orbitals and other intermediate quantities will not be explicitly indicated). In fact, the single-determinant and multideterminant hybrid approximations of Sections~\ref{sec:sdha} and~\ref{sec:mdha} already belong to this family, with the caveat that the orbitals are obtained with a {\it nonlocal} potential. In this section, we are concerned with orbital-dependent exchange-correlation energy functionals with orbitals obtained with a local potential, i.e. staying within the KS scheme\footnote{The boundary between the various single-determinant and multideterminant hybrids of Sections~\ref{sec:sdha} and~\ref{sec:mdha} and the orbital-dependent functionals of the present Section is however thin. For example, it is possible to optimize the orbitals using a local potential in hybrids or range-separated hybrids~\cite{Arb-JSC-07,KimHonHwaRyuChoKim-PCCP-17,SmiCon-JPCA-20}, and in double hybrids or range-separated double hybrids~\cite{SmiFraMusBukGraLupTou-JCP-16,SmiGraWitMusTou-JCTC-20}.}. These approximations tend to be more computationally involved than the approximations previously seen and have thus been much less used so far.

\subsection{Exact exchange}
\label{sec:EXX}

The {\it exact exchange} (EXX) energy functional [Eq.~(\ref{Exn})] can be expressed in terms of the KS orbitals,
\begin{equation}
E_{\x}[\rho] =- \frac{1}{2} \sum_{\sigma\in\{\uparrow,\downarrow\}} \sum_{i=1}^{N_\sigma} \sum_{j=1}^{N_\sigma} \int_{\mathbb{R}^3\times\mathbb{R}^3} \frac{\varphi_{i \sigma}^*(\b{r}_1)\varphi_{j \sigma}(\b{r}_1) \varphi_{j \sigma}^*(\b{r}_2)\varphi_{i \sigma}(\b{r}_2)}{|\b{r}_1-\b{r}_2|}\d\b{r}_1 \d\b{r}_2,
\label{EXX}
\end{equation}
and has exactly the same form as the HF exchange [Eq.~(\ref{ExHF})], but the orbitals used in both expressions are in general different.

Since the exact exchange energy in Eq.~(\ref{EXX}) is not an explicit functional of the density, the corresponding exchange potential $v_\x(\b{r}) = \delta E_\x[\rho]/\delta \rho(\b{r})$ cannot be calculated directly. We can however find an workable equation for $v_\x(\b{r})$ by first considering the functional derivative of $E_{\x}[\rho]$ with respect to the KS potential $v_\s(\b{r})$ and then applying the chain rule:
\begin{equation}
\frac{\delta E_{\x}[\rho]}{\delta v_\s(\b{r})} = \int_{\mathbb{R}^3}  \frac{\delta E_{\x}[\rho]}{\delta \rho(\b{r}')} \frac{\delta \rho(\b{r}')}{\delta v_\s(\b{r})} \d\b{r}'.
\label{OEPEXX1}
\end{equation}
Introducing the non-interacting KS static linear-response function $\chi_0(\b{r}',\b{r}) = \delta \rho(\b{r}')/\delta v_\s(\b{r})$, we can rewrite Eq.~(\ref{OEPEXX1}) as
\begin{equation}
\int_{\mathbb{R}^3} v_\x(\b{r}') \chi_0(\b{r}',\b{r}) \d\b{r}' = \frac{\delta E_{\x}[\rho]}{\delta v_\s(\b{r})},
\end{equation}
which is known as the {\it optimized-effective-potential} (OEP) equation for the exact-exchange potential~\cite{TalSha-PRA-76,GorLev-PRA-94,GorLev-IJQC-95}. 

Using first-order perturbation theory on the KS system, explicit expressions in terms of the orbitals can be derived for $\chi_0(\b{r}',\b{r})$ and $\delta E_{\x}[\rho]/\delta v_\s(\b{r})$. The expression of $\chi_0 (\b{r}',\b{r})$ is
\begin{equation}
\chi_0 (\b{r}',\b{r}) =  -\sum_{\sigma\in\{\uparrow,\downarrow\}} \sum_{i=1}^{N_\sigma} \sum_{a \geq N_\sigma+1} \frac{\varphi_{i\sigma}^*(\b{r}')\varphi_{a\sigma}^*(\b{r}) \varphi_{i\sigma}(\b{r}) \varphi_{a\sigma}(\b{r}')}{\varepsilon_{a\sigma} - \varepsilon_{i\sigma}} + \text{ c.c. },
\end{equation}
where $i$ and $a$ run over occupied and virtual spatial orbitals, respectively, and c.c. stands for the complex conjugate. The expression of $\delta E_{\x}[\rho]/\delta v_\s(\b{r})$ is
\begin{equation}
\frac{\delta E_{\x}[\rho]}{\delta v_\s(\b{r})} = \sum_{\sigma\in\{\uparrow,\downarrow\}} \sum_{i=1}^{N_\sigma} \sum_{j=1}^{N_\sigma} \sum_{a \geq N_\sigma+1} \braket{\varphi_{a\sigma}\varphi_{j\sigma}}{\varphi_{j\sigma}\varphi_{i\sigma}} 
 \frac{\varphi_{a\sigma}(\b{r}) \varphi_{i\sigma}^*(\b{r})}{\varepsilon_{a\sigma} - \varepsilon_{i\sigma}}  + \text{ c.c. },
\end{equation}
where $\braket{\varphi_{a\sigma}\varphi_{j\sigma}}{\varphi_{j\sigma}\varphi_{i\sigma}}$ are two-electron integrals over KS spatial orbitals:
\begin{eqnarray}
\braket{\varphi_{a\sigma}\varphi_{j\sigma}}{\varphi_{j\sigma}\varphi_{i\sigma}} = \int_{\mathbb{R}^3\times \mathbb{R}^3} \frac{\varphi_{a\sigma}^*(\b{r}_1) \varphi_{j\sigma}^*(\b{r}_2) \varphi_{j\sigma}(\b{r}_1) \varphi_{i\sigma}(\b{r}_2)}{|\b{r}_1-\b{r}_2|} \d\b{r}_1 \d\b{r}_2.
\label{2eintorb}
\end{eqnarray}

Applying this OEP method with the EXX energy (and no correlation energy functional) is an old idea~\cite{ShaHor-PR-53,TalSha-PRA-76}, but reasonably efficient calculations for molecules have been possible only relatively recently~\cite{IvaHirBar-PRL-99,Gor-PRL-99}. The EXX occupied orbitals turn out to be very similar to the HF occupied orbitals, and thus the EXX ground-state properties are also similar to the HF ones. However, the EXX virtual orbitals (which see a $-1/r$ asymptotic potential for a neutral system) tend to be much less diffuse than the HF virtual orbitals (which see an exponentially decaying potential for a neutral system), and may be more adapted for calculating excited-state properties.

\subsection{Second-order G\"orling--Levy perturbation theory}
\label{sec:GLPT2}

In 1993, G\"orling and Levy~\cite{GorLev-PRB-93,GorLev-PRA-94} developed a perturbation theory in terms of the coupling constant $\l$ of the adiabatic connection (Section~\ref{sec:adiabatic}) which provides an explicit orbital-dependent second-order approximation for the correlation energy functional. The Hamiltonian along the adiabatic connection [Eq.~(\ref{Hl})] can be written as
\begin{eqnarray}
\hat{H}^{\l} &=& \hat{T} + \l \hat{W}_{\ee} + \hat{V}^\l
\nonumber\\
&=&\hat{H}_\s + \l (\hat{W}_{\ee} - \hat{V}_{\H \x} ) - \hat{V}_{\c}^\l,
\label{HleqHsp}
\end{eqnarray}
where $\hat{H}_\s = \hat{H}^{\l=0} = \hat{T} + \hat{V}_\s$ is the KS non-interacting reference Hamiltonian (which will be assumed to have a nondegenerate ground state). Equation~(\ref{HleqHsp}) was obtained by decomposing the potential operator keeping the density constant as $\hat{V}^\l= \hat{V}_\s - \l \hat{V}_{\H\x} - \hat{V}_{\c}^\l$ where $\hat{V}_\s=\hat{V}^{\l=0}$ is the KS potential operator, $\l \hat{V}_{\H \x}$ is the Hartree-exchange potential operator which is linear in $\l$, and $\hat{V}_{\c}^\l$ is the correlation potential which starts at second order in $\l$, i.e. $\hat{V}_{\c}^\l= \l^2 \hat{V}_{\c}^{(2)} + \cdots$. Using a complete set of orthonormal eigenfunctions $\Phi_n$ and eigenvalues ${\cal E}_n$ of the KS Hamiltonian, $\hat{H}_\s \ket{\Phi_n} = {\cal E}_n \ket{\Phi_n}$, the normalized ground-state wave function of the Hamiltonian $\hat{H}^{\l}$ can be expanded as $\Psi^{\l} = \Phi + \l \Psi^{(1)} + \cdots$ where $\Phi=\Phi_0$ is the ground-state KS single-determinant wave function and $\Psi^{(1)}$ is its first-order correction given by
\begin{eqnarray}
\ket{\Psi^{(1)}} = - \sum_{n\not=0} \frac{\bra{\Phi_n} \hat{W}_\ee - \hat{V}_{\H \x}\ket{\Phi}}{{\cal E}_n -{\cal E}_0} \ket{\Phi_n}.
\label{}
\end{eqnarray}

Using the expression in Eq.~(\ref{Ecl}), the correlation energy functional can also be expanded in powers of $\l$:
\begin{eqnarray}
E_{\c}^{\l}[\rho] &=& \bra{\Psil} \hat{T} + \l \hat{W}_{\ee} \ket{\Psil} - \bra{\Phi} \hat{T} + \l \hat{W}_{\ee} \ket{\Phi}.
\nonumber\\
              &=& E_{\c}^{(0)} + \l E_{\c}^{(1)} + \l^2 E_{\c}^{(2)} + \cdots.
\label{Eclexpand}
\end{eqnarray}
Since $\Psi^{\l=0}=\Phi$, the zeroth-order term vanishes: $E_{\c}^{(0)} =0$. Using the expression of the first-order derivative of $E_{\c}^{\l}$ with respect to $\l$ in Eq.~(\ref{dEcldl}), i.e. $\partial E_{\c}^{\l}/\partial \l=\bra{\Psil} \hat{W}_{\ee} \ket{\Psil} - \bra{\Phi} \hat{W}_{\ee} \ket{\Phi}$, we find that the first-order term vanishes as well: $E_{\c}^{(1)}=0$. The second-order term corresponds to the {\it second-order G\"orling--Levy (GL2) correlation energy} and is given by
\begin{equation}
E_{\c}^{\GL}[\rho] = E_{\c}^{(2)} = \bra{\Phi} \hat{W}_{\ee} \ket{\Psi^{(1)}} = \bra{\Phi} \hat{W}_{\ee} -{V}_{\H \x} \ket{\Psi^{(1)}},
\label{Ec2Phi0WPhi1}
\end{equation}
where the second equality comes from the fact that $\bra{\Phi} \hat{V}_{\H \x} \ket{\Psi^{(1)}} = 0$ since it is the derivative with respect to $\l$ at $\l=0$ of $\bra{\Psi^\l} \hat{V}_{\H \x} \ket{\Psi^{\l}} = \int_{\mathbb{R}^3} v_{\H \x}(\b{r}) \rho(\b{r}) \d \b{r}$ which does not depend on $\l$ by virtue of the fact that the density $\rho(\b{r})$ is constant along the adiabatic connection. Using the last expression in Eq.~(\ref{Ec2Phi0WPhi1}) allows one to express the GL2 correlation energy as
\begin{equation}
E_{\c}^{\GL}[\rho] = - \sum_{n\not=0} \frac{|\bra{\Phi} \hat{W}_\ee - \hat{V}_{\H \x}\ket{\Phi_n}|^2}{{\cal E}_n -{\cal E}_0}.
\label{EcGL2}
\end{equation}
It is instructive to decompose the GL2 correlation energy as
\begin{equation}
E_{\c}^{\GL}[\rho] = E_\c^{\MP} + E_\c^{\text{S}},
\label{EcGL2decomp}
\end{equation}
where $E_\c^{\MP}$ is a MP2-like correlation energy evaluated with KS spin orbitals,
\begin{equation}
E_\text{c}^{\text{MP2}}  = - \frac{1}{4} \sum_{i=1}^{N}\sum_{j=1}^{N} \sum_{a \geq N+1} \sum_{b \geq N+1} \frac{|\bra{\phi_i\phi_j}\ket{\phi_a\phi_b}|^2}{\varepsilon_a+\varepsilon_b-\varepsilon_i-\varepsilon_j},
\label{EcMP2bis}
\end{equation}
and $E_\c^{\text{S}}$ is the contribution coming from the single excitations (which does not vanish here, contrary to HF-based MP2 perturbation theory),
\begin{equation}
E_\text{c}^{\text{S}}  = - \sum_{i=1}^{N} \sum_{a \geq N+1} \frac{|\bra{\phi_i} \hat{V}_\x^{\HF} - \hat{V}_\x \ket{\phi_a}|^2}{\varepsilon_a-\varepsilon_i},
\label{Ecsingles}
\end{equation}
involving the difference between the integrals over the nonlocal HF exchange potential $\bra{\phi_i} \hat{V}_\x^{\HF} \ket{\phi_a} = - \sum_{j=1}^{N} \braket{\phi_i\phi_j}{\phi_j\phi_a}$ and over the local KS exchange potential $\bra{\phi_i} \hat{V}_\x \ket{\phi_a}= \int_{\mathbb{R}^3\times\{\uparrow,\downarrow\}} \phi_i^*(\b{x})v_\x(\b{r}) \phi_a(\b{x}) \d\b{x}$.

Calculations of the GL2 correlation energy using either a non-self-consistent post-EXX implementation or a more complicated OEP self-consistent procedure have been tested (see, e.g., Refs.~\cite{GraHirIvaBar-JCP-02,Eng-INC-03,MorWuYan-JCP-05}) but the results are often disappointing. It is preferable to go beyond second order with random-phase approximations in the adiabatic-connection fluctuation-dissipation approach.

\subsection{Adiabatic-connection fluctuation-dissipation approach}

\subsubsection{Exact adiabatic-connection fluctuation-dissipation expression}

Using the adiabatic-connection formula of Eq.~(\ref{EcintPsi}), the correlation energy functional can be written as
\begin{eqnarray}
E_{\c}[\rho] &=& \int_{0}^{1} \d\l \; \bra{\Psil} \hat{W}_{\ee} \ket{\Psil} - \bra{\Phi} \hat{W}_{\ee} \ket{\Phi}
= \frac{1}{2} \int_{0}^{1} \d\l \int_{\mathbb{R}^3\times\mathbb{R}^3} \frac{\rho_{2,\c}^{\l}(\b{r}_1,\b{r}_2)}{|\b{r}_1 -\b{r}_2|} \d\b{r}_1  \d\b{r}_2,
\label{EcAC}
\end{eqnarray}
where $\rho_{2,\c}^{\l}(\b{r}_1,\b{r}_2) = \rho_{2}^{\l}(\b{r}_1,\b{r}_2) - \rho_{2,\KS}(\b{r}_1,\b{r}_2)$ is the correlation part of the pair density along the adiabatic connection. The pair density $\rho_{2}^{\l}(\b{r}_1,\b{r}_2)$ can be expressed with the pair-density operator $\hat{\rho}_2(\b{r}_1,\b{r}_2) = \hat{\rho}(\b{r}_1) \hat{\rho}(\b{r}_2) - \delta(\b{r}_1-\b{r}_2) \hat{\rho}(\b{r}_1)$ where $\hat{\rho}(\b{r})$ is the density operator,
\begin{eqnarray}
\rho_{2}^{\l}(\b{r}_1,\b{r}_2) = \bra{\Psi^\l} \hat{\rho}_2(\b{r}_1,\b{r}_2) \ket{\Psi^\l} = \bra{\Psi^\l} \hat{\rho}(\b{r}_1) \hat{\rho}(\b{r}_2) \ket{\Psi^\l} - \delta(\b{r}_1-\b{r}_2) \bra{\Psi^\l} \hat{\rho}(\b{r}_1) \ket{\Psi^\l}, 
\end{eqnarray}
and the KS pair density $\rho_{2,\KS}(\b{r}_1,\b{r}_2)$ simply corresponds to the case $\l=0$,
\begin{eqnarray}
\rho_{2,\KS}(\b{r}_1,\b{r}_2) = \rho_{2}^{\l=0}(\b{r}_1,\b{r}_2) = \bra{\Phi} \hat{\rho}(\b{r}_1) \hat{\rho}(\b{r}_2) \ket{\Phi} - \delta(\b{r}_1-\b{r}_2) \bra{\Phi} \hat{\rho}(\b{r}_1) \ket{\Phi}. 
\end{eqnarray}
Since the density does not change with $\l$, i.e. $\bra{\Psi^\l} \hat{\rho}(\b{r}) \ket{\Psi^\l} = \bra{\Phi} \hat{\rho}(\b{r}) \ket{\Phi} = \rho(\b{r})$, the correlation pair density needed in Eq.~(\ref{EcAC}) can thus be expressed as
\begin{eqnarray}
\rho_{2,\c}^{\l}(\b{r}_1,\b{r}_2) = \bra{\Psi^\l} \hat{\rho}(\b{r}_1) \hat{\rho}(\b{r}_2) \ket{\Psi^\l} - \bra{\Phi} \hat{\rho}(\b{r}_1) \hat{\rho}(\b{r}_2) \ket{\Phi}.
\end{eqnarray}

We would like to calculate $\rho_{2,\c}^{\l}(\b{r}_1,\b{r}_2)$ without having to calculate the complicated many-body wave function $\Psi^\l$. For this, we consider the retarded dynamic linear-response function along the adiabatic connection in frequency space (or the so-called Lehmann representation)
\begin{eqnarray}
\chi_\l(\b{r}_1,\b{r}_2;\omega) = \sum_{n\not= 0} \frac{\bra{\Psi^\l} \hat{\rho}(\b{r}_1) \ket{\Psi_n^\l} \bra{\Psi_n^\l}\hat{\rho}(\b{r}_2 ) \ket{\Psi^\l}}{\omega - \omega_n^\l + i0^+} - \frac{\bra{\Psi^\l} \hat{\rho}(\b{r}_2) \ket{\Psi_n^\l} \bra{\Psi_n^\l}\hat{\rho}(\b{r}_1 ) \ket{\Psi^\l}}{\omega + \omega_n^\l + i0^+},
\label{chilLehmann}
\end{eqnarray}
where the sums are over all eigenstates $\Psi_n^\l$ of the Hamiltonian $\hat{H}^\l$, i.e. $\hat{H}^\l \ket{\Psi_n^\l} = E_n^\l \ket{\Psi_n^\l}$, except the ground state $\Psi^\l=\Psi^\l_0$, and $\omega_n^\l = E_n^\l - E_0^\l$ are the corresponding excitation energies. By contour integrating $\chi_\l(\b{r}_1,\b{r}_2,\omega)$ around the right half $\w$-complex plane, we arrive at the (zero-temperature) \textit{fluctuation-dissipation theorem},
\begin{eqnarray}
n_{2,\c}^{\l}(\b{r}_1,\b{r}_2) =  - \int_{-\infty}^{+\infty} \frac{\d\omega}{2\pi} \; [\chi_\l(\b{r}_1,\b{r}_2,i\omega) - \chi_0(\b{r}_1,\b{r}_2,i\omega) ],
\label{FDT}
\end{eqnarray}
which relates ground-state correlations in the time-independent system $\rho_{2,\c}^{\l}(\b{r}_1,\b{r}_2)$ to the linear response of the system due to a time-dependent external perturbation $\chi_\l(\b{r}_1,\b{r}_2,\omega)$.

Combining Eqs.~(\ref{EcAC}) and~(\ref{FDT}), we finally obtain the exact \textit{adiabatic-connection fluctuation-dissipation} (ACFD) formula for the correlation energy~\cite{LanPer-SSC-75,LanPer-PRB-77} (see, also, Ref.~\cite{HarJon-JPF-74}):
\begin{eqnarray}
E_{\c}[\rho] = -\frac{1}{2} \int_{0}^{1} \d\l \int_{-\infty}^{+\infty} \frac{\d\omega}{2\pi} \int_{\mathbb{R}^3\times\mathbb{R}^3} \frac{\chi_\l(\b{r}_1,\b{r}_2;i\omega) - \chi_0(\b{r}_1,\b{r}_2;i\omega) }{|\b{r}_1 -\b{r}_2|} \d\b{r}_1  \d\b{r}_2.
\label{EcACFD}
\end{eqnarray}
The usefulness of the ACFD formula is due to the fact that there are practical ways of directly calculating $\chi_\l(\b{r}_1,\b{r}_2;\omega)$ without having to calculate the many-body wave function $\Psi^\l$. In linear-response time-dependent density-functional theory (TDDFT), one can find a Dyson-like equation for $\chi_\l(\b{r}_1,\b{r}_2;\omega)$,
\begin{eqnarray}
\chi_\l(\b{r}_1,\b{r}_2;\omega) = \chi_0(\b{r}_1,\b{r}_2;\omega) + \int_{\mathbb{R}^3\times\mathbb{R}^3} \chi_0(\b{r}_1,\b{r}_3;\omega) f_\Hxc^\l(\b{r}_3,\b{r}_4;\omega) \chi_\l(\b{r}_4,\b{r}_2;\omega) \d\b{r}_3 \d\b{r}_4,
\label{chilDyson}
\end{eqnarray}
where $f_\Hxc^\l(\b{r}_3,\b{r}_4;\omega)$ is the {\it Hartree-exchange-correlation kernel} associated to the Hamiltonian ${H}^\l$. Here, Eq.~(\ref{chilDyson}) will be considered as the definition for $f_\Hxc^\l$. In principle, the exact correlation energy can be obtained with Eqs.~(\ref{EcACFD}) and (\ref{chilDyson}). In practice, however, we need to use an approximation for $f_\Hxc^\l$.

\subsubsection{Random-phase approximations}

In the \textit{direct random-phase approximation} (dRPA, also just referred to as RPA, or sometimes as time-dependent Hartree), only the Hartree part of the kernel, which is linear in $\l$ and independent from $\omega$, is retained~\cite{LanPer-SSC-75,LanPer-PRB-80},
\begin{eqnarray}
f_\Hxc^{\dRPA,\l}(\b{r}_1,\b{r}_2;\omega) = f_\H^\l(\b{r}_1,\b{r}_2) = \l w_\text{ee}(\b{r}_1,\b{r}_2),
\end{eqnarray}
where $w_\text{ee}(\b{r}_1,\b{r}_2)=1/|\b{r}_1 -\b{r}_2|$ is the Coulomb interaction, and the corresponding dRPA linear-response function then satisfies the equation
\begin{eqnarray}
\chi_{\l}^{\dRPA}(\b{r}_1,\b{r}_2;\omega) = \chi_0(\b{r}_1,\b{r}_2;\omega) + \l \int_{\mathbb{R}^3\times\mathbb{R}^3} \chi_0(\b{r}_1,\b{r}_3;\omega) w_\ee(\b{r}_3,\b{r}_4) \chi_\l^\dRPA(\b{r}_4,\b{r}_2;\omega) \d\b{r}_3 \d\b{r}_4.
\label{childRPADyson}
\end{eqnarray}
The physical contents of this approximation can be seen by iterating Eq.~(\ref{childRPADyson}) which generates an infinite series,
\begin{eqnarray}
\chi_{\l}^{\dRPA}(\b{r}_1,\b{r}_2;\omega) = \chi_0(\b{r}_1,\b{r}_2;\omega) + \l \int_{\mathbb{R}^3\times\mathbb{R}^3} \chi_0(\b{r}_1,\b{r}_3;\omega) w_\ee(\b{r}_3,\b{r}_4) \chi_0(\b{r}_4,\b{r}_2;\omega) \d\b{r}_3 \d\b{r}_4 \;\;\;\;\;\;\;\;\;
\nonumber\\
+ \l^2 \int_{\mathbb{R}^3\times\mathbb{R}^3\times\mathbb{R}^3\times\mathbb{R}^3} \chi_0(\b{r}_1,\b{r}_3;\omega) w_\ee(\b{r}_3,\b{r}_4) \chi_0(\b{r}_4,\b{r}_5;\omega) w_\ee(\b{r}_5,\b{r}_6) \chi_0(\b{r}_6,\b{r}_2;\omega) \d\b{r}_3 \d\b{r}_4 \d\b{r}_5 \d\b{r}_6 + \cdots,
\nonumber\\
\end{eqnarray}
which, after plugging it into Eq.~(\ref{EcACFD}), leads to the dRPA correlation energy as the following perturbation expansion\footnote{Using the operator viewpoint, the series in Eq.~(\ref{EcdRPA}) can be formally summed in the form $E_{\c}^{\dRPA}[\rho] = 1/(4\pi) \int_{-\infty}^{+\infty}  \d\omega \; \Tr[ \ln(1 - \chi_0(i\omega) w_\ee)+ \chi_0(i\omega) w_\ee ]$ (see, e.g., Ref.~\cite{MusRocJanAng-JCTC-16}).}
\begin{eqnarray}
E_{\c}^{\dRPA}[\rho] = -\frac{1}{2} \int_{0}^{1} \d\l \int_{-\infty}^{+\infty} \frac{\d\omega}{2\pi} \Biggl[ \l \int_{\mathbb{R}^3\times\mathbb{R}^3\times\mathbb{R}^3\times\mathbb{R}^3} \frac{\chi_0(\b{r}_1,\b{r}_3;i\omega) \chi_0(\b{r}_4,\b{r}_2;i\omega)}{|\b{r}_1 -\b{r}_2|\; |\b{r}_3 -\b{r}_4|} \d\b{r}_1  \d\b{r}_2 \d\b{r}_3 \d\b{r}_4 \;\;\;\;\;\;\;\;\;
\nonumber\\
+ \l^2 \int_{\mathbb{R}^3\times\mathbb{R}^3\times\mathbb{R}^3\times\mathbb{R}^3\times\mathbb{R}^3\times\mathbb{R}^3} \frac{\chi_0(\b{r}_1,\b{r}_3;i\omega) \chi_0(\b{r}_4,\b{r}_5;i\omega) \chi_0(\b{r}_6,\b{r}_2;i\omega)}{|\b{r}_1 -\b{r}_2|\; |\b{r}_3 -\b{r}_4| \; |\b{r}_5 -\b{r}_6|} \d\b{r}_1  \d\b{r}_2 \d\b{r}_3 \d\b{r}_4 \d\b{r}_5 \d\b{r}_6+ \cdots \Biggl]. \;\;\;
\label{EcdRPA}
\end{eqnarray}
Using now the Lehmann representation [Eq.~(\ref{chilLehmann})] of the KS dynamic linear-response function in terms of the KS orbitals and their energies,
\begin{eqnarray}
\chi_{0}(\b{r}_1,\b{r}_2;\omega) &=& \sum_{\sigma\in\{\uparrow,\downarrow\}} \sum_{i=1}^{N_\sigma}\sum_{a \geq N_\sigma+1}\Biggl[
\frac{\varphi_{i\sigma}^*(\b{r}_1)\varphi_{a\sigma}(\b{r}_1)\varphi_{a\sigma}^*(\b{r}_2)\varphi_{i\sigma}(\b{r}_2) }
{\omega -(\varepsilon_{a\sigma} -\varepsilon_{i\sigma}) +i0^+}
\nonumber\\
&&
\phantom{xxxxxxxxxxxxxxxxxx}
-
\frac{\varphi_{i\sigma}^*(\b{r}_2)\varphi_{a\sigma}(\b{r}_2)\varphi_{a\sigma}^*(\b{r}_1)\varphi_{i\sigma}(\b{r}_1) }
{\omega +(\varepsilon_{a\sigma} -\varepsilon_{i\sigma}) +i0^+} \Biggl],
\label{Chi0r1r2}
\end{eqnarray}
one can obtain, after quite some work,
\begin{eqnarray}
E_{\c}^{\dRPA}[\rho] &=& -\frac{1}{2}  \sum_{i=1}^{N} \sum_{j=1}^{N} \sum_{a \geq N+1} \sum_{b \geq N+1} \frac{|\braket{\phi_i\phi_j}{\phi_a\phi_b}|^2}
{\varepsilon_{a} + \varepsilon_{b} - \varepsilon_{i} - \varepsilon_{j}}
\nonumber\\
&&+ \sum_{i=1}^{N} \sum_{j=1}^{N} \sum_{k=1}^{N} \sum_{a \geq N+1} \sum_{b \geq N+1} \sum_{c \geq N+1} 
 \frac{\braket{\phi_i\phi_j}{\phi_a\phi_b}\braket{\phi_j\phi_k}{\phi_b\phi_c}\braket{\phi_k\phi_i}{\phi_c\phi_a}}{(\varepsilon_{a} + \varepsilon_{b} - \varepsilon_{i} - \varepsilon_{j}) (\varepsilon_{a} + \varepsilon_{c} - \varepsilon_{i} - \varepsilon_{k})} + \cdots.
\nonumber\\
\label{EcdRPA2}
\end{eqnarray}
The dRPA correlation energy is the sum of all the direct terms (i.e., no exchange terms) of the perturbation expansion up to infinite order. In the language of diagrammatic perturbation theory, we say that the dRPA correlation energy is the sum of all direct ring diagrams. Of course, Eq.~(\ref{EcdRPA2}) is not the way to calculate the dRPA correlation energy in practice. This is done by solving the Dyson equation [Eq.~(\ref{childRPADyson})] without explicitly expanding in powers of $\l$, e.g. using matrix equations from linear-response TDDFT~\cite{Fur-PRB-01,TouZhuAngSav-PRA-10} or coupled-cluster theory~\cite{ScuHenSor-JCP-08,TouZhuSavJanAng-JCP-11}.

Most dRPA correlation energy (combined with the EXX energy) calculations are done in a non-self-consistent way, but self-consistent OEP dRPA calculations have also been performed~\cite{HelRohGro-JCP-12,BleHesGor-JCP-13}. One of the main advantage of dRPA is that it accounts for long-range dispersion interactions~\cite{DobMcLRubWanGouLeDin-AJC-01,DobWanDinMclLe-IJQC-05,DobGou-JPCM-12}. However, it shows large self-interaction errors. To overcome the latter drawback and improve the general accuracy, one can add exchange and beyond terms in various ways (see, e.g., Refs.~\cite{TouGerJanSavAng-PRL-09,GruMarHarSchKre-JCP-09,TouZhuAngSav-PRA-10,HelBar-JCP-10,HesGor-MP-10,AngLiuTouJan-JCTC-11,HesGor-PRL-11,Hes-PRA-12,BatFur-JCP-13,ColHelGir-PRB-14,MusRocJanAng-JCTC-16,ErhBleGor-PRL-16,HelColGir-PRB-18,BatSenSenRuz-JCTC-18,HumGruKreZie-JCTC-19,Gor-PRB-19}). This remains an active area of research. For reviews on random-phase approximations, the reader may consult Refs.~\cite{EshBatFur-TCA-12,RenRinJoaSch-JMS-12,CheVooAgeGanBalFur-ARPC-17}.

\section*{Acknowledgements}
I thank \'Eric Canc\`es and Gero Friesecke for discussions and comments on the manuscript. This review chapter grew out of my lecture notes for DFT courses given in several summer schools (ISTPC summer schools in June 2015 and 2017 in France, ICS summer school in July 2015 in France, and ESQC summer school in September 2019 in Italy). 
This work has received funding from the European Research Council (ERC) under the European Union's Horizon 2020 research and innovation programme (grant agreement EMC2 No 810367).

\addcontentsline{toc}{section}{References}

\end{document}